\documentclass[
aps
,prx
,amsmath,amssymb,amsfonts
,floatfix
,longbibliography
,twocolumn,10pt
]{revtex4-1}
\usepackage[utf8]{inputenc}
\usepackage{graphicx}
\usepackage{color}
\usepackage[sf]{subfigure}
\usepackage{fouriernc}
\usepackage[dvipsnames,svgnames,x11names,hyperref]{xcolor}
\usepackage[low-sup]{subdepth}
\usepackage{MnSymbol}

\definecolor{linkcolor}{RGB}{6,69,173} 
\definecolor{diffcolor}{RGB}{175,31,36} 

\usepackage[colorlinks=true,
            linkcolor=linkcolor,
            urlcolor=linkcolor,
            citecolor=linkcolor,
            unicode,
            pdfencoding=auto]{hyperref}
\usepackage{etoolbox}
\usepackage{enumitem}

\usepackage[
]{physpack}

\begin{document}

\title{Topological Superconductivity in Dirac Honeycomb Systems}
\author{Kyungmin Lee}
\thanks{The first two authors contributed equally to this work}
\affiliation{Department of Physics, The Ohio State University, Columbus, Ohio 43210, USA}
\author{Tamaghna Hazra}
\thanks{The first two authors contributed equally to this work}
\affiliation{Department of Physics, The Ohio State University, Columbus, Ohio 43210, USA}
\author{Mohit Randeria}
\affiliation{Department of Physics, The Ohio State University, Columbus, Ohio 43210, USA}
\author{Nandini Trivedi}
\affiliation{Department of Physics, The Ohio State University, Columbus, Ohio 43210, USA}

\begin{abstract}
We predict two topological superconducting phases in microscopic models arising from the Berry phase associated with the valley degree of freedom in gapped Dirac honeycomb systems.
The first one is a topological helical spin-triplet superconductor with a nonzero center-of-mass momentum that does not break time-reversal symmetry. 
We also find a topological chiral-triplet superconductor with Chern number $\pm 1$ with equal-spin-pairing in one valley and opposite-spin-triplet pairing in the other valley.
Our results are obtained for the Kane-Mele model in which we have explored the effect of three different interactions, onsite attraction $U$, nearest-neighbor density-density attraction $V$, and nearest-neighbor antiferromagnetic exchange $J$, within self-consistent Bogoliubov--de Gennes theory.
Transition metal dichalcogenides and cold atom experiments are promising platforms to explore these phases.
\end{abstract}

\maketitle

\section{Introduction}

\para{}
A topological superconductor (SC) has a superconducting gap in the bulk but protected Majorana fermions on the boundaries or in the cores of vortices in an externally applied magnetic field \cite{qi-rmp-2011,sato-rpp-2017}. 
There has been considerable excitement about the search for topological superconductors in recent years.

\para{}
While signatures of topological superconductivity have been observed in one-dimensional chains with proximity-induced superconductivity~\cite{mourik-s-2012,nadj-perge-s-2014}, the experimental search for topological superconductivity in two dimensions is a promising~\cite{lian-pnas-2018} and relatively unexplored territory~\cite{he-science-2017,menard-nc-2017,palacio-morales-a-2018,yin-np-2015}

\para{}
The honeycomb lattice, with special features of Dirac dispersion and opposite Berry curvature around the two inequivalent valleys in the Brillouin zone, has emerged as a paradigmatic system for exploring topological states.
In this paper, we extend these investigations to include attractive interactions between electrons and outline a route to topological superconductivity, highlighting the crucial role played by the Berry phase and valley degree of freedom.

\para{}
Transition metal dichalcogenides (TMDs) with the valley degree of freedom are a viable family of materials in the search for topological superconductivity. 
TMDs are layered materials containing a transition metal layer that form a triangular layer sandwiched between two chalcogen layers.
Based on density functional theory (DFT) calculations that indicate considerable $d$-$p$ mixing between the chalcogen and transition metal ions~\cite{fang-prb-2015},
we expect the effective Hamiltonian to reduce to a honeycomb model, similar to graphene, but with the richness of strong spin-orbit coupling and interactions between electrons.

\para{}
In TMD materials like MoS$_2$ and WS$_2$~\cite{xiao-prl-2012}, superconductivity is observed below ${\sim}$10\,K~\cite{ye-s-2012,lu-s-2015,lu-pnas-2018}, although these appear to be trivial SCs. 
Other TMD materials like $1T'$-WTe$_2$ exhibit gapless edge states, suggesting that they are topological insulators~\cite{fei-np-2017}.
WTe$_2$ is reported to become superconducting under pressure~\cite{kang-nc-2015,pan-nc-2015} and gating~\cite{sajadi-s-2018,fatemi-s-2018}, though whether it is a topological superconductor is still unclear. Also, more recently, magic angle twisted bilayer graphene~\cite{cao-n-2018} has emerged as a model system for understanding superconductivity in the strongly correlated regime.
References~\onlinecite{po-prx-2018,yuan-prb-2018,kang-prx-2018} suggest that, despite the concentration of charge
density on a triangular lattice, the low-energy physics is that of a Dirac honeycomb system.
This is also true for the naturally occurring layered mineral jacutingaite, Pt$_2$HgSe$_3$ where the low-energy physics is dominated by the Hg atoms on a honeycomb lattice \cite{marrazzo-prl-2018}, resulting in a room-temperature quantum spin Hall insulator with a gap of 110\,meV \cite{kandra-a-2019}. 
Preliminary theoretical investigations suggest the possibility of unconventional superconductivity when gated/doped to the van Hove singularities in the band structure~\cite{wu-a-2018}.
The question of intrinsic topological superconductivity in this system is as yet unexplored.

\para{}
Given these motivations, we examine the superconducting states that emerge in
the Kane-Mele model~\cite{kane-prl-2005} as a result of various interactions.
This is the archetypal model on a honeycomb lattice that exhibits
a transition from a topological to a trivial insulator as a function of spin-orbit coupling~(see Fig.~\ref{fig:lattice}).
What are the superconducting instabilities of this gapped Dirac system?
Under what conditions do we get topological superconducting states?
These are the primary questions we address in this paper.

\begin{figure*}[htb!]
\centering
\subfigure[\label{fig:pd-os}]{\includegraphics[height=2.2in]{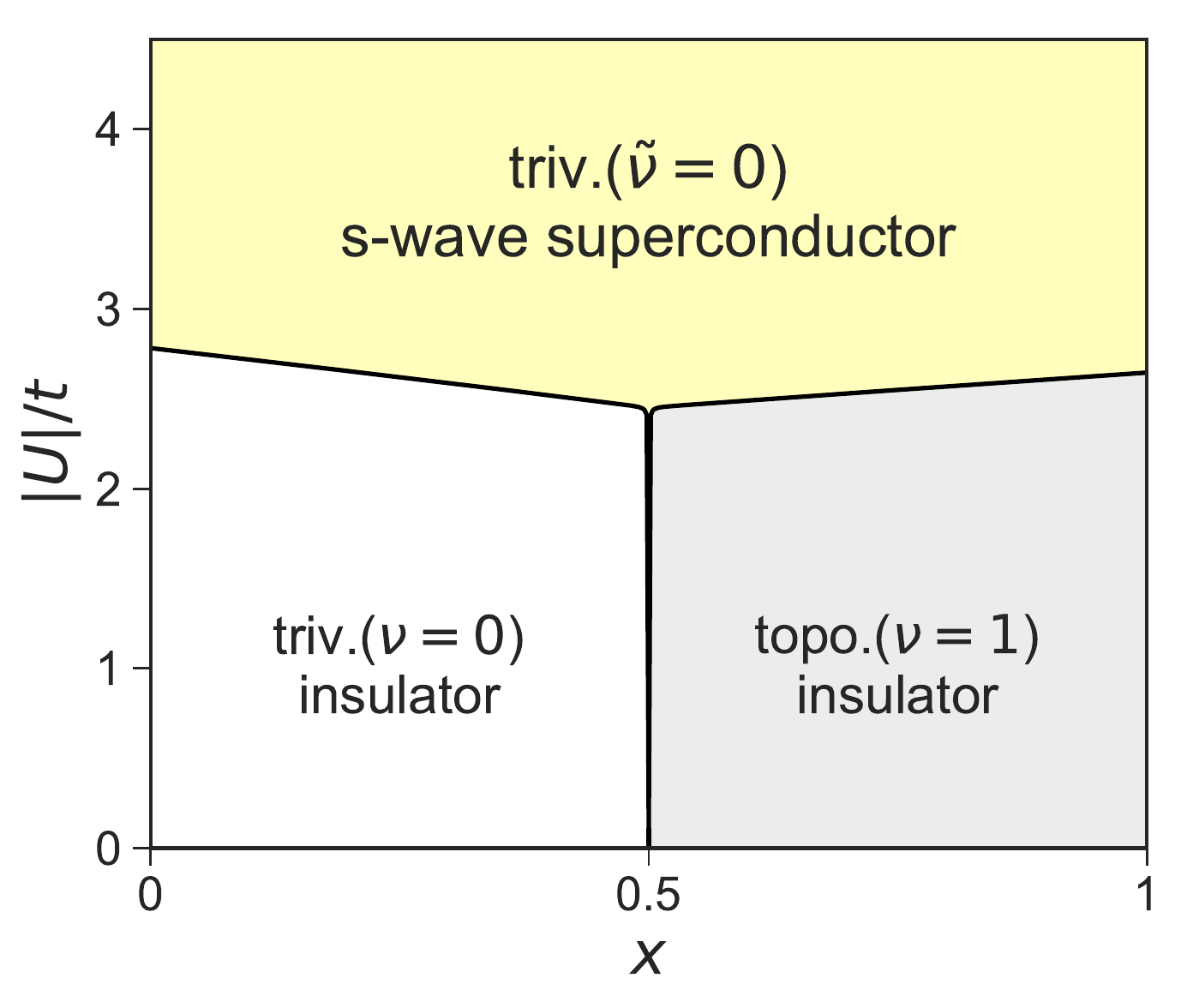}}
\qquad
\subfigure[\label{fig:pd-nn}]{\includegraphics[height=2.2in]{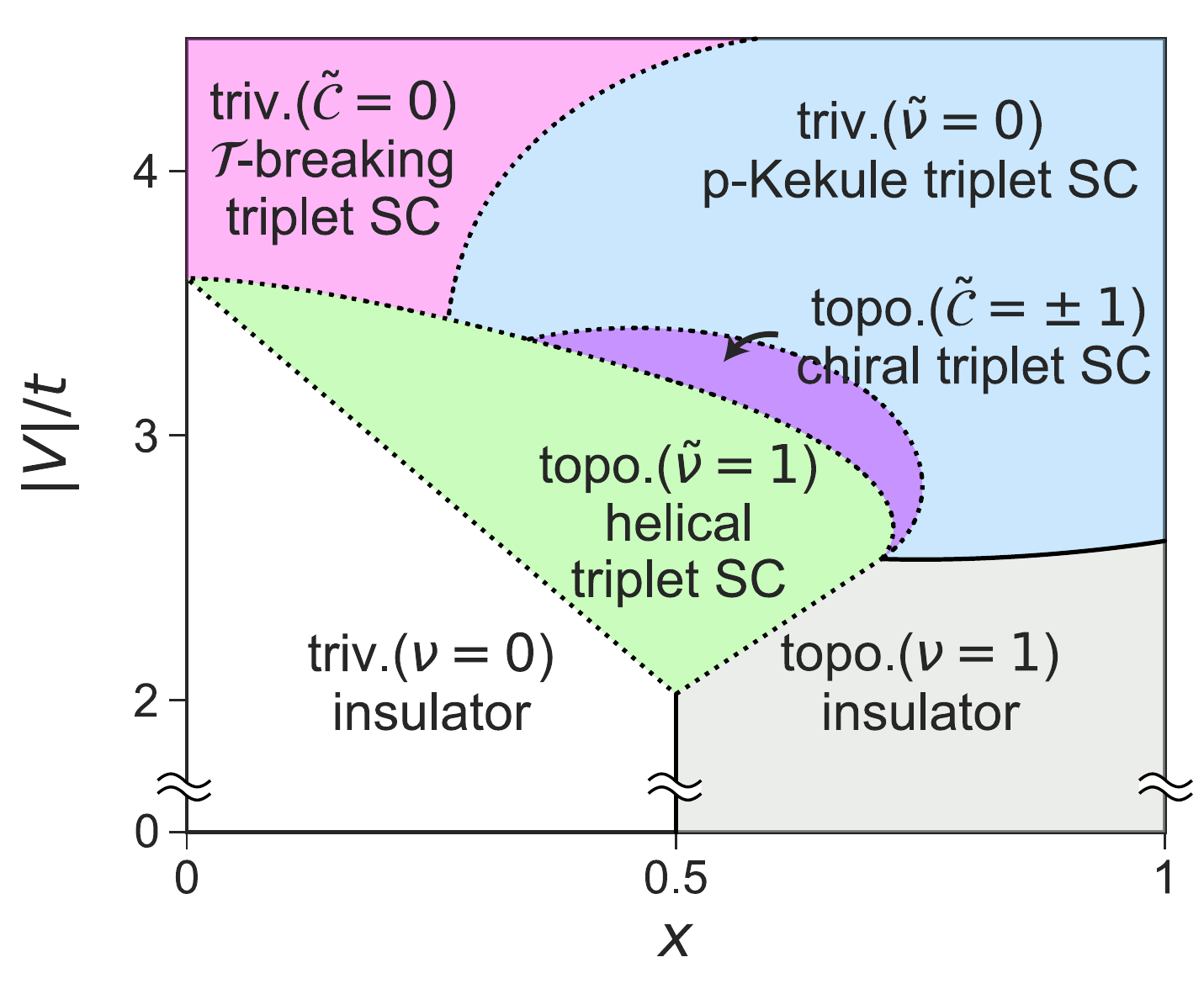}}
\caption{\label{fig:pd}%
Phase diagrams of Kane-Mele model in Eq.~\eqref{eq:kanemele} as functions of the tuning parameter $x = 3\sqrt{3}\lambda_{\text{so}}/(m_{\text{AB}}+3\sqrt{3}\lambda_{\text{so}})$ which interpolates between the trivial and topological insulating band structures, with \subref{fig:pd-os} onsite attractive interaction $U$,
\subref{fig:pd-nn} nearest-neighbor attractive density-density interaction $V$.
Solid lines mark continuous (topological) phase transitions, and the dotted lines mark first order transitions.
\subref{fig:pd-os} With $U$, we find an $s$-wave pairing state that is topologically trivial.
\subref{fig:pd-nn} With $V$, we find more exotic pairing states, two of which are topological:
The topological helical triplet superconductor(SC) (in green) near $x = 1/2$  has equal-spin spin-triplet pairing ($\Delta_{\up\up}, \Delta_{\dn\dn} \neq 0$) and a $\calT$-invariant topological superconducting ground state with $\tilde{\nu}=1$.
The trivial $p$-Kekule triplet SC (in blue) near $x=1$ has spin-triplet pairing between opposite spins ($d^z \neq 0$), is $\calT$-invariant and is topologically trivial.
Both of these states have nonzero center-of-mass momentum pairs, with non-trivial real-space patterns in the pairing order parameters, shown in Fig.~\ref{fig:nn-pairing-pattern}.
The other two superconducting phases (shown in purple and in pink) have a mixture of both types of triplet pairing and are $\calT$-breaking.
The topological chiral triplet SC (in purple) is a topological state with Chern number $\tilde{\calC}=\pm 1$.
The trivial $\calT$-breaking triplet SC (in pink) on the other hand is topologically trivial with $\tilde{\calC}=0$.
}
\end{figure*}

\para{}
We use self-consistent Bogoliubov--de Gennes theory to map out the phase diagrams of the Kane-Mele model with three different types of interactions, and analyze the topological invariants associated with the resulting superconducting phases.
Throughout this paper, we will use the terms ``trivial,'' and ``topological'' to refer to zero and nonzero topological invariants of the corresponding symmetry class.
For the three types of interactions, we find the following:
\begin{enumerate}[label=(\roman*)]
\item We show that onsite attraction, irrespective of whether the parent insulator is topological or trivial, the resulting superconductor is non-topological~[see Fig.~\ref{fig:pd-os}].
\item For nearest-neighbor attraction, topological superconductivity can arise from both the trivial as well as the topological insulator, and is most prominent near the transition~[see Fig.~\ref{fig:pd-nn}].
\item With antiferromagnetic nearest-neighbor interaction, we find exotic singlet states with broken rotation, translation, and time-reversal symmetries; however, none of these states are topological (see Fig.~\ref{fig:pd-heis}).
\end{enumerate}

\para{}
Our most significant results on topological superconducting states pertain to 
Fig.~\ref{fig:pd-nn}] where we 
find that two of the four superconducting states are topological, a time-reversal-symmetric helical superconductor and a chiral superconductor with Chern number $\pm 1$ that breaks time-reversal.
These topological states involve pairing within the same Dirac cone, and are stabilized when the underlying band structure is close to the transition between the topological and the trivial insulating phases.

\para{}
The topological superconducting states we find are different from those discussed in the literature.
For example, unlike $^3$He-B, the helical superconductor we predict has a nonzero center-of-mass (c.m.) momentum due to the valley degree of freedom.
The chiral superconductor too is different from the proposed paired state for the spinless $\nu=\frac{5}{2}$ quantum Hall state with Chern number 1, or the $p\pm ip$ superconducting state in spinful Sr$_2$RuO$_4$ or in $^3$He-A that have a Chern number of $\pm$2.
The chiral SC we predict is composed of a condensate of equal-spin pairs with nonzero c.m. momentum, and another condensate of opposite-spin pairs with the c.m. momentum reversed.

\para{}
In the final section, we compare our results with previous theoretical works on superconductivity in TMDs, and also comment on the implications of our results for cold atom experiments.

\section{Kane-Mele Model with Interactions}
\label{sec:model}

\begin{figure}\centering 
\subfigure[\label{fig:honeycomb}]{%
\includegraphics[width=1.25in]{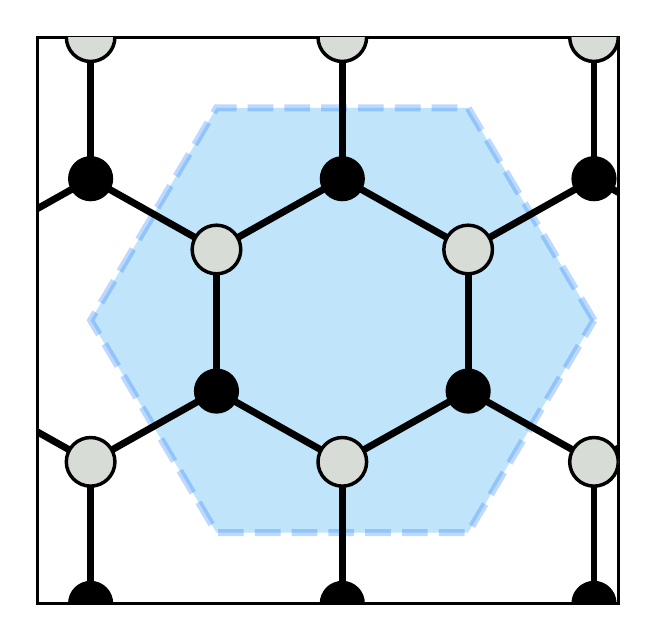}%
}\qquad%
\subfigure[\label{fig:brillouinzone}]{%
\includegraphics[width=1.25in]{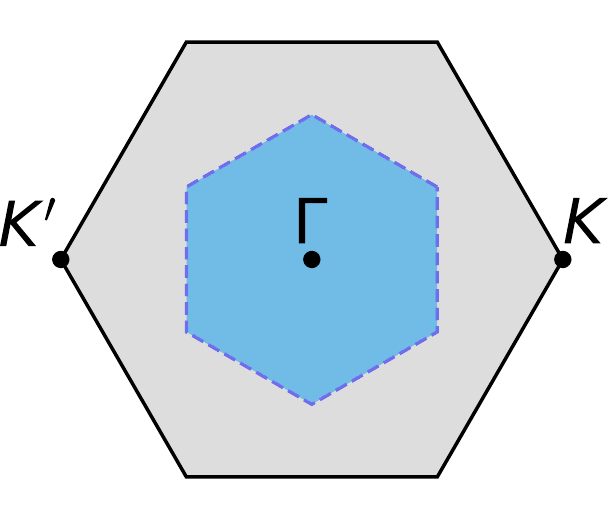}%
}
\subfigure[\label{fig:dispersion-trivial}]{%
\frame{\includegraphics[height=0.82in]{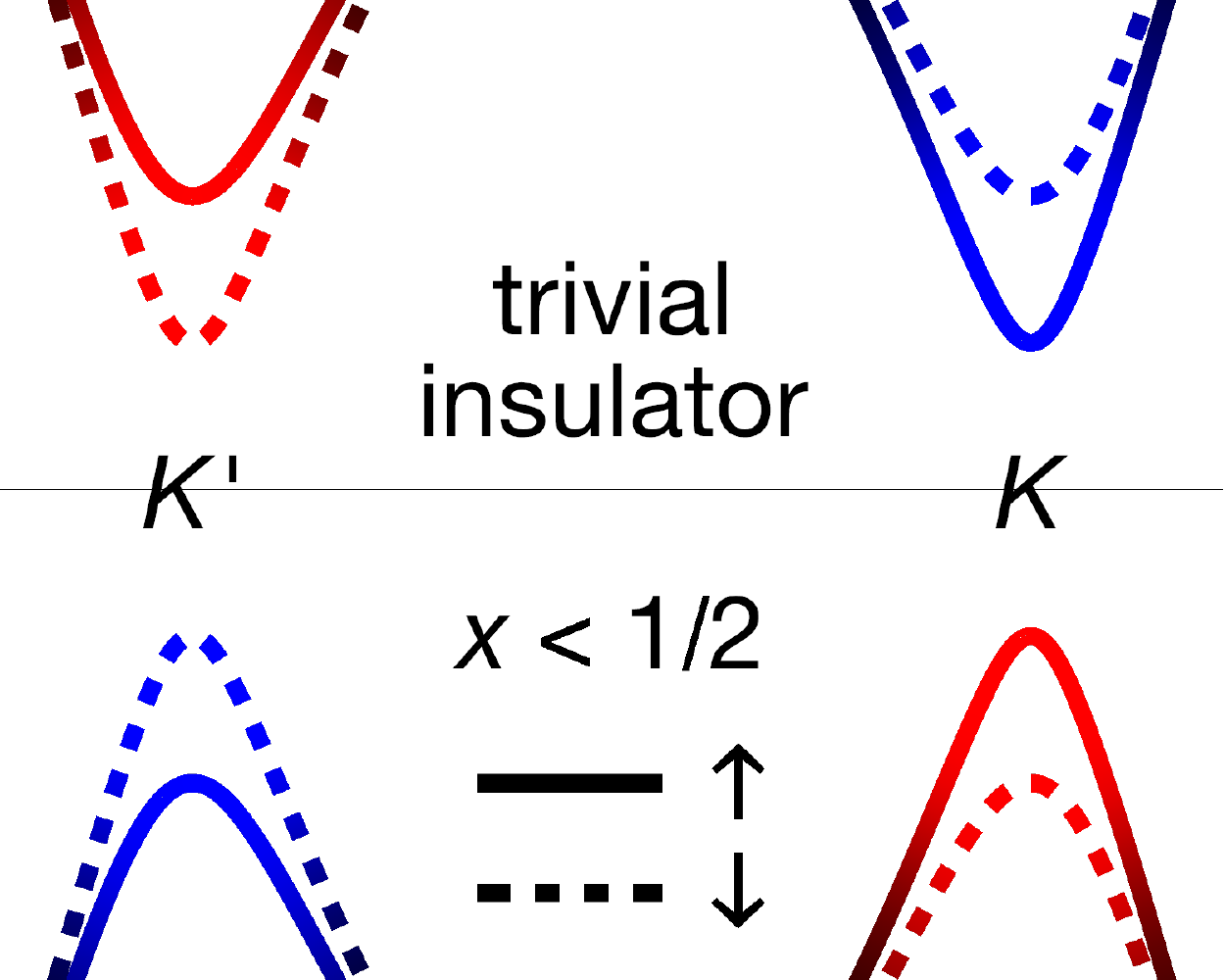}}%
}
\subfigure[\label{fig:dispersion-gapclosing}]{%
\frame{\includegraphics[height=0.82in]{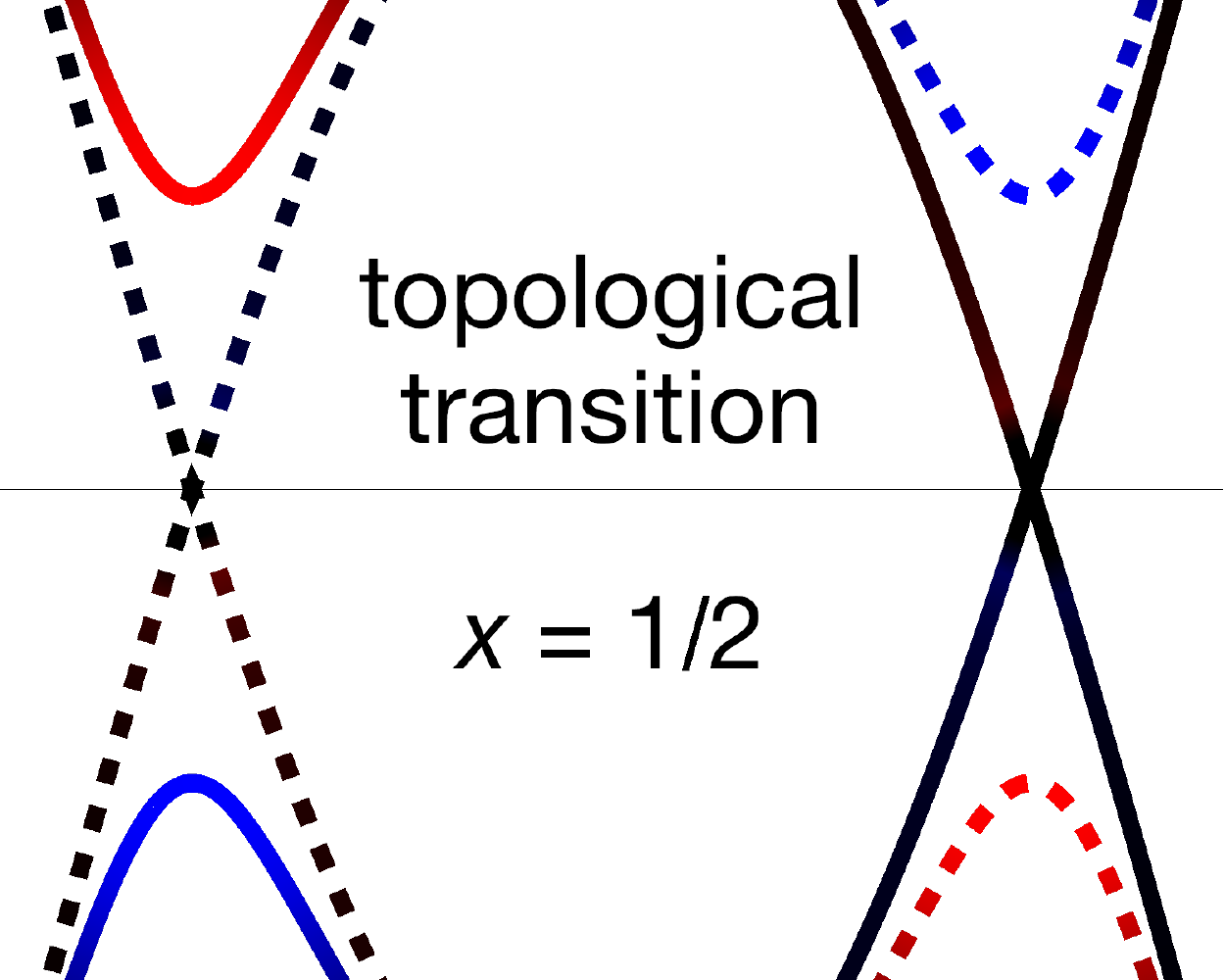}}%
}
\subfigure[\label{fig:dispersion-topological}]{%
\frame{\includegraphics[height=0.82in]{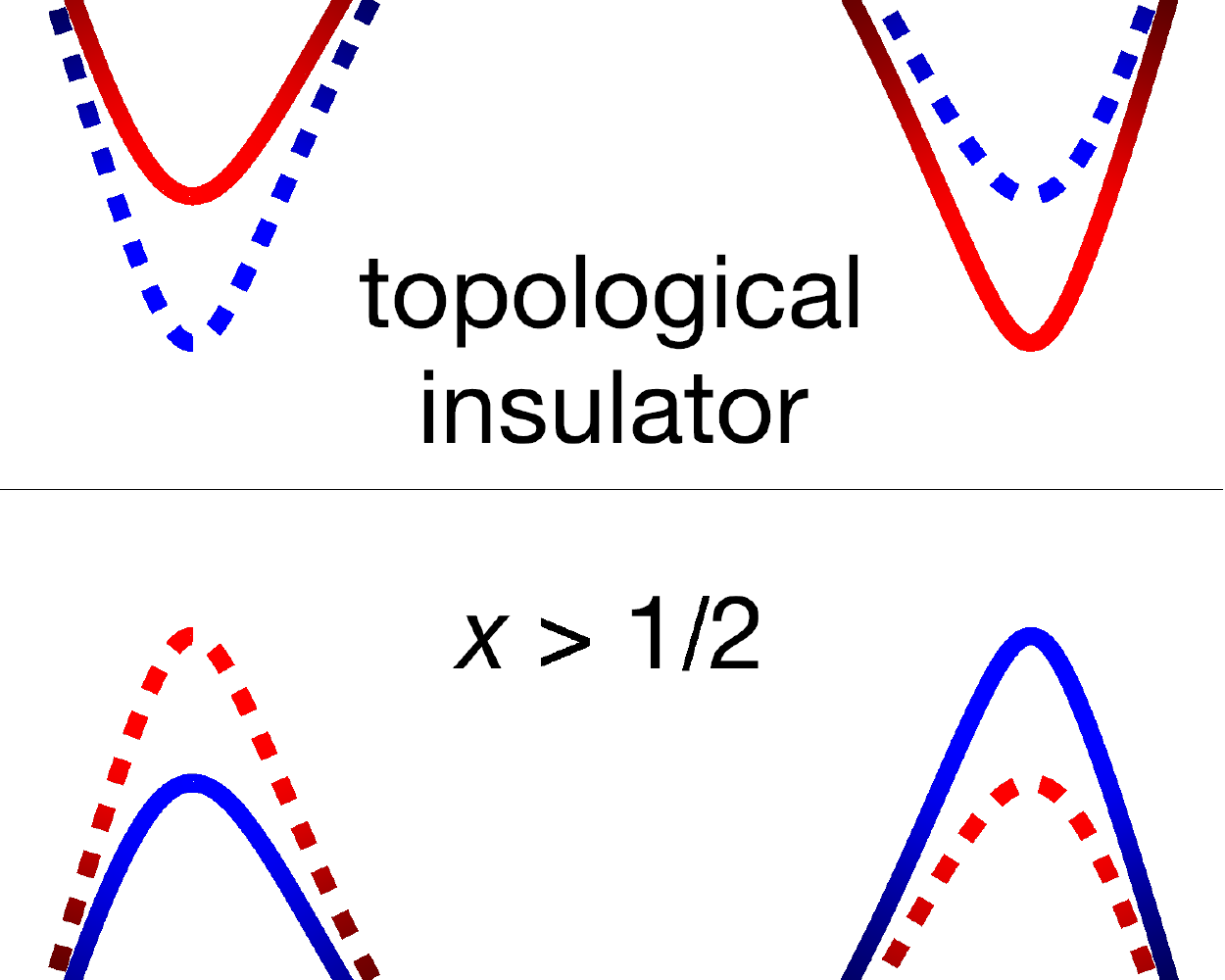}}%
}
\includegraphics[height=0.82in]{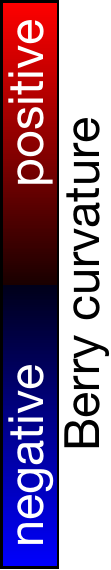}%
\caption{\label{fig:lattice}%
\subref{fig:honeycomb}
Honeycomb lattice on which the Hamiltonian in Eq.~\eqref{eq:kanemele} is defined.
The blue hexagon marks the $\sqrt{3}\times\sqrt{3}$ supercell used in our study, which allows pairing with nonzero center-of-mass (c.m.) crystal momentum $K$ and $K'$ of Cooper pairs in addition to $\Gamma$.
\subref{fig:brillouinzone}
Brillouin zone of the honeycomb lattice.
The inner blue hexagon represents the reduced Brillouin zone of the supercell;
both $K$ and $K'$ defined for the original Brillouin zone are folded to the $\Gamma$ point in the reduced Brillouin zone.
\subref{fig:dispersion-trivial}--\subref{fig:dispersion-topological}
Dispersions of the non-interacting Kane-Mele model defined in Eq.~\eqref{eq:kanemele}.
The solid (dashed) curves show the dispersion of electrons with spin up (down).
The parameter $x = 3\sqrt{3}\lambda_{\text{so}}/(m_{\text{AB}}+3\sqrt{3}\lambda_{\text{so}})$ that represents the relative strength of the Ising spin-orbit coupling is varied between \subref{fig:dispersion-trivial} $0 \le x<1/2$ in the trivial insulator phase, \subref{fig:dispersion-gapclosing} $x=\frac{1}{2}$ at the topological transition, and \subref{fig:dispersion-topological} $\frac{1}{2} < x \le 1$ in the topological insulator phase.
The color of the curves indicates the sign of the Berry curvature:
In each spin sector, the signs of the Berry curvature at $K$ and $K'$ are opposite in the trivial phase, and the same in the topological phase.
At the topological transition ($x=\frac{1}{2}$), there is a single Dirac cone in each spin sector in the corresponding valley.
}
\end{figure}

\para{}
To study the pairing instability of a two-dimensional Dirac system across the topological phase transition between topological and trivial insulating phases, we take the Kane-Mele model defined on a honeycomb lattice [Fig.~\ref{fig:honeycomb}] as the underlying band structure~\cite{kane-prl-2005}:
\begin{align}
\label{eq:kanemele}
  \calH_{\mathrm{KM}}
    &=
      - t
      \sum_{\langle i, j \rangle} 
          \psi_{i}^{\dagger} \psi_{j}
      - \mu \sum_{i}  \psi_{i}^\dagger \psi_{i} \nonumber\\
    &\quad 
      - i \lambda_{\text{so}}
       \sum_{\llangle i, j \rrangle}
          \nu_{ij} \psi_{i}^{\dagger} \sigma^{z} \psi_{j} 
      +  m_{\text{AB}} \sum_{i} \xi_{i} \psi_{i}^{\dagger} \psi_{i}
\end{align}
where $\psi_{i}^{\dagger} \equiv (c_{i\up}^{\dagger}, c_{i\dn}^{\dagger})$ is the electron creation operator at site $i$, and $\langle \cdot, \cdot \rangle$ and $\llangle \cdot, \cdot \rrangle$ represent nearest-neighbor and next-nearest-neighbor pairs of sites.
Here, $t$ is the nearest-neighbor hopping amplitude,
$\mu$ the chemical potential,
$\lambda_{\text{so}}$ the strength of Ising spin-orbit coupling, with $\nu_{ij} = \mathrm{sgn} ( \hat{z} \cdot (\bfv_1 \times \bfv_2) )$ where $\bfv_1$ and $\bfv_2$ are nearest-neighbor vectors that connect an electron hop from site $i$ to site $j$, and $m_{\text{AB}}$ the sublattice potential, with $\xi_{i}=1$ ($-1$) if the site $i$ belongs to the sublattice $A$ ($B$).
The sublattice potential breaks inversion symmetry and reduces the symmetry group of the Hamiltonian to $D_{3h}$.
For the sake of simplicity, we do not include the Rashba spin-orbit coupling in our analysis.
Our main results, nevertheless, remain the same for a small Rashba coupling, as we discuss later.

\para{Symmetry and topology}
The topology of a non-interacting (or mean-field) Hamiltonian is characterized by different topological indices depending on the dimensionality and the symmetry of the system~\cite{schnyder-prb-2008,kitaev-acp-2009}.
The band structure $H_{\mathrm{KM}}$ has time-reversal symmetry ($\calT$ symmetry) with $\calT^2=-1$, and thus belongs to the class AII~\cite{altland-prb-1997}.
In two dimensions, this class has two distinct topological phases characterized by a $\bbZ_2$ topological index $\nu=0$ or 1.
To take the system across the topological phase transition, we introduce a parameter $x$ between 0 and 1, which is related to the spin-orbit coupling and sublattice potential by $3 \sqrt{3} \lambda_{\text{so}} = E_g x$ and $m_{\text{AB}} = E_g (1-x)$.
$H_{\mathrm{KM}}$ has a topological (trivial) ground state for $x>\frac{1}{2}$ ($x<\frac{1}{2}$).
The low-energy degrees of freedom involve two massive spin-polarized Dirac cones at each ``valley'' centered at $K$ and $K'$ [Figs.~\ref{fig:brillouinzone}--\subref{fig:dispersion-topological}]. 
At $x=\frac{1}{2}$, the band structure is at a topological phase transition, with one of the Dirac cones in each valley being massless.
The mass of the other Dirac cones remains constant at $E_g$ throughout the transition for all values of $x$. 
For the purpose of our calculation we have chosen $E_g = t/2$.
Adding a small Rashba spin-orbit coupling does not affect the topology of the system, as long as the bulk gap remains finite \cite{kane-prl-2005}.

\para{Interactions}
We study the pairing instability of the Hamiltonian $H_{\mathrm{KM}}$ with three different types of interactions:
(1) attractive onsite interaction $-U \sum_{i} n_{i\up} n_{i\dn}$,
(2) attractive nearest-neighbor density-density interaction $-V \sum_{\langle i j \rangle} n_{i} n_{j}$,
or (3) antiferromagnetic nearest-neighbor Heisenberg interaction $J \sum_{\langle i j \rangle} \boldsymbol{\sigma}_{i} \cdot \boldsymbol{\sigma}_{j}$,
where $n_{i\sigma} \equiv c_{i\sigma}^{\dagger} c_{i\sigma}$, $n_{i} \equiv \psi^{\dagger}_{i} \psi_{i} $, and ${\sigma}_{i}^{\mu} \equiv \psi_{i}^{\dagger} \sigma^{\mu} \psi_{i}$ for $\mu=x,y,z$.
In each case, we decouple the interaction in the pairing channel and find the Bogoliubov--de Gennes (BdG) ground states.
All the superconducting states that emerge self-consistently in this analysis are fully gapped. 
This allows us to calculate the relevant topological index in each phase corresponding to its symmetry class (see Appendix~\ref{sec:computetopoindex}).
Once again, Rashba spin-orbit coupling does not qualitatively affect the results, as long as it is weak compared to the Bogoliubov quasiparticle gap.

\begin{table}[ht!]
\centering
\caption{Summary of the superconducting phases in Fig.~\ref{fig:pd-nn} found with attractive nearest-neighbor density-density interaction $V$.
$\Phi^{K}$ and $\Phi^{K'}$ are spatial form factors defined by
$\Phi^{\bfQ}_{ij} = e^{i \bfQ \cdot (\bfr_{i} + \bfr_{j})}$,
representing pairing of two electrons at $K$ and $K'$ valleys, respectively.
}
\label{tab:phases-nn}
\begin{ruledtabular}
\begin{tabular}{cllcl}
  & Superconducting & Order parameter $\Delta$                         & $\calT$-sym. & Topo. \\
  & phase           & 
  &              & index
  \\
  \hline
  \fcolorbox[rgb]{0,0,0}{0.79, 0.98, 0.74}{\rule{0ex}{0.5ex}\rule{0.5ex}{0ex}}
    & Topo. helical triplet
    & $\Delta_{\up\up} \sim \Phi^{K}$, $\Delta_{\dn\dn} \sim \Phi^{K'}$
    & $\checkmark$
    & $\tilde{\nu}=1$
    \\
  \fcolorbox[rgb]{0,0,0}{0.79, 0.91, 0.99}{\rule{0ex}{0.5ex}\rule{0.5ex}{0ex}}
    & Triv. $p$-Kekule triplet
    & $d^z \sim \Phi^{K} - \Phi^{K'}$
    & $\checkmark$
    & $\tilde{\nu} = 0$
    \\
  \fcolorbox[rgb]{0,0,0}{0.78, 0.58, 1.0}{\rule{0ex}{0.5ex}\rule{0.5ex}{0ex}}
    & Topo. chiral triplet
    & $\Delta_{\up\up} \sim \Phi^{K}$, $d^z \sim \Phi^{K'}$
    & $\bigtimes$
    & $\tilde{\calC} = \pm 1$
    \\
    & 
    & (or its $\calT$ partner)
    &
    \\
  \fcolorbox[rgb]{0,0,0}{0.99, 0.71, 0.97}{\rule{0ex}{0.5ex}\rule{0.5ex}{0ex}}
    & Triv. $\calT$-breaking
    & $\Delta_{\up\up}, \Delta_{\dn\dn} \sim \Phi^{K}-\Phi^{K'}$,
    & $\bigtimes$
    & $\tilde{\calC} = 0 $
    \\
    & triplet
    & $d^z \sim \Phi^{K} + \Phi^{K'}$
    &
\end{tabular}
\end{ruledtabular}
\end{table}

\begin{figure}[ht!]\centering%
\includegraphics[width=3in]{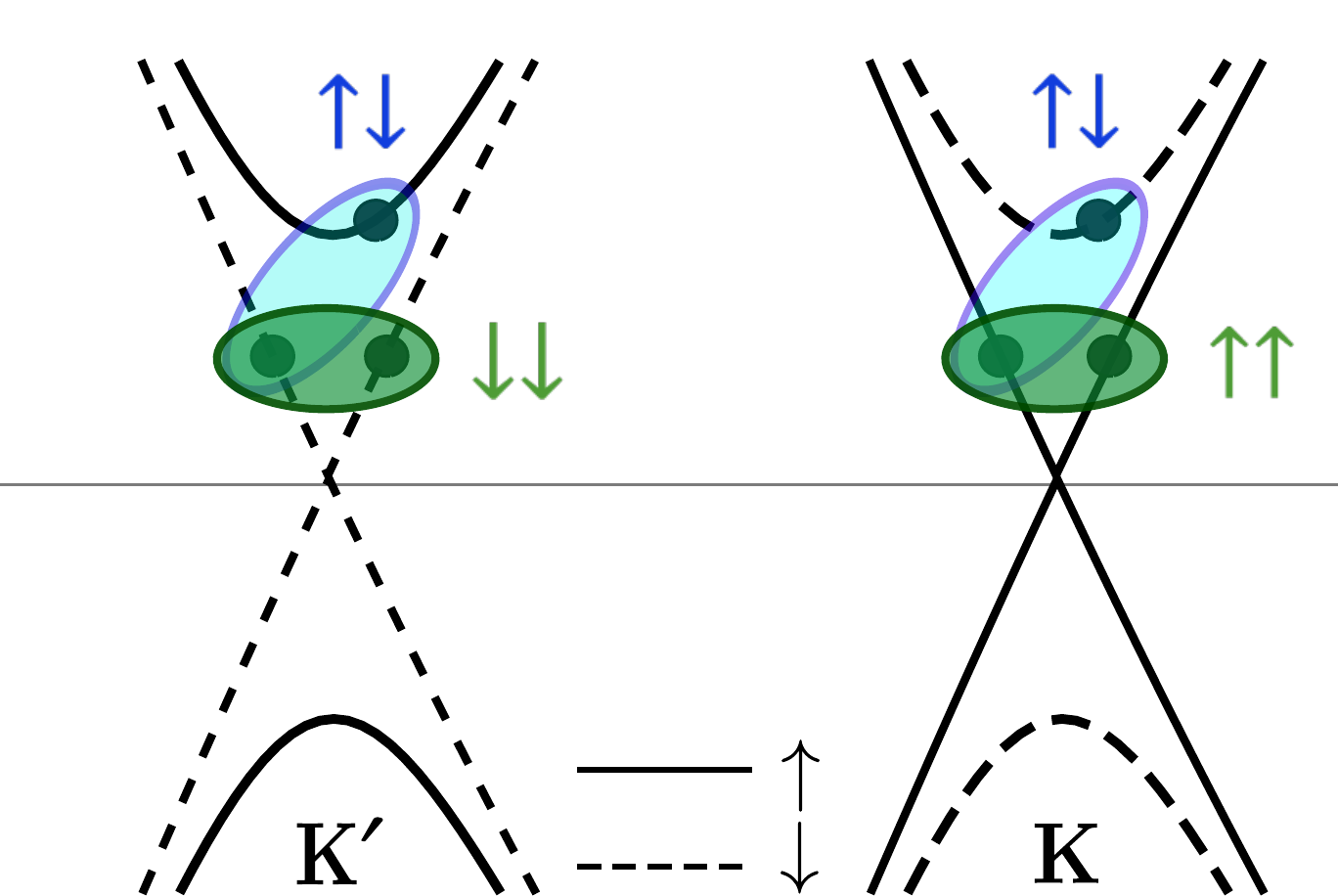}
\caption{\label{fig:DiracConePairing}%
As described in Table \ref{tab:phases-nn}, the phase diagram for nearest neighbor attractive interaction $V$ is understood in terms of six order parameters, corresponding to triplet pairing between up-up, down-down, and up-down pairs of fermions on each valley. (See  Appendix~\ref{sec:kspacedescription} for details.) 
The real-space pattern $\Phi^K_{ij}$ corresponding to the condensate at the valley $K$ is shown in Fig.~\ref{fig:nn-pairing-pattern-planewave}. 
The $p$-Kekule pair potential exhibits interference between the two condensates $\Phi^K_{ij}$ and $\Phi^{K'}_{ij}$. 
}
\end{figure}

\para{Nonzero c.m. momentum pairs}
Since the low-energy electronic degrees of freedom lie at valleys near $K$ and $K'$ [see Figs.~\ref{fig:brillouinzone}--\subref{fig:dispersion-topological}], we also allow pairing of two electrons from the same valley.
To incorporate such pairing with Cooper pairs having nonzero center-of-mass momentum $2K\equiv K'$ or $2K' \equiv K$, we use a supercell with six sites [blue hexagon in Fig.~\ref{fig:honeycomb}], whose reduced Brillouin zone folds the $K$ and $K'$ to $\Gamma$ [blue hexagon in Fig.~\ref{fig:brillouinzone}].
This introduces 6 onsite pairing order parameters and 36 nearest-neighbor pairing order parameters.
We then minimize the ground-state energy within this exhaustive parameter space averaging over $24\times 24$ momentum grid.
Note that we are not imposing a particular structure of the pairing order parameter;
we are allowing the self-consistency loop to pick the lowest-energy configuration in the space of 42 complex pairing order parameters.

\section{Various Superconducting Phases and Their Topology}


\para{Onsite attraction $U$}
In the Kane-Mele model at $\mu=0$ with onsite attractive interaction $U$, we find three different phases as shown in Fig.~\ref{fig:pd-os}.
Away from $x=\frac{1}{2}$, the system is an insulator for weak interaction due to the nonzero band gap:
its topological property is completely determined by the underlying band structure parametrized by $x$.
For strong enough interaction, we find a continuous transition to a uniform $s$-wave spin-singlet superconducting phase.
Note that in the presence of spin-orbit coupling (at $x \neq 0$), spin-singlet and spin-triplet are not symmetry-distinct, and pair amplitudes $\langle c_{i \sigma} c_{j \sigma'} \rangle$ in both spin channels can be nonzero in general.
The onsite interaction, however, allows pair potential $\Delta$ only in the spin-singlet channel.
Throughout this paper, we use the terms spin-singlet and spin-triplet pairings to refer to the spin component of $\Delta$ and not necessarily the pair amplitude.

\para{}
Since the pairing leaves the $\calT$ symmetry intact, the Bogoliubov--de Gennes Hamiltonian is in the class DIII, with a $\bbZ_2$ topological index $\tilde{\nu}=0$ or 1, defined analogously to the $\bbZ_2$ topological index $\nu$ of class AII topological insulator, but in terms of the Bogoliubov quasiparticles in Nambu space.
The superconducting state that arises from either the topological insulator or the trivial insulator is a trivial superconductor with $\tilde{\nu}=0$.
This can be understood in the following way:
The insulating phase can be seen as a $\calT$-invariant superconductor with zero pair potential.
Such a ``superconducting state'' is trivial since $\tilde{\nu}=2\nu=0 \text{ (mod 2)}$ independent of $\nu$; (the factor of 2 is due to the particle-hole redundancy of Nambu spinors).
At a continuous transition to a superconducting state, $\tilde{\nu}$ cannot change since the single-particle gap does not close.
Thus, it is natural that the superconductor that emerges from a continuous transition from a trivial or topological time-reversal-invariant insulator, is topologically trivial.
Conversely, a topological superconductor must be separated from a time-reversal-invariant insulator either by a discontinuous transition, or an intervening state where the single-particle gap closes.


\begin{figure}\centering%
\subfigure[\label{fig:nn-pairing-pattern-planewave}%
]{\includegraphics[width=1.75in]{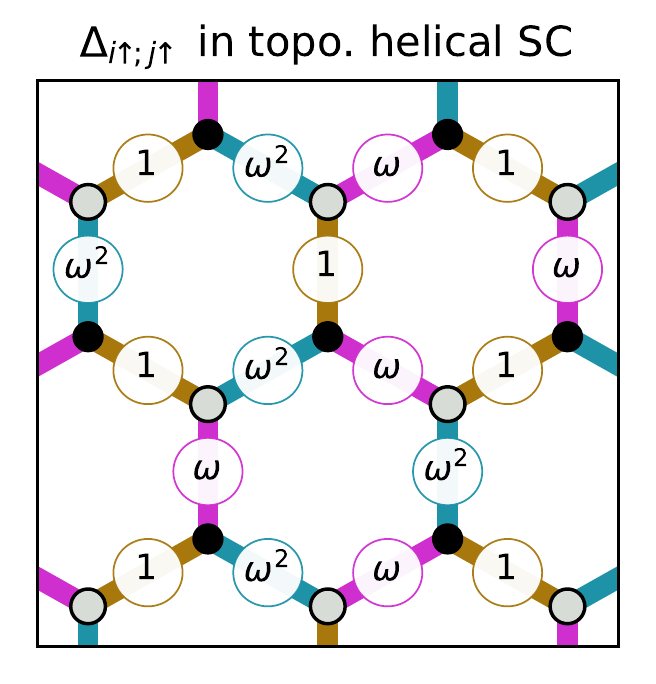}}%
\subfigure[\label{fig:nn-pairing-pattern-pkekule}%
]{\includegraphics[width=1.75in]{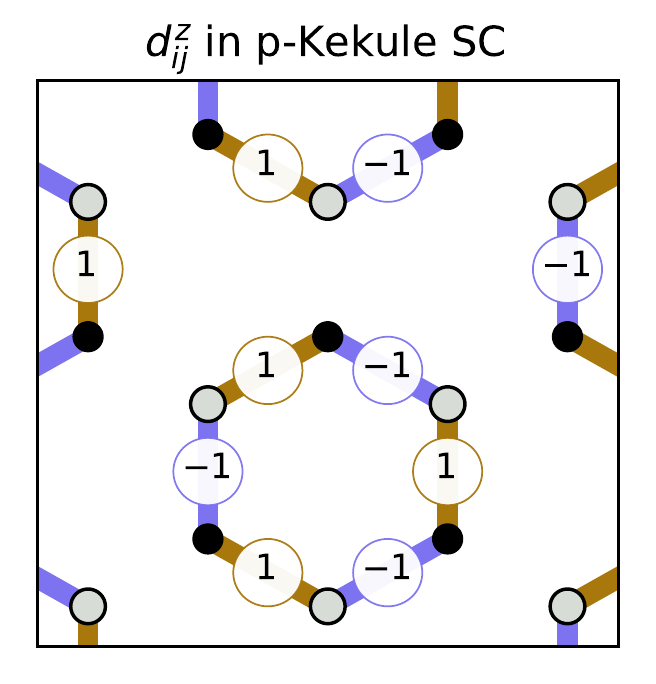}}%
\caption{\label{fig:nn-pairing-pattern}%
Real-space patterns of the pairing order parameters that we find with nearest-neighbor attractive density-density interaction.
A bond between sites $i$ and $j$ represents pair potential
\subref{fig:nn-pairing-pattern-planewave}
$\Delta_{i\up; j\up}$ of the ``topological helical triplet SC,'' which is $\sim \Phi^K_{ij}$, and
\subref{fig:nn-pairing-pattern-pkekule}
$d^{z}_{ij}$ of the ``$p$-Kekule SC,'' which is $\sim \Phi^{K}_{ij} - \Phi^{K'}_{ij}$.
The color of a bond marks the phase of the order parameter, which is also indicated $1$, $-1$, $\omega$, and $\omega^2$ on the bonds ($\omega \equiv e^{2\pi i / 3}$).
Since both $\Delta_{i\up; j\up}$ and $d^{z}_{ij}$ are antisymmetric under $i \leftrightarrow j$,
we choose a convention for the phases:
$i$ is always chosen from the A sublattice, and $j$ from the B sublattice.
}
\end{figure}

\para{Nearest-neighbor density-density attraction $V$}
With attractive nearest-neighbor density-density interaction $V$, we find a much richer phase diagram shown in Fig.~\ref{fig:pd-nn}.
(We have implicitly assumed the presence of long-range Coulomb repulsion to prevent phase separation at stronger interaction.)
Unlike $U$ which only allows spin-singlet pairing, $V$ also allows spin-triplet pairing channels.
The pair potential $\Delta_{i\sigma;j\sigma'}$ between electrons at sites $i$ and $j$ with spins $\sigma$ and $\sigma'$ can thus be decomposed into spin-singlet and three spin-triplet channels as
\begin{align}
  \Delta_{i\sigma; j\sigma'}
    &=
      \left[
        ( \psi_{ij} \sigma^0 + \bfd_{ij} \cdot \boldsymbol{\sigma} ) i \sigma^y
      \right]_{\sigma\sigma'},
\end{align}
where $\sigma^\mu$ for $\mu=0,x,y,z$ are the identity and the Pauli matrices in spin space.
Since, however, the Hamiltonian $H_{\mathrm{KM}}$ only has a U(1) spin rotation symmetry related to the $S_z$ conservation rather than the full SU(2) spin rotation symmetry, it is more convenient to decompose the pairing channels into $\psi$ (Cooper pairs with spin $S=0$), $\Delta_{\up\up}$ ($S=1$, $S_z=1$), $d^z$ ($S=1$, $S_z=0$), and $\Delta_{\dn\dn}$ ($S=1$, $S_z=-1$).
We find four distinct superconducting phases, all of which have $\Delta$ purely in the spin-triplet channel (with $\psi_{ij} = 0$).
These phases and their order parameters are summarized in Table~\ref{tab:phases-nn},
and can be understood in terms of spin and valley degrees of freedom, as shown in Fig.~\ref{fig:DiracConePairing}.
(Appendix~\ref{sec:symop} discusses how these order parameters transform under symmetry operations.)

\para{Topological helical SC}
Around $x=\frac{1}{2}$ at weaker interaction strength, we find a helical spin-triplet superconductor, which is $\calT$ invariant and characterized by a non-trivial topological $\bbZ_2$ index $\tilde{\nu}=1$ [green region in Fig.~\ref{fig:pd-nn}].
The pairing in this state is in the equal-spin channel ($\Delta_{\up\up}, \Delta_{\dn\dn} \neq 0$), with nonzero momentum Cooper pairs, as indicated by the real-space pattern of $\Delta_{i\up;j\up}$ shown in Fig.~\ref{fig:nn-pairing-pattern-planewave}, which goes as $\Delta_{i\up;j\up} \sim \Phi^{K}$, where $\Phi^{\bfQ}_{ij} \equiv e^{i \bfQ \cdot (\bfr_{i} + \bfr_{j})}$, for $i$ in sublattice A and $j$ in sublattice B.
$\Phi^{\bfQ}$ represents pairing with center-of-mass momentum $2\bfQ$.
The magnitude of the pair potential is uniform across all unit cells and only the phase modulates.

\para{}
This $\calT$-invariant superconducting state, whose non-trivial topology is characterized by the $\bbZ_2$ topological index $\tilde{\nu}=1$, can be understood in terms of the Dirac dispersions at each valley.
When $x \approx \frac{1}{2}$, the low energy electronic degrees of freedom are spin-valley locked [see Fig.~\ref{fig:dispersion-gapclosing}].
The order parameters $\Delta_{i\up;j\up} \sim \Phi^{K}_{ij}$ and $\Delta_{i\dn;j\dn} \sim \Phi^{K'}_{ij}$, therefore represent pairing between two electrons of the same spin from the same valley, which can be written in momentum space as
\begin{align}
  \sum_{\bfq}
    \Delta_{K+\bfq} c_{K + \bfq,\up}^{\dagger} c_{K - \bfq, \up}^{\dagger} + 
    \Delta_{K'+\bfq} c_{K' + \bfq,\dn}^{\dagger} c_{K' - \bfq, \dn}^{\dagger}
    + \mathrm{H.c.}
\end{align}
For small $\bfq$, $\Delta_{K+\bfq} \approx \Delta_{K} + O(q^2)$ with $\Delta_{K} \neq 0$.
The nonzero momentum pair potential $\Delta_{K+\bfq}$ thus plays the role of ``uniform $s$-wave'' gap within the Dirac cone at the $K$ valley (and similarly $\Delta_{K'+\bfq}$ for the $K'$ valley), which effectively becomes $p_x \pm i p_y$ pairing in the band basis \cite{sato-plb-2003,fu-prb-2006}.
This results in a nonzero Chern number $\tilde{\calC}=\pm1$ in each spin sector, leading to a non-trivial $\bbZ_2$ index $\tilde{\nu}=1$.

\para{}
As we have argued previously for the onsite attraction, a transition from an insulator to a topological superconductor must either involve an intermediate trivial superconducting phase if it is continuous, or be first order.
Within our exploration of the phase diagram, we have not found any intermediate phase between the insulating phases, both trivial and topological, and the topological helical superconducting phase.
Is the transition first order, or have we simply missed the intermediate phase?
In Appendix~\ref{sec:firstorder} we present a more careful study of the nature of this transition, where we identify a jump in the order parameter, a clear sign of a first-order transition.

\para{$p$-Kekule SC}
At $x=1$ and nearby where the underlying band structure is in the topological insulator phase, we find a $\calT$-invariant triplet SC which is topologically trivial ($\tilde{\nu}=0$) [blue region in Fig.~\ref{fig:pd-nn}].
The pairing in this state is in the opposite-spin spin-triplet channel ($d^z \neq 0$), and also has nonzero momentum Cooper pairs, forming the ``$p$-Kekule'' pattern in real space [see Fig.~\ref{fig:nn-pairing-pattern-pkekule}], which was originally discussed in the context of graphene~\cite{roy-prb-2010}.
This phase was previously found by \textcite{tsuchiya-prb-2016} who studied the same Hamiltonian ($H_{\text{KM}}$ with $V$) in the $x=1$ limit.

\para{Topological chiral SC}
In a thin region between the topological helical SC and the $p$-Kekule SC, we also find a $\calT$-breaking topological triplet SC with nonzero Chern number $\tilde{\calC} = \pm 1$ [purple region in Fig.~\ref{fig:pd-nn}].
We refer to this state as topological chiral SC, following Ref.~\onlinecite{qi-prl-2009}.
In this state, one of the valleys develops equal-spin pairing gap within the same cone, while the other valley develops an opposite-spin spin-triplet pairing gap across the two Dirac cones in the same valley.
This results in a nonzero Chern number with unequal contribution from the two valleys.

\para{Trivial $\calT$-breaking SC}
At $x \approx 0$ and at larger interaction strength,
the system favors a pairing state which is $\calT$ breaking with a mixture of equal-spin and opposite-spin pairing channels in both valleys [pink region in Fig.~\ref{fig:pd-nn}].
This is distinct from the chiral SC in that it is topologically trivial ($\tilde{\calC} = 0$).
(See Appendix~\ref{sec:tbreakingphase} for discussions on the structure of the order parameter in this phase.)


\begin{figure*}
\centering
\subfigure[\label{fig:pd-os-mu}]{\includegraphics[height=2.2in]{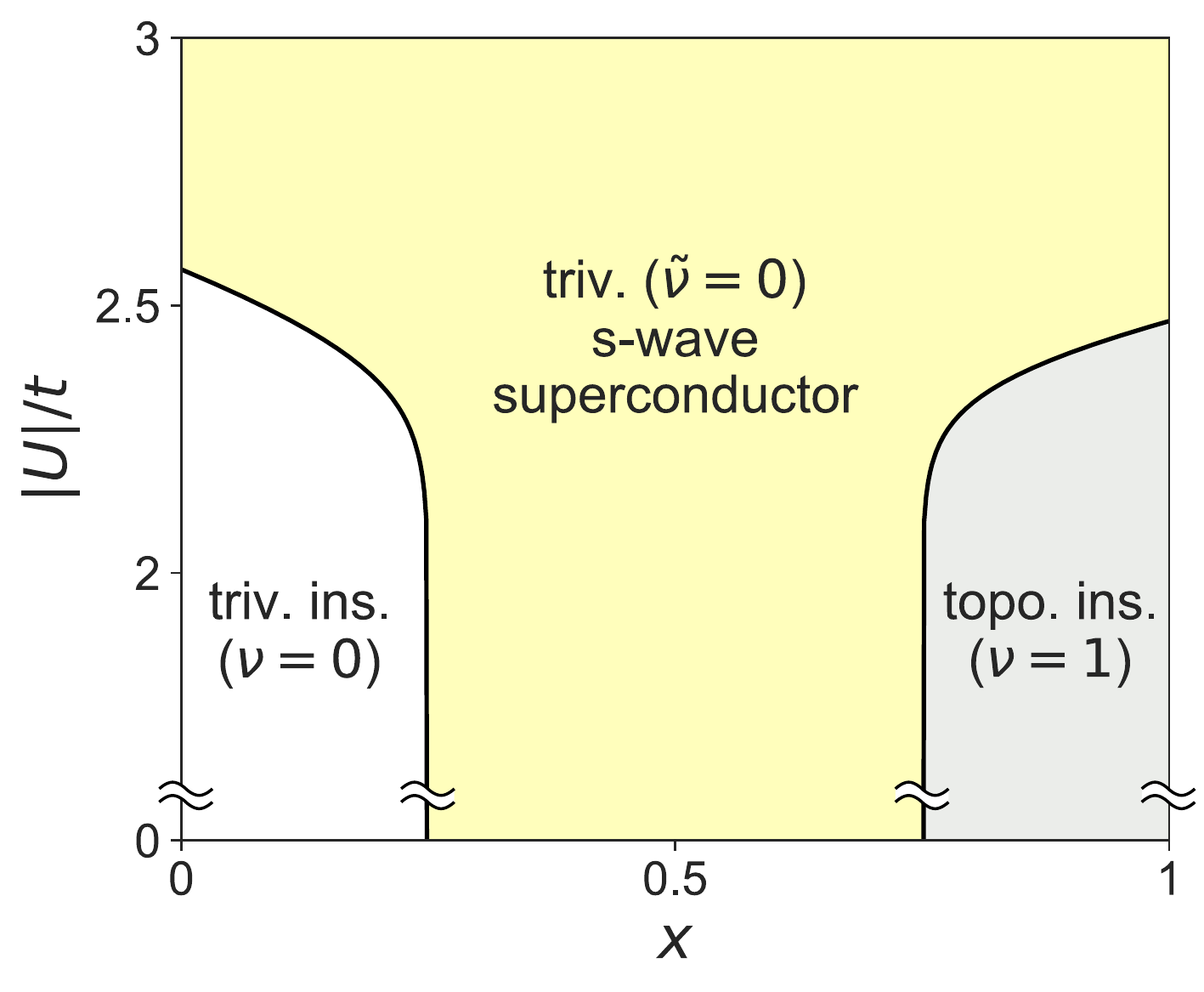}}
\qquad
\subfigure[\label{fig:pd-nn-mu}]{\includegraphics[height=2.2in]{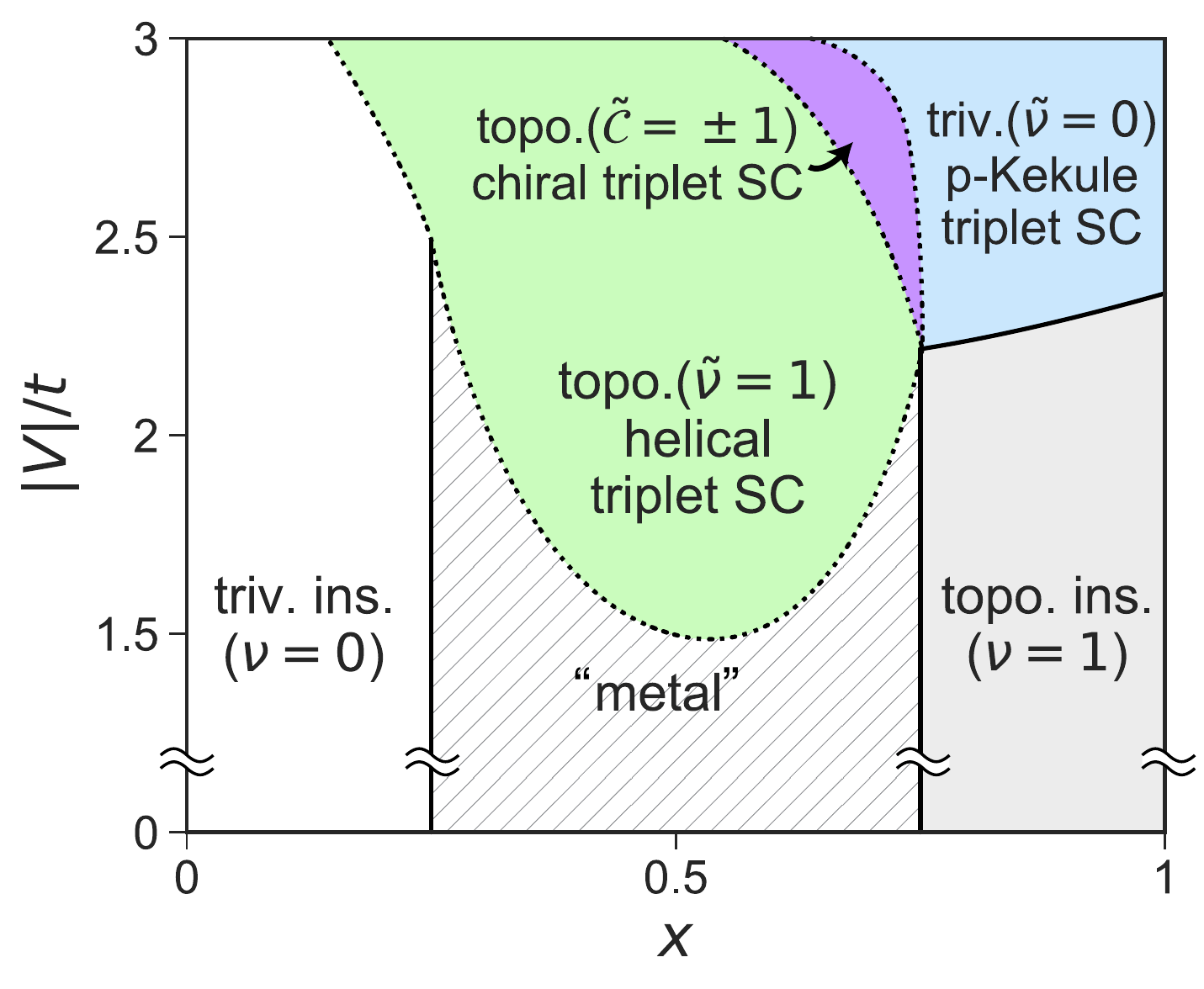}}
\\
\subfigure[\label{fig:pd-mu-scan}]{\quad\includegraphics[height=0.90in]{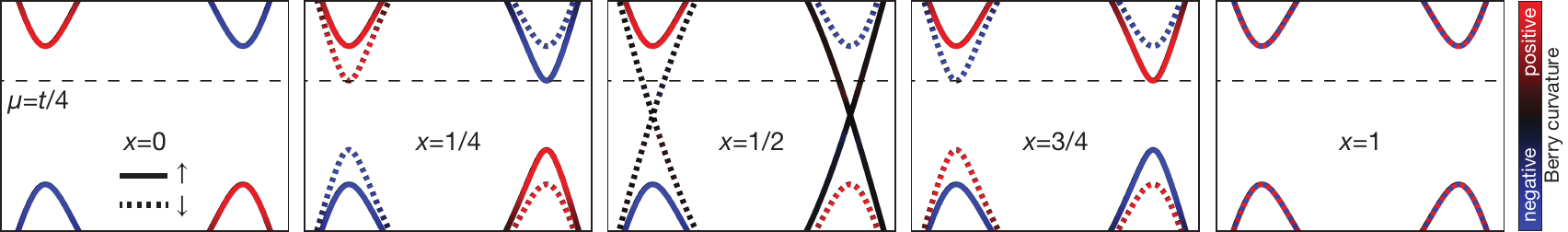}}
\caption{%
\label{fig:pd-mu}%
\subref{fig:pd-os-mu}, \subref{fig:pd-nn-mu}
Phase diagrams at chemical potential $\mu=t/4=E_g/2$ away from half-filling, with 
  \subref{fig:pd-os-mu} attractive onsite interaction $U$ and
  \subref{fig:pd-nn-mu} attractive nearest-neighbor density-density interaction $V$.
\subref{fig:pd-mu-scan} Dispersions of the non-interacting band structure at different values of $x$, with the chemical potential $\mu$ marked by the horizontal dashed lines.
Within the range $\frac{1}{4} < x < \frac{3}{4}$, the normal-state band structure contains a non-spin-degenerate Fermi surface in each valley.
With $U$, we find $s$-wave superconducting phase as in Fig.~\ref{fig:pd-os}.
When there are Fermi surfaces ($\frac{1}{4} < x < \frac{3}{4}$), pairing amplitude should develop with infinitesimal $U$.
With $V$, we find similar phases as to Fig.~\ref{fig:pd-nn}, in addition to the ``metal'' phase near $x=\frac{1}{2}$.
The ``metal'' phase is defined to be regions with a very small pair amplitude ($\langle c_{i\sigma} c_{j\sigma'} \rangle <10^{-6}$), which is numerically difficult to distinguish from zero.
Unlike the trivial $s$-wave superconductivity, the nonzero center-of-mass momentum pairing is not necessarily an infinitesimal instability even in the presence of Fermi surfaces, due to their trigonal warping.
}%
\end{figure*}

\begin{figure}
\centering%
\includegraphics[width=3in]{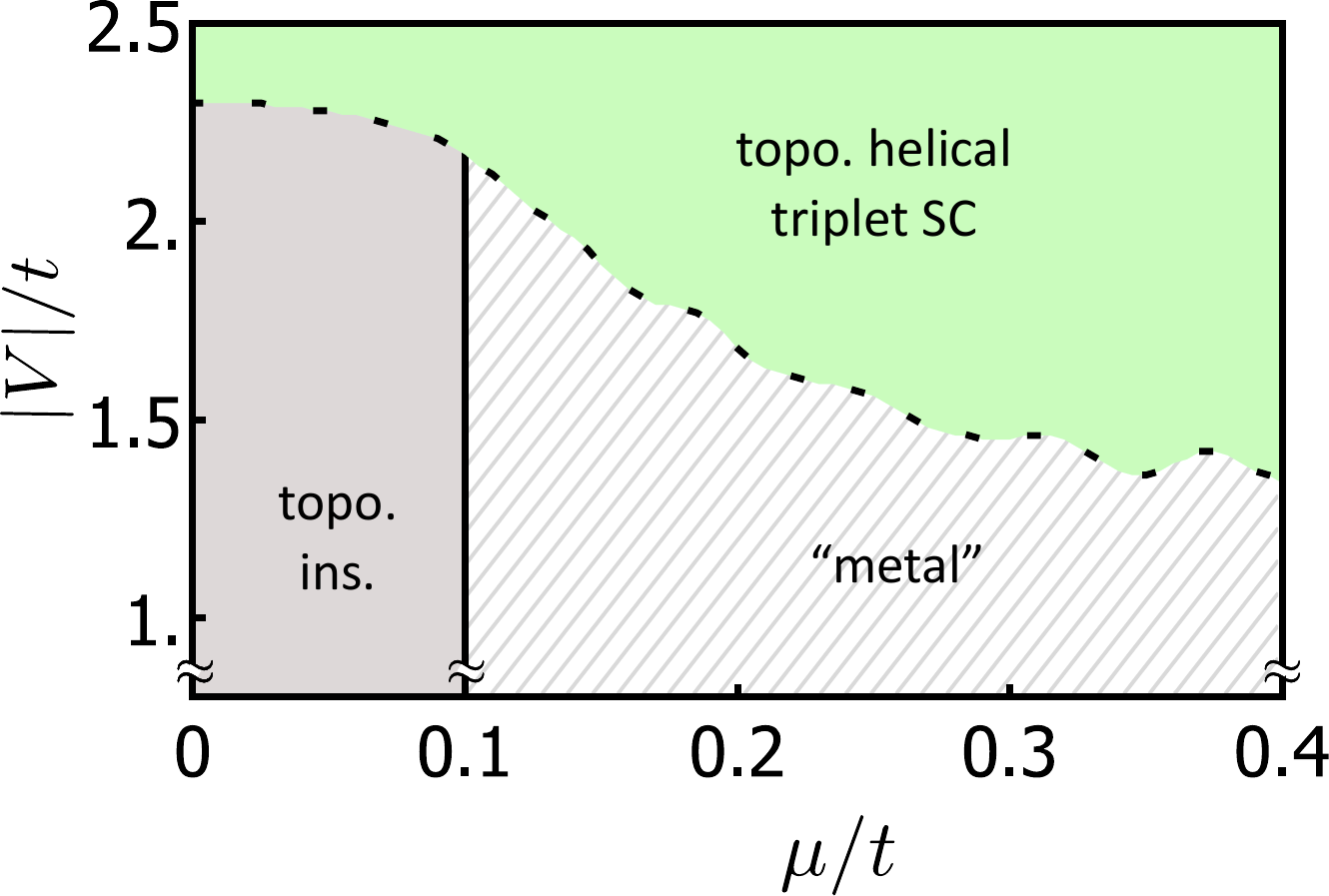}
\caption{\label{fig:doping}%
The critical interaction strength for the transition to the topological helical SC is lowered by increasing $\mu$. 
We show a doping-driven transition at $x=0.6$ on a $90\times90$ lattice with a temperature of $T=t/100$.
}
\end{figure}

\para{Finite doping $\mu\neq 0$}
So far, we have considered the band structure at half filling with $\mu=0$, and found topological superconducting phases with $V$.
Do these topological phases exist even when the underlying band structure is metallic?
Figures~\ref{fig:pd-mu} and Fig.~\ref{fig:doping} summarize the phase diagrams at nonzero chemical potential $\mu=t/4$.
Note that $E_g=t/2$, and therefore the band structure is metallic with a single non-spin-degenerate Fermi surface in each valley within the range $\frac{1}{4} < x < \frac{3}{4}$ [see Fig.~\ref{fig:pd-mu-scan}].
As shown in Fig.~\ref{fig:pd-mu}, the $\mu \neq 0$ phase diagrams contain the same superconducting phases as the $\mu = 0$ ones, in both cases of $U$ and of $V$.
The topological indices of these phases remain identical to the $\mu=0$ counterparts.
Importantly, we find that the topological helical superconductor that we find with $V$ is accessible at lower interaction strength with increasing $\mu$, as shown in Fig.~\ref{fig:doping}.

\para{}
Within the range $\frac{1}{4} < x < \frac{3}{4}$, where the normal state band structure contains Fermi surfaces, the $s$-wave superconductivity with $U$ becomes an infinitesimal instability.
For the superconducting phases that we find with $V$, all of which have spatially modulating pair potential, the electrons that form a Cooper pair are not time-reversal partners:
They reside at momenta opposite of $K$ or $K'$ (e.g. $c_{K+\bfq,\sigma}$ and $c_{K-\bfq,\sigma'}$).
Because of the trigonal warping of the Fermi surfaces, these two electrons cannot both be at the Fermi level, except on a finite number of $k$-points.
Therefore, such nonzero momentum pairings are no longer infinitesimal instabilities, even in the presence of Fermi surfaces, and requires finite interaction strength.
Following this argument, we mark the region near $x=\frac{1}{2}$ in Fig.~\ref{fig:pd-nn-mu} with very small pair potential (numerically indistinguishable from zero) as ``metal.''
The warping is minimal near the metal-insulator transition in the underlying band structure, but in spite of the finite density of states in this limit, intravalley pairing is still not an infinitesimal instability because the low energy fermions exactly at $K$ and $K'$ are sublattice polarized, and the nearest-neighbor interaction pairs fermions from opposite sublattices.
Nevertheless, this does not rule out the possibility that the underlying metallic state is unstable to other pairing channels, such as spin-singlet extended $s$-wave.


\begin{figure}[t]\centering
\subfigure[\label{fig:pd-haldane}]{\includegraphics[height=1.6in]{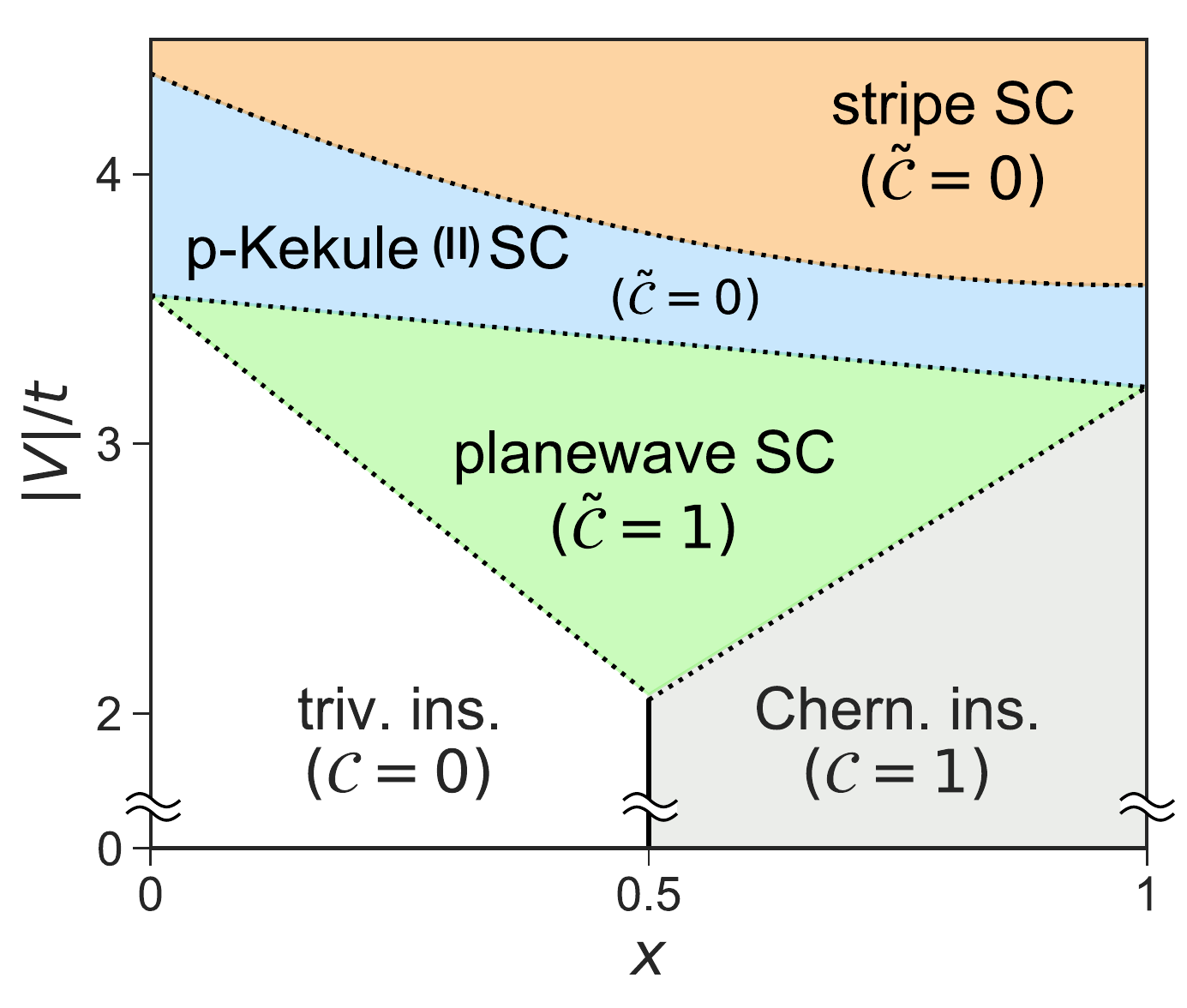}}
\subfigure[\label{fig:nn-pairing-pattern-stripe}]{%
\parbox[b][1.63in][t]{1.48in}{
\includegraphics[height=1.45in]{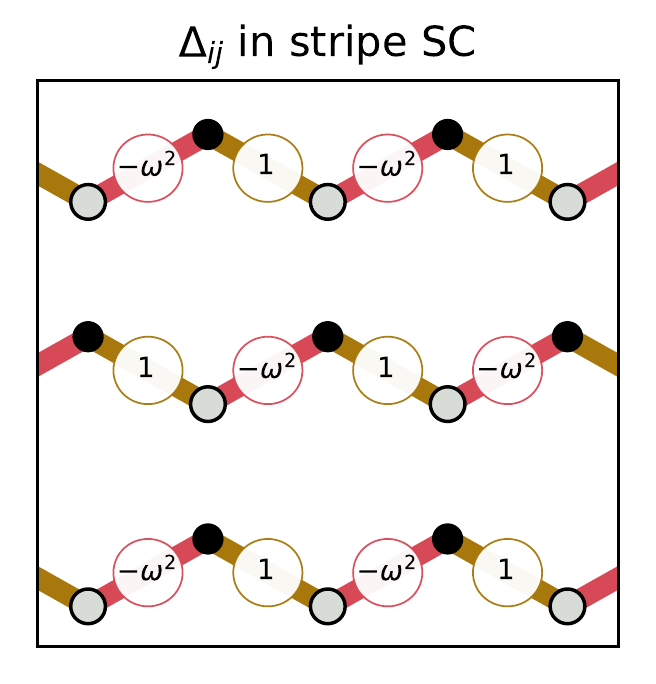}%
}%
}%
\caption{
\subref{fig:pd-haldane}
Phase diagram of Haldane model with nearest-neighbor attractive interaction $V$.
\subref{fig:nn-pairing-pattern-stripe}
The real-space pattern of the pairing gap $\Delta_{ij}$ of the ``stripe SC'' phase.
For the same reason as in Fig.~\ref{fig:nn-pairing-pattern}, $i$ is always chosen from the A sublattice and $j$ from the B sublattice.
}
\end{figure}

\para{Haldane model}
A natural corollary of the topological helical SC is that if we were to consider only one spin species, as in the Haldane model~\cite{haldane-prl-1988}, we expect a chiral SC near the topological transition in the band structure at $\mu=0$.
This turns out to be true:
By solving the self-consistent Bogoliubov--de Gennes equation of the following Hamiltonian,
\begin{align}
  H_{\mathrm{Haldane}-V}
    &=
      - t
       \sum_{\langle i, j \rangle}
          c_{i}^{\dagger} c_{j}
      - i \lambda 
       \sum_{\llangle i, j \rrangle}
          \nu_{ij} c_{i}^{\dagger} c_{j} \nonumber\\ 
    &\quad
      +  m_{\mathrm{AB}} \sum_{i} \xi_{i} c_{i}^{\dagger} c_{i}
      - V \sum_{\langle i, j \rangle} n_{i} n_{j}
\end{align}
as a function of $V$ and $x$ defined analogously to that of the Kane-Mele model above, we get a phase diagram shown in Fig.~\ref{fig:pd-haldane}.
For smaller values of $V$ we find the topological ``plane-wave SC,'' whose $\Delta_{ij}$ is equivalent to the $\Delta_{i\up;j\up}$ of the helical SC in Fig.~\ref{fig:pd-nn} and thus has Chern number $\tilde{\calC}=1$.
The chirality is determined by the underlying band structure, since the time-reversal symmetry is explicitly broken at the band-structure level, even without the interaction.
Due to the reduced degrees of freedom and thus less number of competing orders, the topological plane-wave SC phase expands and spans the whole range of $x$.

\para{}
At stronger $V$, we find two more superconducting phases, which we refer to as the ``$p$-Kekule (II) SC'' and ``stripe SC,'' both of which have zero Chern number.
Note that the ``$p$-Kekule (II) SC'' phase in the Haldane model is different from the
$p$-Kekule triplet SC phase of the Kane-Mele model:
the spatial structure of $\Delta_{ij}$ of ``$p$-Kekule (II) SC'' is identical to that of $d_{ij}^{z}$ of ``$p$-Kekule SC.''
However, while $p$-Kekule triplet SC pairs two electrons from different Dirac cones in the same valley, 
``$p$-Kekule (II) SC'' pairs two electrons from the same Dirac cone in the same valley, due to the lack of the other Dirac cone.
The ``stripe SC,'' whose spatial structure of this phase is shown in Fig.~\ref{fig:nn-pairing-pattern-stripe}, breaks the $C_3$ rotation symmetry, but preserves the original translation symmetry of the lattice.
This state pairs electrons from the opposite valleys.

\para{}
The Haldane model has been experimentally realized with ultracold atoms~\cite{jotzu-n-2014} and there are proposals to engineer near-neighbor interactions~\cite{anisimovas-pra-2016}.
Based on our calculation, we predict that the resulting superconductivity with attractive interactions should be topological with a Chern number of $\tilde{\calC}=\pm1$.


\begin{figure}
\centering
\includegraphics[height=2.2in]{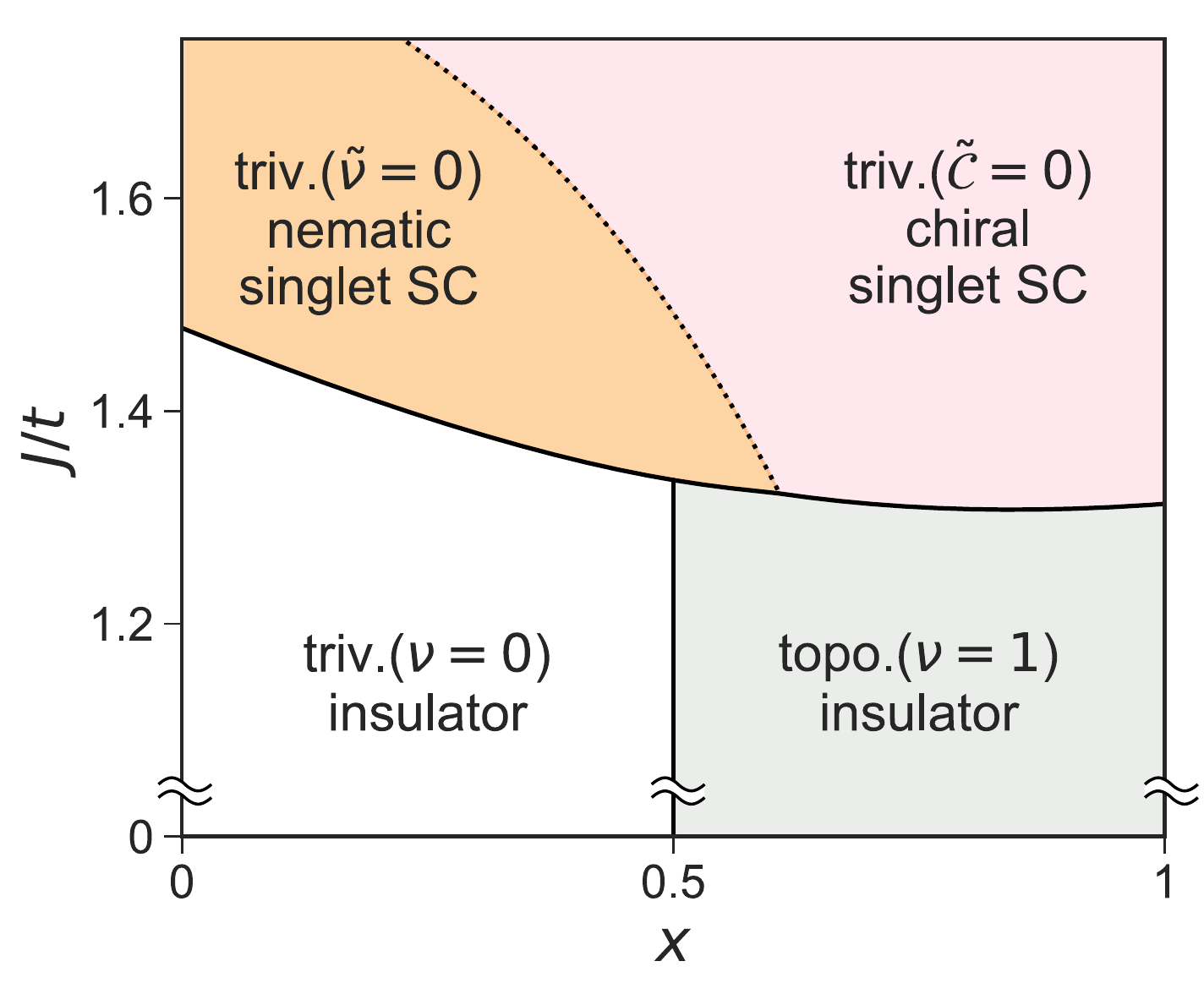}
\caption{\label{fig:pd-heis}%
Phase diagrams of Kane-Mele model in Eq.~\eqref{eq:kanemele} as functions $x$ with nearest-neighbor antiferromagnetic Heisenberg interaction $J$.
We find two distinct topologically trivial singlet pairing states.
Near $x=0$ we find a topologically trivial nematic singlet SC that is $\calT$ invariant, and breaks the $C_3$ rotation symmetry of the system.
Near $x=1$ we find a topologically trivial chiral singlet SC, which is $\calT$ breaking with pairing in the spin-singlet channel.
}
\end{figure}

\begin{figure}
\centering
\subfigure[\label{fig:heisenberg-pairing-pattern-nematicvbs}%
]{\includegraphics[width=1.75in]{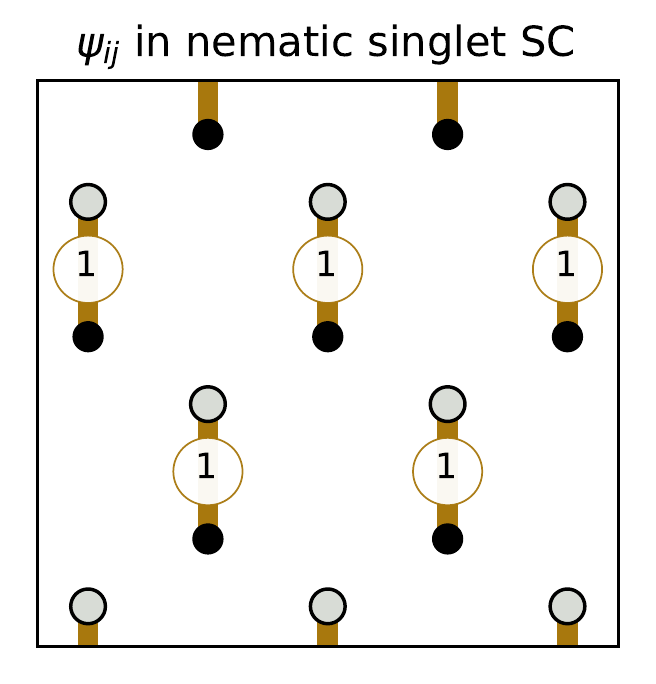}}%
\subfigure[\label{fig:heisenberg-pairing-pattern-chiralvbs}%
]{\includegraphics[width=1.75in]{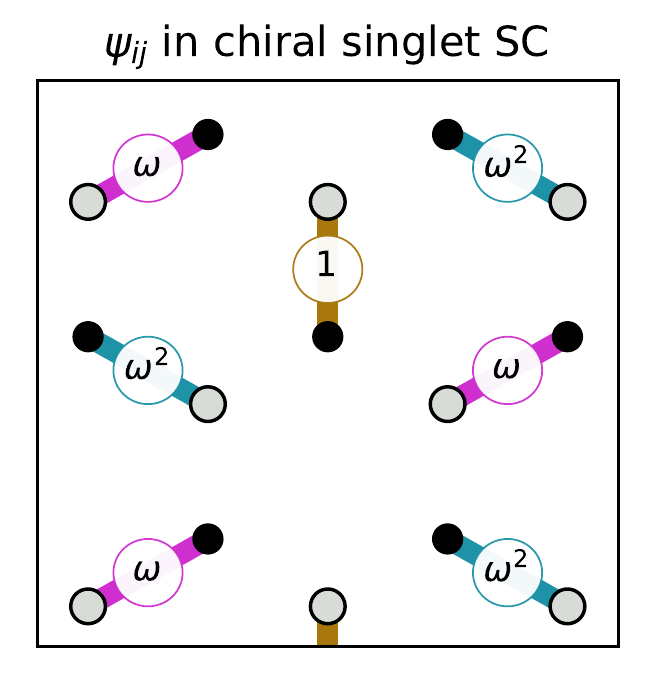}}%
\caption{Real-space patterns of the spin-singlet pair potential $\psi_{ij}$ of
\subref{fig:heisenberg-pairing-pattern-nematicvbs} the nematic singlet SC and
\subref{fig:heisenberg-pairing-pattern-chiralvbs} the chiral singlet SC phases that we find with nearest-neighbor antiferromagnetic Heisenberg exchange $J$.
}
\label{fig:heisenberg-pairing-pattern}
\end{figure}

\para{Antiferromagnetic Heisenberg exchange $J$}
With antiferromagnetic Heisenberg exchange $J$ between nearest-neighboring sites at $\mu=0$, we find two distinct superconducting states as shown in Fig.~\ref{fig:pd-heis}.
Both of these states are topologically trivial, but have exotic characteristics:
The pairing state for $x \lesssim \frac{1}{2}$ is a nematic singlet SC, which is $\calT$ invariant but breaks rotation symmetry.
The pairing state for $x \gtrsim \frac{1}{2}$, on the other hand, is a chiral singlet SC, which is in the spin-singlet channel, yet is $\calT$ breaking and also breaks translation symmetry. 
The real-space patterns of the singlet order parameter $\psi_{ij}$ in these phases are shown in Fig.~\ref{fig:heisenberg-pairing-pattern}.

\section{Discussion and Outlook}

\para{Summary}
To summarize, we have derived the phase diagram of the Kane-Mele model across its trivial-insulator-to-topological-insulator transition, with various interactions using the Bogoliubov--de Gennes framework.
With attractive onsite interaction $U$, we find trivial $s$-wave superconductivity as expected.
With nearest-neighbor interactions, both the attractive density-density interaction $V$, and the antiferromagnetic Heisenberg exchange $J$, we find exotic superconducting phases with finite Cooper-pair momentum.
Especially with $V$, we find two distinct topological superconducting phases, one $\calT$ invariant and one $\calT$ breaking, near the trivial-insulator-to-topological-insulator transition, where one pair of the Dirac cones become gapless.

\para{New route to topological superconductivity}
While the models we have solved are specific, the broad lessons we have learned are applicable to a more general class of phenomena.
The central thrust of our work is to understand the conditions under which we get topological superconductivity in a Dirac system.
Through our study of the Kane-Mele model, we have identified two crucial ingredients for obtaining a topological superconductor.
First, there needs to be uniform pairing within a Dirac cone~\cite{sato-plb-2003,fu-prl-2008}.
Second, such pairing must manifest on a single time-reversed pair of non-degenerate Dirac cones for $\calT$-invariant helical SC.
This corresponds to ``topological helical triplet SC'' in Fig.~\ref{fig:pd-nn} that is characterized by a $\bbZ_2$ topological index $\tilde{\nu}=1$.
If the intra-cone pairing is nonzero only on one Dirac cone, we have a chiral superconductor characterized by a nonzero Chern number $\tilde{\calC}$.
This corresponds to the purple region in Fig.~\ref{fig:pd-nn}, which is $\calT$ breaking.

\para{}
A single time-reversed pair of spin-polarized Dirac cones appears naturally at the topological transition of the Kane-Mele model at $x=\frac{1}{2}$.
Pairing internal to each of these Dirac cones is necessarily between equal-spin electrons.
It is only with nearest-neighbor density-density attraction that the equal-spin pairing channel is allowed.
Both onsite attraction and antiferromagnetic Heisenberg exchange enable pairing in the singlet channel, we therefore find no topological superconductivity with these interactions.

\para{}
Thus far, the search for topological superconductivity has been driven largely by one theme:
break $\calT$ and get effectively spinless fermions, and then induce (effective) $p$-wave pairing between them.
This originates from work by Kitaev in 1D~\cite{kitaev-pu-2001} and $\calT$ breaking is central to this quest.
One of the strengths of the work presented here is a route to 2D topological superconductivity in presence of $\calT$ invariance and an explicit demonstration in the context of the Kane-Mele model.

\para{BCS-BEC crossover and connection with topology}
The intuition from the $p+ip$ superconductors is that the strong coupling BEC regime is trivial whereas topological superconductivity only arises in the weak coupling BCS regime. We note, based on our studies, that 
such a demarcation does not apply to the honeycomb Dirac system.
The most obvious difference is that in our model, the Fermi energy is in the middle of the band gap so that we have both electron and hole bands, each with nontrivial Berry phase.
Unlike the $p+ip$ superconductors where the sense of ``winding'' is related to the winding of the order parameter along the Fermi surface, in a Dirac system the winding is related to the Berry phase of the underlying band structure.
This makes our normal state qualitatively different from a trivial vacuum.
Therefore, upon including interaction in an otherwise insulating state, the system can enter topological superconducting state even in the BEC regime.

\para{Comparison with previous theoretical studies}
In previous theoretical studies, 
pairing in the TMD materials has hitherto been studied without incorporating the full effect of the honeycomb lattice~\cite{yuan-prl-2014,hsu-nc-2017}, ignoring the Dirac physics and the $\pi$ Berry phase around the valley.
\textcite{yuan-prl-2014} considered onsite and nearest-neighbor attraction on a \emph{triangular} lattice, and found $\calT$-breaking topological superconductivity only in the presence of Rashba spin-orbit coupling.
We note that the phases discussed there are, in principle, included in our mean-field study and turn out to be energetically less favored than the finite momentum paired states that we encounter.
\textcite{hsu-nc-2017} used renormalization group analysis to explore the leading instability of one spin-polarized circular Fermi surface at $K$ and $K'$ with onsite repulsive interactions.
They found several degenerate paired states: an interpocket chiral SC, an intrapocket chiral SC and an intrapocket helical SC similar to our topological helical triplet SC phase.

\para{Experimental probes}
We expect that the theoretical phase diagrams and general principles for topological superconductivity that we have unearthed from simple models are relevant for the low-energy physics of monolayer TMD materials, such as $\mathrm{MoS_2}$, $\mathrm{WS_2}$, $\mathrm{WTe_2}$.

\para{}
Recent experiments on monolayer WTe$_2$ \cite{sajadi-s-2018,fatemi-s-2018} have observed gating-driven transition from quantum spin Hall insulator to superconductor.
The type of superconductivity induced in this system, and its topological properties, are not yet known.
If the superconductivity is driven by electron-phonon interaction, where the attractive onsite $U$ is the most relevant effective interaction, we can place the system in Fig.~\ref{fig:pd-os-mu} across the topological insulator and trivial $s$-wave superconductor phases.
If, on the other hand, the superconductivity is driven by electron-electron interaction, where the onsite pairing is suppressed by strong short-range repulsion, phase diagrams with $V$ [Fig.~\ref{fig:pd-nn}] or with $J$ (Fig.~\ref{fig:pd-heis}) may be relevant to superconductivity in these systems.

\para{}
The phases we have described could be experimentally identified by establishing signatures of spin-triplet pairing, of spatially modulated superconductivity, and of the Majorana edge modes characteristic of the topological superconductors.
The spin susceptibility measured using Knight shift and relaxation rates may be used to identify triplet pairing and discern whether it is equal-spin or opposite-spin pairing.
The $p$-Kekule SC with $S_z=0$ would exhibit a suppression of spin susceptibility to zero, with out-of-plane fields, unlike the other phases.
The equal-spin paired helical superconductor would have spin-polarized Majorana modes counterpropagating along the edges of the sample, which would contribute to a finite quantized thermal Hall conductivity in the superconducting state.
Time-reversal breaking in the chiral superconductor states could be identified by polar Kerr effect~\cite{kapitulnik-njp-2009} or muon spin rotation spectroscopy.

\para{}
Detecting the spatial modulation of the phase in the helical superconductor is possible using the dc-SQUID setup outlined in Ref.~\onlinecite{hsu-nc-2017}.
In addition, in realistic samples we expect finite Rashba spin-orbit coupling to result in a singlet order parameter derived from both the up-spin condensate with momentum $2K$ and the down-spin condensate with momentum $2K'$.
The resulting pair density wave in the singlet channel would be observable by scanning Josephson tunneling microscopy (SJTM)~\cite{hamidian-n-2016} with a superconducting tip with singlet pairs.

\para{}
The pair density wave nature of the $p$-Kekule SC would be expected to show up both in STM and in SJTM experiments with a tip exfoliated from the substrate.
However, as we show in Appendix~\ref{sec:sVspKekule}, this might require going to extremely low temperatures to prevent tunneling between the three equivalent $p$-Kekule configurations related to each other by a lattice translation.

\para{}
Spatial modulation of the order parameter is a direct consequence of intravalley pairing.
In the TMDs, it is now well established that circularly polarized light can be used to selectively excite fermions from one valley.
An  observable consequence of intravalley pairing would then be a suppression of the cooperon energy observed with circularly polarized light as we approach the superconducting transition by lowering temperature.

\begin{acknowledgments}
We thank P. Coleman and Y.-T. Hsu for useful discussions.
K. L. and N. T. acknowledge support from the National Science Foundation Grant No. DMR-1629382.
T. H. and M. R. are supported by National Science Foundation Grant No. DMR-1410364.
\end{acknowledgments}

\appendix

\section{Computing topological indices}
\label{sec:computetopoindex}

\para{}
The various phases that we find as solutions to the Bogoliubov--de Gennes equations have different broken symmetries, time reversal, particle hole, space group, and spin rotation.
To characterize the topology of these phases requires calculating topological indices that correspond to the symmetry class.
For insulators, an efficient numerical method of calculating Chern number $\calC$ has been presented by \textcite{fukui-jpsj-2005}, and a related method of calculating the $\bbZ_2$ index $\nu$ by \textcite{fukui-prb-2007}.
Here, we summarize these numerical methods, and their extension to superconducting systems.
For superconductors, we denote the Chern number of the Bogoliubov quasiparticle dispersion by $\tilde{\calC}$ and the corresponding $\bbZ_2$ index by $\tilde{\nu}$.
We  note that for the special case of equal-spin pairing on a $S_z$-conserving band structure (such as $H_{\mathrm{KM}}$), the $\bbZ_2$ invariant is simply the difference of the Chern numbers in the two time-reversed spin sectors $\tilde{\nu}=(\tilde{C}_\uparrow - \tilde{C}_\downarrow )/2$.

\subsection{Chern number in insulating and superconducting states}

\begin{figure}\centering%
\subfigure[\label{fig:plaquette}]{\includegraphics[height=1.5in]{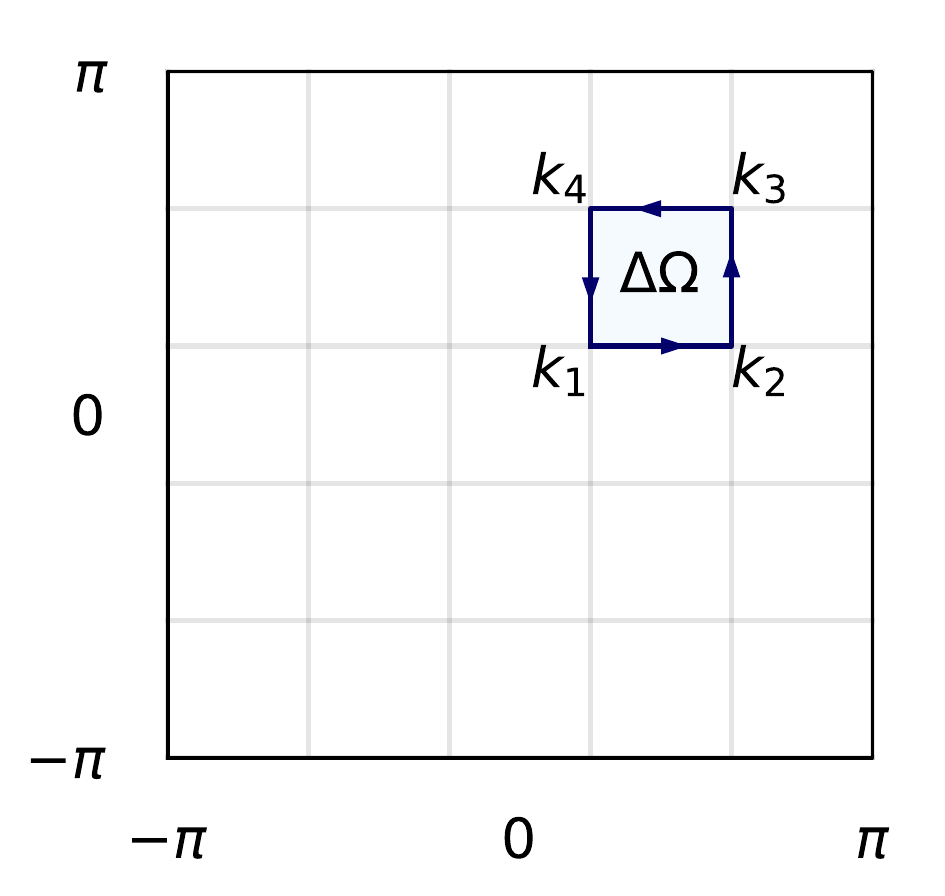}} 
\subfigure[\label{fig:halfbz}]{\includegraphics[height=1.5in]{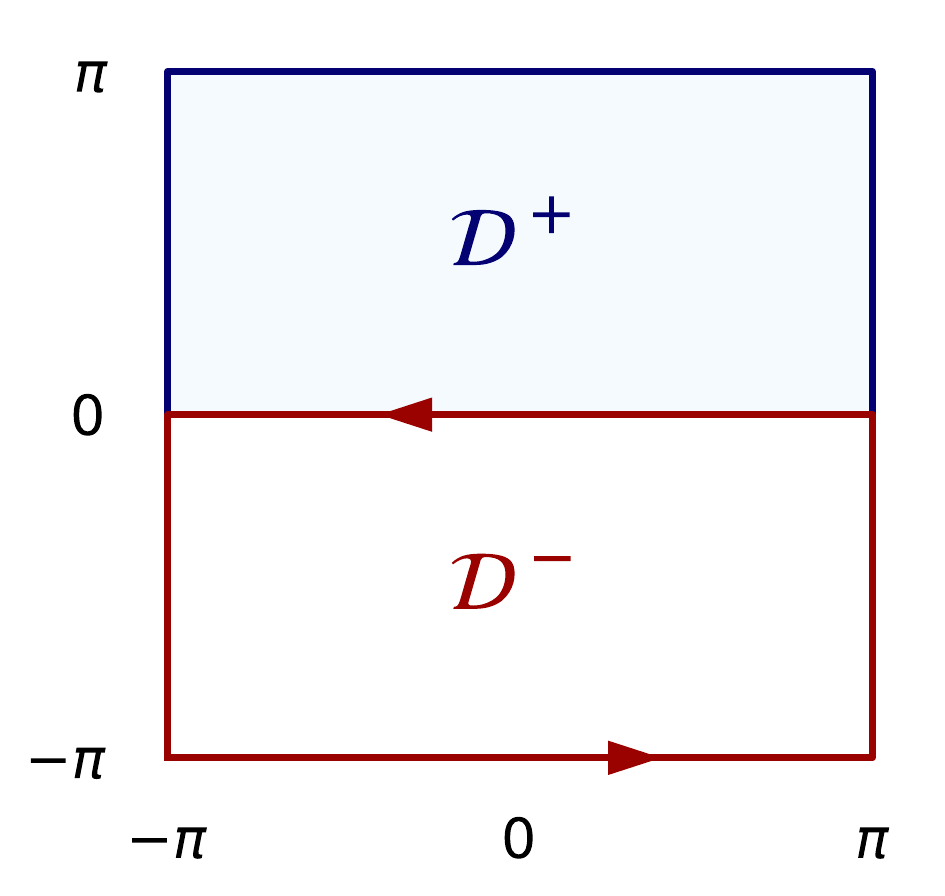}}
\caption{%
\subref{fig:plaquette}
Berry flux through a plaquette in momentum space.
The Chern number can be calculated numerically by collecting the Berry flux through all the plaquettes in the Brillouin zone.
\subref{fig:halfbz}
The partitioning of the Brillouin zone into two domains that are time-reversal partners of each other.
The $\bbZ_2$ topological invariant $\nu$ of a time-reversal-invariant insulator (or $\tilde{\nu}$ of a time-reversal-invariant superconductor) can be calculated 
as the sum of the Berry flux through $\calD^{+}$ and the Berry phase around the boundary of $\calD^-$.
}
\end{figure}

\para{}
For an insulating band structure in two dimensions with broken time-reversal symmetry, the topological index which characterizes the topological class is the Chern number $\calC \in \bbZ$, also referred to as the Thouless, Kohmoto, Nightingale, and den Nijs (TKNN) invariant~\cite{thouless-prl-1982}.
Analytical calculation of $\calC$ depends on fixing the gauge such that the Bloch wave function is smooth in the entire Brillouin zone, except at a number of points.
Calculation of $\calC$ from the numerically calculated eigenstates is, however, ill behaved;
a more efficient numerical method which does not require gauge fixing has been presented by \textcite{fukui-jpsj-2005}.

\para{}
Given a set of Bloch wave functions $\ket{n(\bfk)}$ defined on the Brillouin zone, its Berry connection defined as $\bfA_n(\bfk) = i \bracket{n(\bfk), \nabla_{\bfk} , n(\bfk) }$ can be expressed as
\begin{align}
\bfA_{n}(\bfk) \cdot \delta \bfk
    \approx \mathrm{arg} \left( \bracket{n(\bfk + \delta \bfk), n(\bfk)} \right)
\end{align}
Thus, the line integral of the Berry connection around a plaquette [Fig.~\ref{fig:plaquette}] can be written as
\begin{align}
  \int_{1 \rightarrow 2 \rightarrow 3 \rightarrow 4 \rightarrow 1}
  \bfA \cdot \mathrm{d} \bfk 
    &=
        \mathrm{arg}
        \left[
            \bracket{1,4}
            \bracket{4,3}
            \bracket{3,2}
            \bracket{2,1}
        \right]
\end{align}
where $\ket{i}$ is a shorthand for $\ket{n (\bfk_{i})}$.
This is also the Berry flux  $\Delta \Omega_{\square}$
through the plaquette, modulo $2\pi$.
Defining $U_{ij} \equiv \bracket{i, j}$ (Wilson line between sites $i$ and $j$), we can write the total flux as
\begin{align}
\Delta \Omega_{\square}
    &=
        \mathrm{arg} ( U_{14} U_{43} U_{32} U_{21} )
\end{align}
The Chern number is the total number of fluxes through the whole Brillouin zone, which thus can be calculated as
\begin{align}
\calC
    &\equiv
        \frac{1}{2\pi} \int \! \Omega \; \mathrm{d}^{2} k
    =
        \frac{1}{2\pi} \sum_{\square}
        \Delta \Omega_{\square}
\end{align}
where
$\Delta \Omega_{\square}
    \equiv \mathrm{Arg} ( U_{14} U_{43} U_{32} U_{21} )$,
assuming that the Berry curvature is a smooth function of $\bfk$ and the plaquettes are small enough such that the flux through every plaquette is smaller than $\pi$.

\para{}
This method of calculating the Chern number can be extended to multi-band systems.
The total Chern number of a set of bands can be calculated simply by summing the Chern numbers of all the band.
In general, however, there can be band crossings which introduce degeneracies at certain $\bfk$ points.
In such a case the Berry curvature of a single band is not well defined.
It is thus necessary, for the numerical calculation of Berry flux, to work with objects which are invariant under unitary transformation within the manifold defined by the select bands.
It is easy to see that the following Wilson line between sites $i$ and $j$,
\begin{align}
  U_{ij} \equiv \det_{n,m} \left\langle i, n \middle\vert j, m \right\rangle
\end{align}
with $n,m$ being the select band indices, is invariant under unitary transformation within the manifold defined by the select bands.

\para{}
In the superconducting phase with broken time-reversal symmetry, the relevant topological index is the superconducting Chern number $\tilde{\calC}$.
This is computed exactly as shown above, except that  $\ket{n(\bfk)}$ now refers to the Bogoliubov quasiparticle wave function in Nambu space. 
In this work, unless otherwise specified, $\calC$ and $\tilde{\calC}$ denote the total Chern number of the bands with negative energy eigenvalues.

\subsection{\texorpdfstring{$\bbZ_2$}{Z2} topological index in insulating state}
\label{sec:z2index}

\para{}
For two-dimensional band structures with time-reversal symmetry with $\calT^2=-1$ (class AII), a $\bbZ_2$ index $\nu$ characterizes the symmetry-protected topological phases~\cite{fu-prb-2007,fu-prl-2008}, rather than the Chern number which is zero by symmetry.
A numerical method for the calculation of $\nu$ has been presented by \textcite{fukui-prb-2007}.

\para{}
The time-reversal operator $\calT$ can be written as a product of complex-conjugation operator $\calK$ and a unitary matrix $U_{T}$: $\calT = \calK U_{T}$.
For example, for spin-$\frac{1}{2}$ fermions, we can set $U_{T} = i \sigma_{2}$.
For a time-reversal-invariant Hamiltonian, its momentum-space representation $H_{\bfk}$ transforms under $U_{T}$ as 
\begin{align}
    U_{T} H_{\bfk} U_{T}^{\dagger} &= H_{-\bfk}^{*}
\end{align}
This places a constraint on the eigenstates of the Hamiltonian:
If $u_{\bfk}$ is an eigenstate of $H_{\bfk}$ ($H_{\bfk} u_{\bfk} = E_{\bfk} u_{\bfk}$), then
\begin{align}
U_{T} H_{\bfk} U_{T}^{\dagger} U_{T} u_{\bfk}
  &= 
  H_{-\bfk}^{*} U_{T} u_{\bfk}
  =
    E_{\bfk} U_{T} u_{\bfk}
\end{align}
thus $(U_{T} u_{\bfk})^{*}$ is an eigenstate of $H_{-\bfk}$ with eigenvalue $E_{\bfk}$.
From this we can connect eigenstates at $\bfk$ with eigenstates at $-\bfk$.
We cannot, however, enforce $u_{-\bfk} = ( U_{T} u_{\bfk} )^{*}$ for all $\bfk$ since it is inconsistent with $\calT^{2}=-1$.

\para{}
We can nevertheless choose a gauge convention in the following way.
Let us first consider a single time-reversal pair of bands, and label the Bloch eigenstates by the band indices I and II, where $u_{\bfk}^{\mathrm{I}}$ and $u_{-\bfk}^{\mathrm{II}}$ are Kramer's pairs.
Then, we can enforce the following relationship between the two:
\begin{align}
    \label{eq:timereversalgaugechoice}
    U_{T} u_{\bfk}^{\mathrm{I }} &\equiv   {u_{-\bfk}^{\mathrm{II}}}^{*}, \qquad \text{ and } \qquad
    U_{T} u_{\bfk}^{\mathrm{II}} \equiv - {u_{-\bfk}^{\mathrm{I }}}^{*}
\end{align}
The time-reversal symmetry guarantees $\Omega_{\bfk}^{\mathrm{II}} = - \Omega_{-\bfk}^{\mathrm{I}}$
and the total Chern number is always zero.
However, the $\bbZ_2$ topological index $\nu$ can be written in terms of the Chern number of each band as
\begin{align}
    \nu 
      &= 
        \frac{\calC^{\mathrm{I}} - \calC^{\mathrm{II}}}{2}
      = \calC^{\mathrm{I}}
      = \frac{1}{2\pi} \int_{\mathrm{BZ}} \Omega^{\mathrm{I}} \; \mathrm{d}^{2} k
        \qquad\text{(mod 2)}
\end{align}
From Eq.~\eqref{eq:timereversalgaugechoice} we can relate the Berry connection in the two bands as
\begin{align}
    \bfA_{\bfk}^{\mathrm{II}} &= + \bfA_{-\bfk}^{\mathrm{I}}.
\end{align}
(Note the plus sign.)
Now, the following expression written in terms of the total Berry curvature and the total Berry connection evaluated over half of the Brillouin zone [see Fig.~\ref{fig:halfbz}] can be written in terms of those of band I:
\begin{align}
&
\int_{D^+}\!\!\Omega_{\bfk} \;\mathrm{d}^{2} k
+
\int_{\partial D^{-}}\!\!\bfA_{\bfk} \cdot \mathrm{d} \bfk
    \\
    &\quad=
        \int_{D^+}\!\!\!\! (\Omega_{\bfk}^{\rmI} + \Omega_{-\bfk}^{\rmI}) \mathrm{d}^{2} k
        + 2 \left(
        - \int_{D^-}\!\!\!\! \Omega_{\bfk}^{\rmI} \mathrm{d}^{2} k
        +
        \int_{\partial D^{-}}\!\!\!\! \bfA_{\bfk}^{\rmI} \cdot \mathrm{d} \bfk
        \right).
    \nonumber
\end{align}
The first term on the right hand side is the total Berry flux of the band I;
the expression in the parentheses evaluates to an integer multiple of $2\pi$, since both of its integrals evaluate the flux through $\calD^{-}$, modulo $2\pi$.
Therefore, the Chern number of band I, and thus $\nu$, satisfies
\begin{align}
\nu
  &=
    \calC^{\mathrm{I}}
  =
    \frac{1}{2\pi}
    \int_{D^+} \!\! \Omega_{\bfk} \;\rmd^{2} k
    +
    \frac{1}{2\pi}
    \int_{\partial D^{-}} \!\! \bfA_{\bfk} \cdot \rmd \bfk
  \quad\text{(mod 2)}
\end{align}
which can be calculated numerically as
\begin{align}
  \nu
    &=
      \frac{1}{2\pi} \sum_{\square \in \calD^+} \Omega_{\square}
      + \frac{1}{2\pi} \sum_{ \overleftarrow{ij} \in \partial \calD^{-}} \mathrm{Arg} U_{ij}  
\end{align}

\begin{figure*}\centering%
\subfigure[\label{fig:topoindex-vorticity-chern}]{\includegraphics[height=1.8in]{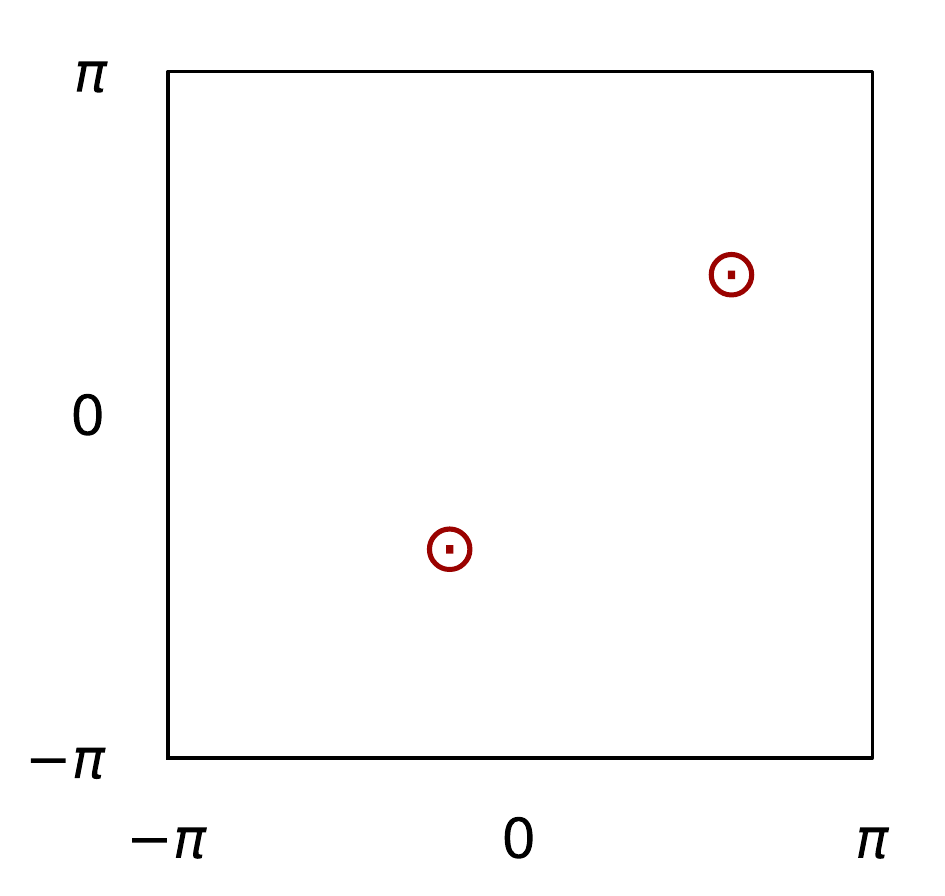}}\qquad
\subfigure[\label{fig:topoindex-vorticity-z2}]{\includegraphics[height=1.8in]{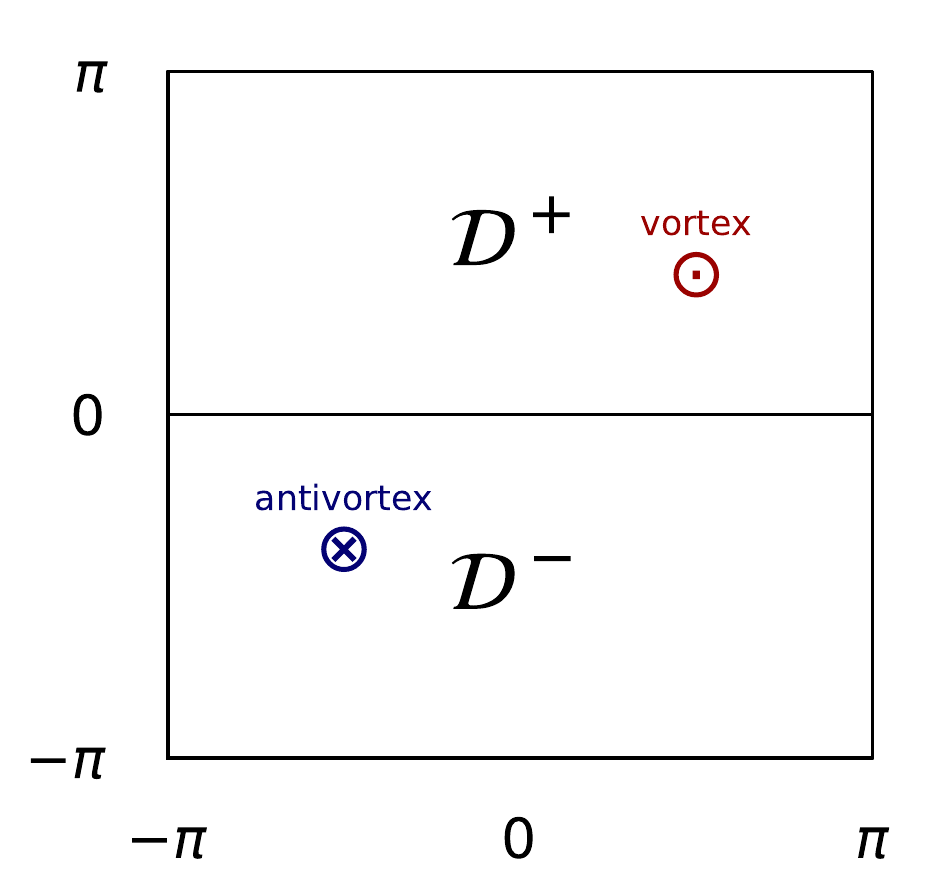}}\qquad
\subfigure[\label{fig:topoindex-vorticity-vanish}]{\includegraphics[height=1.8in]{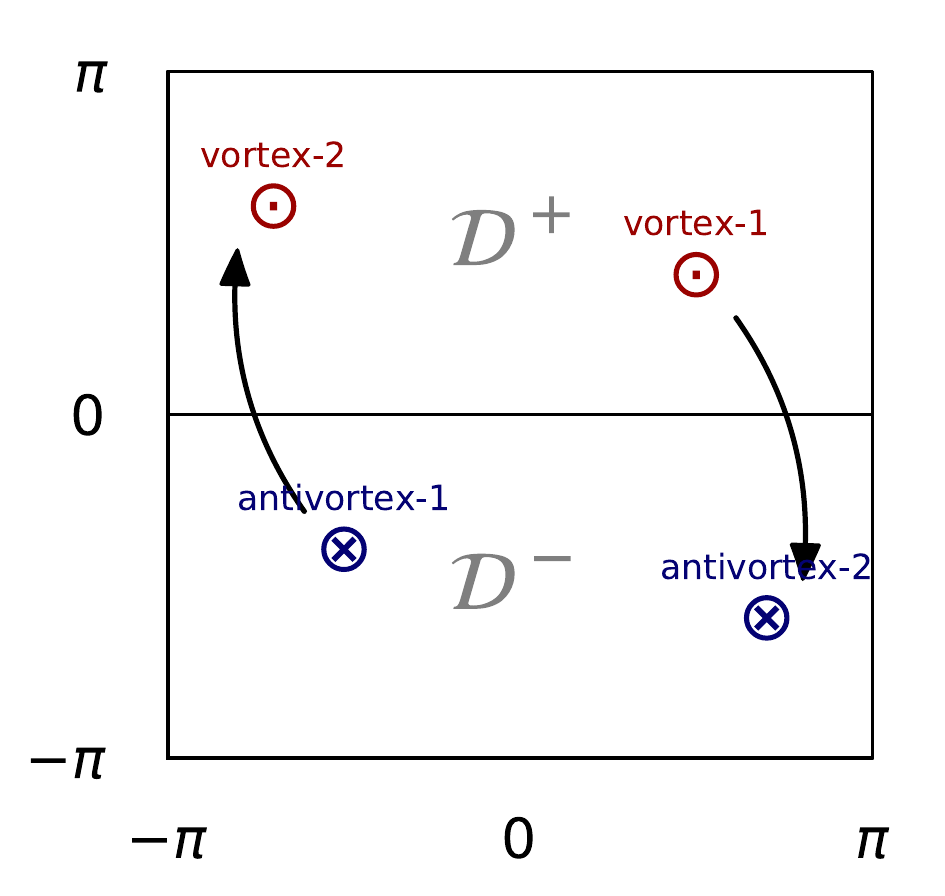}}
\caption{\label{fig:topoindex-vorticity}%
\subref{fig:topoindex-vorticity-chern}
Vortices of Bloch wave function within Brillouin zone in a band structure with Chern number $\calC=2$.
\subref{fig:topoindex-vorticity-z2}.
In a time-reversal-invariant band structure, vortex and antivortex always come in pairs.
Their locations are gauge dependent; the gauge choice Eq.~\eqref{eq:timereversalgaugechoice} ensures that they are at opposite momenta.
\subref{fig:topoindex-vorticity-vanish}
Even vorticity is equivalent to zero, since Eq.~\eqref{eq:timereversalgaugechoice} does not constrain the relative positions of different vortex-antivortex pairs.
}
\end{figure*}

\para{}
As pointed out by \textcite{kohmoto-ap-1985}, the Chern number of a band structure is the total vorticity of its Bloch wave function in the entire Brillouin zone [see Fig.~\ref{fig:topoindex-vorticity-chern}];
while choosing a different gauge can move the positions of the vortices, or create vortex-antivortex pairs, the total vorticity remains independent of the gauge choice.
For a time-reversal-invariant band structure, however, the total vorticity is zero, since vortex and antivortex always come in pairs.
We can, nevertheless, ensure that the vortex and its time-reversal partner antivortex lie at opposite momenta (and also away from time-reversal-invariant momenta), by enforcing the condition \eqref{eq:timereversalgaugechoice} [see Fig.~\ref{fig:topoindex-vorticity-z2}].
Then, the vorticity in half of the Brillouin zone gives us the topological index.

\para{}
From this argument it is also easy to see that the topological classes for $\calT$-invariant Hamiltonians in 2D form a $\bbZ_2$ group and not $\bbZ$, i.e., even vorticity is equivalent to trivial.
Consider the case where we have vorticity of $+2$ in half of the Brillouin zone as in Fig.~\ref{fig:topoindex-vorticity-vanish}, with vortex-antivortex pairs 1 and 2.
Since Eq.~\eqref{eq:timereversalgaugechoice} only constrains the relative positions of vortex and antivortex that are time-reversal partners and not the relative positions of different pairs,
we can move the positions of the vortex-1 and antivortex-1 and annihilate them by combining them, respectively with antivortex-2 and vortex-1.

\subsection{Time reversal operator in Nambu space}

\para{}
In a time-reversal-invariant superconductor with $\calT^2=-1$ (class DIII), the $\bbZ_2$ topological index $\tilde{\nu}$ can be calculated using the same method as $\nu$ as we have described so far.
A crucial step is to fix the gauge of the wave function at $\bfk$ relative to its time-reversed partner at $-\bfk$ [Eq.~\eqref{eq:timereversalgaugechoice}].
As we show below, this is non-trivial when the U(1) gauge symmetry is broken in a superconductor and forces us to address what time-reversal invariance means in this situation.
For our purposes, we consider a Hamiltonian to be $\calT$ invariant if there exists a gauge in which $[ \calT,H ]=0$.
Here, we outline a prescription to identify this gauge.

\para{}
If a normal-state Hamiltonian $H_0$ is invariant under time-reversal operator $\calT = \calK U_{T}$, $H_0$ should satisfy $U_{T} H_0 U_{T}^{\dag} = H_{0}^{*}$.
For a BdG Hamiltonian
\begin{align}
  \tilde{H} &= \begin{pmatrix} H_0 & \Delta \\ \Delta^{\dag} & -H_{0}^{\intercal} \end{pmatrix},
\end{align}
we can na\"ively extend the time-reversal operator to Nambu space as $\tilde{\calT} = \calK \tilde{U}_{T}$, where
\begin{align}
  \label{eq:nambu-tr-0}
  \tilde{U}_{T} &= \begin{pmatrix} U_T & 0 \\ 0 & U_T^{*} \end{pmatrix}.
\end{align}
Then, under time reversal, $\tilde{H}$ transforms as
\begin{align}
  \label{eq:nambu-tr-trans-0}
  \tilde{U}_{T} \tilde{H} \tilde{U}_{T}^{\dag}
    &=
        \begin{pmatrix}
            H^{*}
                & U_{T} \Delta U_{T}^{\intercal} \\
                  \left( U_{T} \Delta U_{T}^{\dag} \right)^{\dag}
                & - H^{\dag}
        \end{pmatrix}.
\end{align}
Therefore, if 
\begin{align}
  \label{eq:nambu-tri-delta-0}
  U_T \Delta U_{T}^{\intercal} = \Delta^{*},
\end{align}
$\tilde{H}$ satisfies
\begin{align}
  \label{eq:nambu-tri-whole}
  \tilde{U}_{T} \tilde{H} \tilde{U}_{T}^{\dag} &= \tilde{H}^{*}
\end{align}
which appears identical to the time-reversal invariance of an insulating Hamiltonian.

\para{}
The problem, however, is that the overall phase of $\Delta$ is not a physical quantity, and the time-reversal invariance should not depend on it.
To resolve this, we introduce a phase $\phi$ to the time-reversal operator
\begin{align}
  \label{eq:nambu-tr-1}
  \tilde{U}_{T} &= \begin{pmatrix} U_T & 0 \\ 0 & e^{i\phi} U_T^{*} \end{pmatrix}
  \tag{\ref*{eq:nambu-tr-0}$'$}
\end{align}
under which $\tilde{H}$ transforms as
\begin{align}
  \tilde{U}_{T} \tilde{H} \tilde{U}_{T}^{\dag}
    &=
      \begin{pmatrix}
        H^{*}
          & U_{T} \Delta U_{T}^{\intercal} e^{-i \phi} \\
        e^{i \phi} \left( U_{T} \Delta U_{T}^{\dag} \right)^{\dag}
          & - H^{\dag}
        \end{pmatrix}
  \tag{\ref*{eq:nambu-tr-trans-0}$'$}
\end{align}
and the condition for time-reversal invariance is
\begin{align}
  \label{eq:nambu-tri-delta}
    U_T \Delta U_{T}^{\intercal} e^{-i \phi} = \Delta^{*}.
    \tag{\ref*{eq:nambu-tri-delta-0}$'$}
\end{align}
For $\Delta=0$, this is satisfied for any value of $\phi \in [0. 2\pi)$, and the time-reversal invariance only depends on how $H_0$ transforms under $U_{T}$.
For $\Delta \neq 0$, on the other hand, there is a unique value of $\phi$ which satisfies Eq.~\eqref{eq:nambu-tri-delta}, given an instance of $\Delta$.

\para{}
For a self-consistently found $\Delta_{ij}$, where $i$ and $j$ represent all local degrees of freedom including site, orbital, and spin, this phase $\phi$, if it exists, needs to satisfy
\begin{align}
  \left[
    \sum_{k,l} [U_{T}]_{ik} \Delta_{kl} [U_T]_{lj}
  \right]
  e^{-i \phi} = \Delta_{ij}^{*}
\end{align}
for all $i, j$.
Thus, we can first choose $\phi$ as
\begin{align}
  \phi &=
    \mathrm{Arg}
      \left[
        \sum_{i,j}
        \sum_{k,l} [U_{T}]_{ik} \Delta_{kl} [U_T]_{lj}
        \Delta_{ij}
      \right],
\end{align}
and use this $\phi$ to construct the time-reversal operator $\tilde{U}_{T}$.
We can then check whether Eq.~\eqref{eq:nambu-tri-whole} is satisfied, after which we can compute the $\bbZ_2$ topological index.

\section{Structure of the Bogoliubov--de Gennes Hamiltonian}

\begin{figure*}
\subfigure[\label{fig:supercell-sitelabel}]{\includegraphics[height=1.6in]{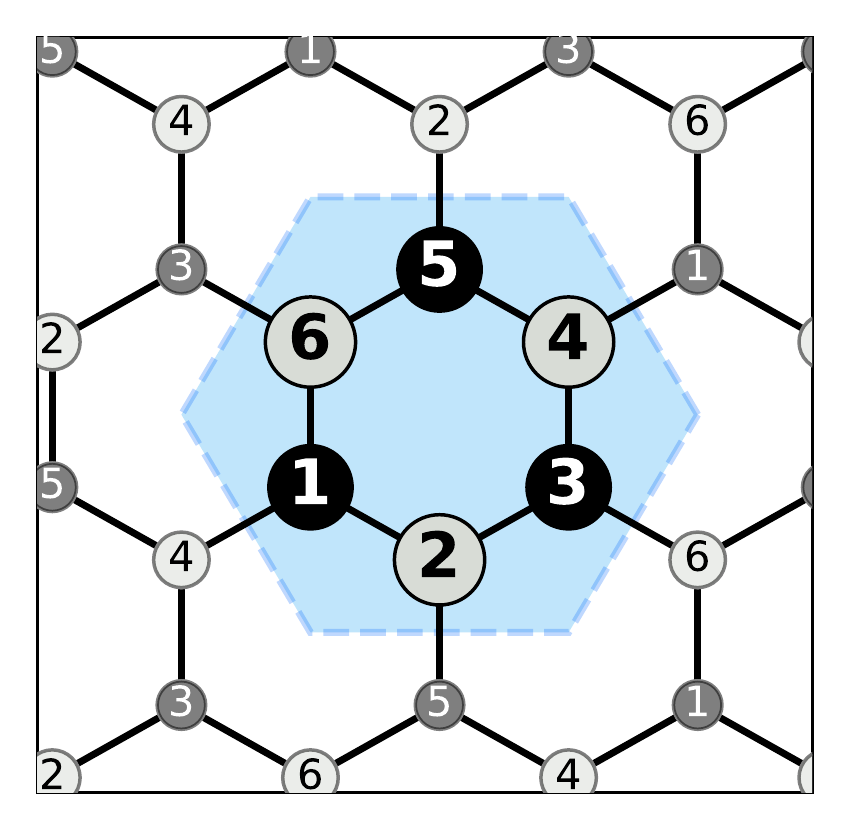}}%
\subfigure[\label{fig:honeycomb-vectors}]{\includegraphics[height=1.6in]{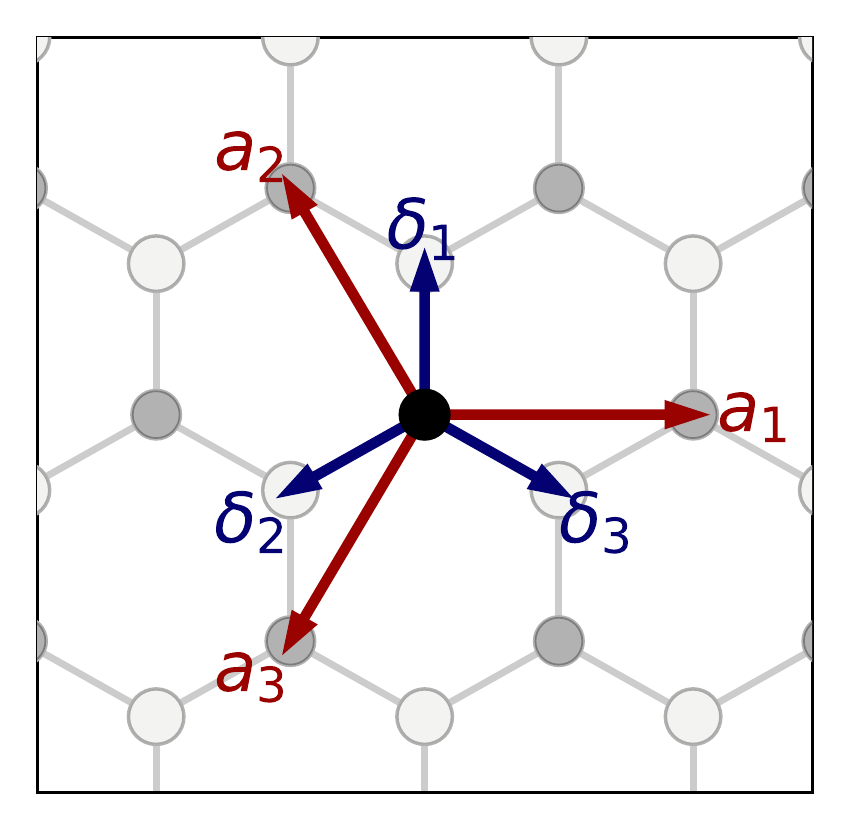}}
\subfigure[\label{fig:bzsplit}
]{\includegraphics[height=1.3in]{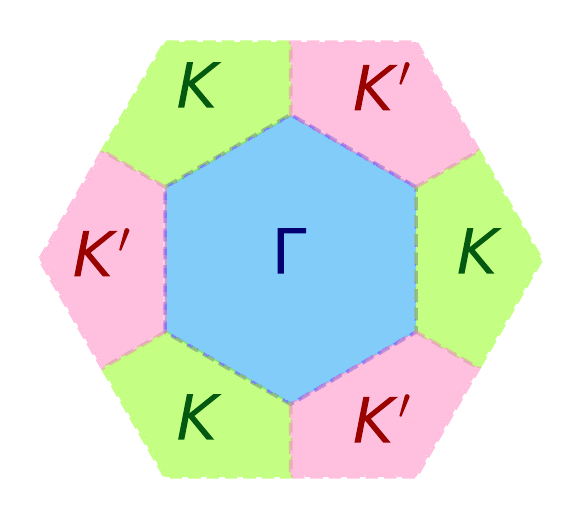}}
\caption{%
\subref{fig:supercell-sitelabel}
$\sqrt{3} \times \sqrt{3}$ supercell containing six sites, which are labeled from 1 to 6.
\subref{fig:honeycomb-vectors}
Vectors $\boldsymbol{\delta}_{i}$ and $\bfa_{i}$ for $i=1,2,3$ connecting nearest- and next-nearest-neighboring sites, respectively.
\subref{fig:bzsplit}
The original Brillouin zone of the honeycomb lattice (the large outer hexagon) can be partitioned into three regions around $\Gamma$, $K$, and $K'$;
the region near the $\Gamma$ is the reduced first Brillouin zone of the $\sqrt{3} \times \sqrt{3}$ supercell.
}
\end{figure*}

\para{}
As explained in Sec.~\ref{sec:model}, we work with a supercell containing six sites, which are labeled in Fig.~\ref{fig:supercell-sitelabel}.
The Bogoliubov--de Gennes Hamiltonian in this basis, at each crystal momentum $\bfk$ in the reduced Brillouin zone, is a $24\times24$ matrix: 6 for sites, 2 for spins, and 2 for Nambu space:
\begin{widetext}
\begin{align}
H_{\mathrm{BdG}}(\bfk)
  &=
    \begin{pmatrix}
      H_{\mathrm{KM},\up}^{(6)}(\bfk) &
        0 &
        \Delta_{\up\up}^{\mathrm{nn}}(\bfk) &
        \psi^{\mathrm{os}} + \Delta_{\up\dn}^{\mathrm{nn}}(\bfk)
    \\
        0 &
        H_{\mathrm{KM},\dn}^{(6)} (\bfk) &
        -\psi^{\mathrm{os}} + \Delta_{\dn\up}^{\mathrm{nn}}(\bfk) &
        \Delta_{\dn\dn}^{\mathrm{nn}}(\bfk)
    \\
      [{\Delta_{\up\up}^{\mathrm{nn}}}]^{\dagger}(\bfk) &
        [ -{\psi^{\mathrm{os}}} + {\Delta_{\dn\up}^{\mathrm{nn}}}(\bfk)]^{\dagger} &
            -[H_{\mathrm{KM},\up}^{(6)}(-\bfk)]^{\mathsf{T}} &
                0
    \\
      [{\psi^{\mathrm{os}}} + {\Delta_{\up\dn}^{\mathrm{nn}}}(\bfk)]^{\dagger} &
        [{\Delta_{\dn\dn}^{\mathrm{nn}}}(\bfk)]^{\dagger} &
            0 &
                -[H_{\mathrm{KM},\dn}^{(6)}(-\bfk)]^{\mathsf{T}} 
    \\
    \end{pmatrix}.
\end{align}
Here, $H_{\mathrm{KM},\sigma}^{(6)}(\bfk)$ is a $6\times6$ matrix representing the Kane-Mele Hamiltonian in the six-site supercell basis $(c_{\bfk 1\sigma}, c_{\bfk 2\sigma},~...~c_{\bfk 6\sigma})$
for spin $\sigma=\up,\dn$ at momentum $\bfk$ in the reduced Brillouin zone, which can be written in terms of the hopping parameter $t$, sublattice potential $m_{\mathrm{AB}}$, and spin-orbit coupling $\lambda_{\text{so}}$ as
\begin{align}
H_{\mathrm{KM},\sigma}^{(6)}(\bfk) &=
  \begin{pmatrix}
    +m_{\mathrm{AB}} &
        -t e^{ i \bfk \cdot \boldsymbol{\delta}_3} &
        -i \lambda_{\text{so}} \; \sigma \; \varphi_{\text{nnn}} ( \bfk) &
        -t e^{ i \bfk \cdot \boldsymbol{\delta}_2} &
        +i \lambda_{\text{so}} \; \sigma \; \varphi_{\text{nnn}} (-\bfk) &
        -t e^{ i \bfk \cdot \boldsymbol{\delta}_1}
    \\
         &
    -m_{\mathrm{AB}} &
        -t e^{-i \bfk \cdot \boldsymbol{\delta}_2} &
        +i \lambda_{\text{so}} \; \sigma \; \varphi_{\text{nnn}} (-\bfk) &
        -t e^{-i \bfk \cdot \boldsymbol{\delta}_1} &
        -i \lambda_{\text{so}} \; \sigma \; \varphi_{\text{nnn}} ( \bfk) &
    \\
         &
         &
    +m_{\mathrm{AB}} &
        -t e^{ i \bfk \cdot \boldsymbol{\delta}_1} &
        -i \lambda_{\text{so}} \; \sigma \; \varphi_{\text{nnn}} ( \bfk) &
        -t e^{ i \bfk \cdot \boldsymbol{\delta}_3} &
    \\
         &
         &
         &
    -m_{\mathrm{AB}} &
        -t e^{-i \bfk \cdot \boldsymbol{\delta}_3} &
        +i \lambda_{\text{so}} \; \sigma \; \varphi_{\text{nnn}} (-\bfk) &
    \\
         &
        \text{H.c.} &
         &
         &
    +m_{\mathrm{AB}} &
        -t e^{+i \bfk \cdot \boldsymbol{\delta}_2} &
    \\
         &
         &
         &
         &
         &
    -m_{\mathrm{AB}} 
  \end{pmatrix},
\end{align}
\end{widetext}
where
$\varphi_{\mathrm{nnn}}(\bfk) \equiv \sum_{i=1}^{3} e^{i \bfk \cdot \bfa_i}$, and $\sigma=+1(-1)$ for up (down) spin.
$\boldsymbol{\delta}_{i}$ and $\bfa_{i}$ are the vectors connecting the nearest- and next-nearest-neighboring sites, respectively, as defined in Fig.~\ref{fig:honeycomb-vectors}.

\para{}
With onsite interaction $U$, only the onsite spin-singlet component $\psi^{\mathrm{os}}$ is allowed, defined on every site:
$\psi^{\mathrm{os}}
=\mathrm{diag}(
    \psi_{1}^{\mathrm{os}}, 
    \psi_{2}^{\mathrm{os}}, 
    \ldots,
    \psi_{6}^{\mathrm{os}}
    )
$.
With nearest-neighbor density-density interaction $V$ or the Heisenberg interaction $J$, pair potential is defined on every nearest-neighbor bonds:
\begin{widetext}
\begin{align}
\Delta_{\sigma, \sigma'}^{\mathrm{nn}}(\bfk)
  &=
\begin{pmatrix}
         0 &
         \Delta_{1\sigma,2\sigma'}^{\mathrm{nn}} e^{+i \bfk \cdot \boldsymbol{\delta}_3} &
         0 &
         \Delta_{1\sigma,4\sigma'}^{\mathrm{nn}} e^{+i \bfk \cdot \boldsymbol{\delta}_2} &
         0 &
         \Delta_{1\sigma,6\sigma'}^{\mathrm{nn}} e^{+i \bfk \cdot \boldsymbol{\delta}_1}
    \\
         \Delta_{2\sigma,1\sigma'}^{\mathrm{nn}} e^{-i \bfk \cdot \boldsymbol{\delta}_3} &
         0 &
         \Delta_{2\sigma,3\sigma'}^{\mathrm{nn}} e^{-i \bfk \cdot \boldsymbol{\delta}_2} &
         0 &
         \Delta_{2\sigma,5\sigma'}^{\mathrm{nn}} e^{-i \bfk \cdot \boldsymbol{\delta}_1} &
         0
    \\
         0 &
         \Delta_{3\sigma,2\sigma'}^{\mathrm{nn}} e^{+i \bfk \cdot \boldsymbol{\delta}_2} &
         0 &
         \Delta_{3\sigma,4\sigma'}^{\mathrm{nn}} e^{+i \bfk \cdot \boldsymbol{\delta}_1}&
         0 &
         \Delta_{3\sigma,6\sigma'}^{\mathrm{nn}} e^{+i \bfk \cdot \boldsymbol{\delta}_3}
    \\
         \Delta_{4\sigma,1\sigma'}^{\mathrm{nn}} e^{-i \bfk \cdot \boldsymbol{\delta}_2} &
         0 &
         \Delta_{4\sigma,3\sigma'}^{\mathrm{nn}} e^{-i \bfk \cdot \boldsymbol{\delta}_1} &
         0 &
         \Delta_{4\sigma,5\sigma'}^{\mathrm{nn}} e^{-i \bfk \cdot \boldsymbol{\delta}_3} &
         0
    \\
         0 &
         \Delta_{5\sigma,2\sigma'}^{\mathrm{nn}} e^{+i \bfk \cdot \boldsymbol{\delta}_1} &
         0 &
         \Delta_{5\sigma,4\sigma'}^{\mathrm{nn}} e^{+i \bfk \cdot \boldsymbol{\delta}_3}&
         0 &
         \Delta_{5\sigma,6\sigma'}^{\mathrm{nn}} e^{+i \bfk \cdot \boldsymbol{\delta}_2}
    \\
         \Delta_{6\sigma,1\sigma'}^{\mathrm{nn}} e^{-i \bfk \cdot \boldsymbol{\delta}_1} &
         0 &
         \Delta_{6\sigma,3\sigma'}^{\mathrm{nn}} e^{-i \bfk \cdot \boldsymbol{\delta}_3} &
         0 &
         \Delta_{6\sigma,5\sigma'}^{\mathrm{nn}} e^{-i \bfk \cdot \boldsymbol{\delta}_2} &
         0 
\end{pmatrix}.
\end{align}
\end{widetext}
Fermion anticommutation requires $\Delta_{i\sigma,j\sigma'}^{\mathrm{nn}} = - \Delta_{j\sigma',i\sigma}^{\mathrm{nn}}$.
The nearest-neighbor pair potential can contain both spin-singlet and spin-triplet components.
Within our study, however, we have found only the spin-triplet components of the $\Delta^{\mathrm{nn}}$ to be nonzero with attractive density-density interaction $V$, and only the spin-singlet components to be nonzero for antiferromagnetic Heisenberg interaction $J$.

\para{}
Equivalently, we can express the BdG Hamiltonian of the $\sqrt{3}\times\sqrt{3}$ supercell completely in momentum space.
In terms of the basis
$\psi_{\bfk}^{\dagger} = (c_{\bfk}^{\dagger}, c_{K + \bfk}^{\dagger}, c_{-K + \bfk}^{\dagger},
c_{-\bfk}, c_{-K - \bfk}, c_{K - \bfk})$,
where $c_{\bfk}^{\dagger} = (c_{\bfk,A,\up}^{\dagger},c_{\bfk,B,\up}^{\dagger}, c_{\bfk,A,\dn}^{\dagger},c_{\bfk,B,\dn}^{\dagger})$,
the BdG Hamiltonian is written as
\begin{widetext}
\begin{align}
\label{eq:hamiltonianblock-kspace}
H_{\mathrm{BdG}}(\bfk) &=
    \begin{pmatrix}
      H_{\mathrm{KM}}^{(2)}(\bfk) &
        0 &
        0 &
        \Delta^{\Gamma \Gamma}(\bfk) &
        \Delta^{\Gamma K}(\bfk) &
        \Delta^{\Gamma K'}(\bfk)
        \\
        0 &
        H_{\mathrm{KM}}^{(2)}(K+\bfk) &
        0 &
        \Delta^{K \Gamma}(\bfk) &
        \Delta^{K K}(\bfk) &
        \Delta^{K K'}(\bfk)        
        \\
        0 &
        0 &
        H_{\mathrm{KM}}^{(2)}(-K+\bfk) &
        \Delta^{K' \Gamma}(\bfk) &
        \Delta^{K' K}(\bfk) &
        \Delta^{K' K'}(\bfk)
        \\
        &
        &
        &
        -[H_{\mathrm{KM}}^{(2)}(-\bfk)]^{\mathsf{T}} &
        0 &
        0 
        \\
        &
        \text{H.c.} &
        &
        0 &
        -[H_{\mathrm{KM}}^{(2)}(-K-\bfk)]^{\mathsf{T}} &
        0         
        \\
        &
        &
        &
        0 &
        0 &
        -[ H_{\mathrm{KM}}^{(2)}(K-\bfk)]^{\mathsf{T}}
    \end{pmatrix},
\end{align}
\end{widetext}
where $\bfk$ spans the reduced Brillouin zone [blue region in Fig.~\ref{fig:bzsplit}] and the degrees of freedom in the rest of the Brillouin zone appear in $H_{\mathrm{BdG}}(\bfk)$ as additional bands. 
$H_{\mathrm{KM}}^{(2)}(\bfk)$ is a $4\times4$ matrix representing the Kane-Mele Hamiltonian in the two-site unit-cell basis.
This is block-diagonal in spin-space with the following representation in the basis of sublattice eigenstates created by $c_{\bfk A\sigma},c_{\bfk B\sigma}$:
\begin{align}
    \begin{pmatrix}
             m_{\rm AB} + \lambda_{\mathrm{so}} \sigma \tilde{\varphi}_{\mathrm{nnn}}(\bfk)  - \mu
             & 
             -t \gamma(\bfk)
             \\
             -t \gamma^*(\bfk)
             &
             -m_{\rm AB} - \lambda_{\mathrm{so}} \sigma \tilde{\varphi}_{\mathrm{nnn}}(\bfk)  - \mu
    \end{pmatrix},
\end{align}
where $\tilde{\varphi}_{\mathrm{nnn}}(\bfk) \equiv 2\sum_{i=1}^{3} \sin(\bfk \cdot \bfa_i)$, and $\gamma(\bfk)=\sum_{l=1}^{3} e^{i \bfk \cdot \boldsymbol{\delta}_{l}}$ and $\sigma=+1(-1)$ for up (down) spin.

\para{}
The diagonal blocks of the pair potential $\Delta^{\Gamma\Gamma}$, $\Delta^{KK}$, and $\Delta^{K'K'}$ represent pairing  between valleys (zero c.m. momentum), while the off-diagonal blocks represent pairing within the valley (nonzero c.m. momentum).
Especially since the low-energy fermionic degrees of freedom lie in the region near $K$ and $K'$, we expect that pairing will develop within and between these regions:
\begin{align}
    \begin{pmatrix}
        \Delta^{\Gamma \Gamma}(\bfk) &
        \Delta^{\Gamma K}(\bfk) &
        \Delta^{\Gamma K'}(\bfk)
        \\
        \Delta^{K \Gamma}(\bfk) &
        \Delta^{K K}(\bfk) &
        \Delta^{K K'}(\bfk)        
        \\
        \Delta^{K' \Gamma}(\bfk) &
        \Delta^{K' K}(\bfk) &
        \Delta^{K' K'}(\bfk)
    \end{pmatrix}
    \approx
    \begin{pmatrix}
        0 &
        0 &
        0 &
        \\
        0 &
        \Delta^{K K}(\bfk) &
        \Delta^{K K'}(\bfk)        
        \\
        0 &
        \Delta^{K' K}(\bfk) &
        \Delta^{K' K'}(\bfk)
    \end{pmatrix}.
\end{align}
We describe the momentum-space representation in more detail in the next appendix.

\section{Momentum space description of intra-valley pairing}
\label{sec:kspacedescription}

\para{}
Considering the pairing Hamiltonian in momentum space illuminates several interesting subtleties of the exotic paired states we have described.
For instance, with nearest-neighbor attraction $V$, the order parameter in each spin channel has nine degrees of freedom  corresponding to the nearest-neighbor bonds in the supercell.
However, it turns out as we have shown in Table~\ref{tab:phases-nn} that the rich phase diagram of Fig.~\ref{fig:pd-nn} is described by only two spatial form factors $\Phi^K$ and $\Phi^{K'}$.
In Appendix~\ref{sec:MomentumSpaceDecoupling}, we show why this is the natural choice for spatial order parameter. In general, there can be a relative phase between the two condensates described by these order parameters, when they coexist in any spin sector.
In Appendix~\ref{sec:sVspKekule}, we show how this relative phase distinguishes the $p$-Kekule and $s$-Kekule form factors, and demonstrate that the energy barrier between these configurations is quite small.
In Appendix~\ref{sec:swaveToP+IP}, we show that in the basis of band eigenstates in the vicinity of the valleys, these order parameters $\Phi^K$ and $\Phi^{K'}$ have the form of a $p \pm ip$ pair potential.

\subsection{Mean-field decomposition in momentum space}
\label{sec:MomentumSpaceDecoupling}

\para{}
We have considered a large set of pairing order parameters to exhaust all possible symmetry-breaking configurations of $\Delta_{i\sigma;j\sigma'}$ within the unit cell. A complementary, but equivalent, description of the mean-field order parameter involves decoupling the interaction in momentum space. For the Hamiltonian with nearest-neighbor density-density attraction,
\begin{align}
    \calH
    =
        \calH_{\rm KM} - 
        \frac{|V|}{N} 
        \sum_{\bfQ \bfk \bfk' \boldsymbol{\delta}_l,\sigma\sigma} 
                &c_{\bfQ+\bfk,A\sigma}^\dagger c_{\bfQ-\bfk,B\sigma'}^\dagger
                e^{i \bfk \cdot \boldsymbol{\delta}_l} 
                \nonumber \\
                & c_{\bfQ-\bfk',B\sigma'}^{\phantom\dagger}
                c_{\bfQ+\bfk',A\sigma}^{\phantom\dagger}
                e^{i \bfk' \cdot \boldsymbol{\delta}_l}
\end{align}
the mean-field Hamiltonian  takes the form
\begin{align}
    \calH_{\mathrm{BdG}}
    &= 
        \calH_{\mathrm{KM}} +
        \sum_{\bfQ \bfk \sigma \sigma'}
            \Big[
                \Delta_{\bfQ \sigma \sigma'} (\bfk)
                c_{\bfQ+\bfk,A\sigma}^\dagger c_{\bfQ-\bfk,B\sigma'}^\dagger
                + \mathrm{H.c.}
            \Big]
        \nonumber\\
    &\qquad\quad\;
        + \frac{N}{|V|}
        \sum_{ \bfQ \boldsymbol{\delta}_l \sigma \sigma'}
            |\tilde{\Delta}_{\bfQ \boldsymbol{\delta}_l \sigma \sigma'}|^2,
\end{align}
where $2\bfQ$ is the center-of-mass momentum of the pairs with $\bfQ=\Gamma$,$K$, or $K'$;
$2N$ is the number of sites and $\boldsymbol{\delta}_l$ are the vectors corresponding to the nearest neighbor bonds in Fig.~\ref{fig:honeycomb-vectors}.
The pair potential $\Delta_{\bfQ\sigma\sigma'}(\bfk)$ is defined in terms of the order parameters $\tilde{\Delta}_{\bfQ\boldsymbol{\delta}_{l}\sigma\sigma'}$ along a bond direction $\boldsymbol{\delta}_{l}$ by 
\begin{align}
    \Delta_{\bfQ\sigma\sigma'}(\bfk)
    =&
    \sum_{\boldsymbol{\delta}_{l}}
    \tilde{\Delta}_{\bfQ\boldsymbol{\delta}_{l}\sigma\sigma'}
    e^{i \bfk\cdot\boldsymbol{\delta}_{l}}
    \\
    \tilde{\Delta}_{\bfQ\boldsymbol{\delta}_{l}\sigma\sigma'}
    = -\frac{|V|}{N}
    &\sum_{\bfp} \langle c_{\bfQ-\bfp,B\sigma'} c_{\bfQ+\bfp,A\sigma}
    \rangle e^{-i \bfp\cdot\boldsymbol{\delta}_{l}}
\end{align}
Hereafter, spin indices are dropped whenever the statements apply to order parameters in all spin channels.

\para{}
For the low-energy fermions living at the valleys,
$\tilde{\Delta}_{\bfQ=\Gamma}$ corresponds to inter-valley pairing
and $\tilde{\Delta}_{\bfQ=K(K')}$ to pairing within the $K(K')$ valley. 
The order parameters $\tilde{\Delta}_{\bfQ\boldsymbol{\delta}_{l}}$ are related to the real-space order parameters $\Delta_{i,j}=\langle c_{i} c_{j} \rangle$ by 
\begin{align}
  \Delta_{ij}
  &=
    \frac{1}{3}
    \sum_{\bfQ} \tilde{\Delta}_{\bfQ,\bfr_j-\bfr_i} e^{i \bfQ \cdot (\bfr_i+\bfr_j)},
\end{align}
for $i \in \rmA$ and $j \in \rmB$. 
The dependence of the center-of-mass coordinate and the relative coordinate is explicitly shown. 
In the basis $c_{\bfk}^{\dagger} = (c_{\bfk,A,\up}^{\dagger},c_{\bfk,B,\up}^{\dagger}, c_{\bfk,A,\dn}^{\dagger},c_{\bfk,B,\dn}^{\dagger})$, the pairing terms in the Hamiltonian take the form $c_{\bfQ+\bfk}^{\dagger} \hat{\Delta}_\bfQ(\bfk) c_{\bfQ-\bfk}^{\dagger}$ with 
\begin{align}
    \hat{\Delta}_\bfQ(\bfk) =
    \begin{pmatrix}
        0 &
        \Delta_{\bfQ \uparrow\uparrow}(\bfk) &
        0 &
        \Delta_{\bfQ \uparrow\downarrow}(\bfk)
        \\
        -\Delta_{\bfQ \uparrow\uparrow}(-\bfk) &
        0 &
        -\Delta_{\bfQ \downarrow\uparrow}(-\bfk) &
        0
        \\
        0 &
        \Delta_{\bfQ \downarrow\uparrow}(\bfk) &
        0 &
        \Delta_{\bfQ \downarrow\downarrow}(\bfk) &
        \\
        -\Delta_{\bfQ \uparrow\downarrow}(-\bfk) &
        0 &
        -\Delta_{\bfQ \downarrow\downarrow}(-\bfk) &
        0
    \end{pmatrix}
\end{align}
This is the same as the matrices $\Delta^{\bfP \bfP'}(\bar{\bfk})$ in Eq.~\eqref{eq:hamiltonianblock-kspace} if we identify the c.m. momentum as $2\bfQ = \bfP - \bfP'$ and the relative momentum as $2\bfk = 2\bar{\bfk} + \bfP -\bfP'$.
The Bogoliubov quasiparticle wave functions $\ket{\Psi_{\bfk n}}$ are obtained as the eigenvectors of $H_{\mathrm{BdG}}(\bfk)$ in Eq.~\eqref{eq:hamiltonianblock-kspace} with the self-consistent order parameters obtained by solving
\begin{align}
  \tilde{\Delta}_{\bfQ \boldsymbol{\delta}_l \sigma\sigma'}
    &=
      -\frac{|V|}{2N} 
      \sum_{\bfk} 
        \left\langle \Psi_{\bfk n} \middle\vert
        \frac{\partial H_{\rm BdG}(\bfk)}{\partial \tilde{\Delta}_{\bfQ \boldsymbol{\delta}_l \sigma\sigma'}^*}
        \middle\vert \Psi_{\bfk n} \right\rangle 
\end{align}
with the $\bfk$ sum spanning the (blue) reduced Brillouin zone in Fig.~\ref{fig:bzsplit}.

\para{}
For nearest-neighbor density-density attraction, the four superconducting phases that we find are all described by $\tilde{\Delta}_{\bfQ\boldsymbol{\delta}_1}=\tilde{\Delta}_{\bfQ\boldsymbol{\delta}_2}=\tilde{\Delta}_{\bfQ\boldsymbol{\delta}_3}$ with $\bfQ=K$ or $K'$. 
This results in a pair potential $\Delta_{\bfQ}(\bfk)$ which is $\bfk$ independent for small $\bfk$, where low-energy fermionic excitations live within a valley.
In real space, this corresponds to a form factor $\Phi^\bfQ_{ij}=e^{i\bfQ\cdot(\bf{r}_i+\bf{r}_j)}$ characteristic of pairing within the valley at $\bfQ$.
It is possible to see that, without breaking $C_3$ rotation, the other two possibilities for pairing within a given valley lead to vanishing pair potential for small $\bfk$ and are therefore energetically unfavorable.

\para{}
The $p$-Kekule SC [blue region in Fig.~\ref{fig:pd-nn}] corresponds to triplet opposite-spin pairing (OSP) at both valleys $\tilde{d}^{z}_{K}, \tilde{d}^{z}_{K'} \neq 0$, resulting in a Larkin-Ovchinnikov--type~\cite{larkin-spj-1965} pair density wave due to interference of the two form factors $\Phi^{\mathrm{LO}}=\Phi^{K}-\Phi^{K'}$ in real space. 
The topological helical SC (green region) corresponds to equal-spin pairing (ESP) at both valleys $\tilde{\Delta}_{K,\up\up}=e^{i\phi}\tilde{\Delta}_{K',\dn\dn}\neq0$, leading to a Fulde-Ferrell--type~\cite{fulde-pr-1964} phase-modulating pair potential in real space for each spin sector. 
The relative phase $e^{i\phi}$ between the condensates does not affect the ground-state energy.

\para{}
We emphasize that although we have considered 36 order parameters to rule out all kinds of symmetry-breaking paired states, the entire phase diagram Fig.~\ref{fig:pd-nn} is described in the six-parameter space spanned by triplet pairing with the spatial form factors $\Phi^{K}$ and $\Phi^{K'}$.

\subsection{\texorpdfstring{Relative phase between condensates at $K$ and $K'$: $s$-Kekule vs. $p$-Kekule}{Relative phase between condensates at K and K': s-Kekule vs. p-Kekule}}
\label{sec:sVspKekule}

\para{}
As we point out in Appendix~\ref{sec:MomentumSpaceDecoupling}, the helical SC has an additional Goldstone mode corresponding to the relative phase of the condensates at $K$ and $K'$. This is because, in addition to the total charge $N_\up+N_\dn$, the charge in each spin sector fluctuates independently. As a result, the ground state breaks an additional U(1) symmetry corresponding to the conservation of $S_z=N_\up-N_\dn$.

\begin{figure}
\centering
\subfigure[\label{fig:FvsDeltaKekule-Polar}%
]{\includegraphics[width=3in]{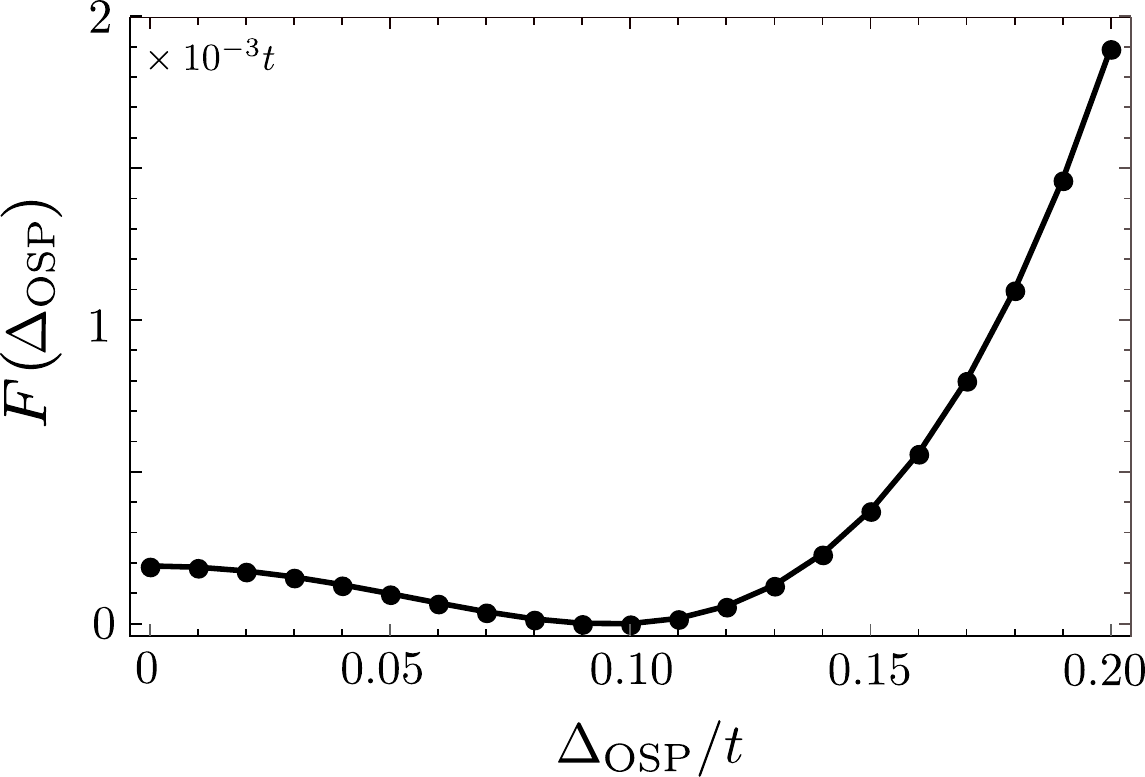}}
\subfigure[\label{fig:FvsDeltaKekule-Cut}%
]{\includegraphics[width=3in]{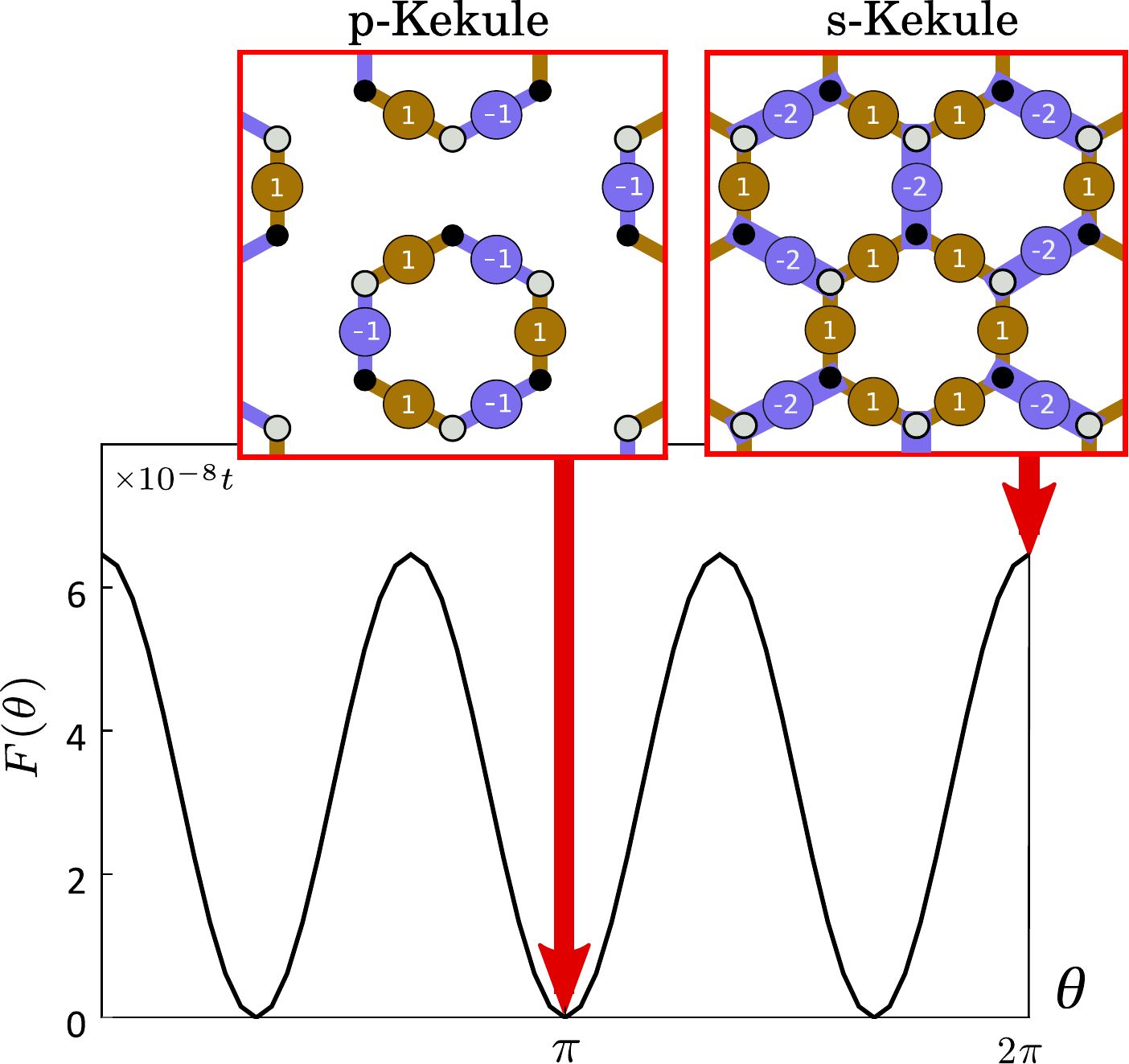}}%
\caption{\label{fig:FvsDeltaKekule}%
Relative phase between the condensates at the two valleys breaks the near degeneracy between the $s$-Kekule and $p$-Kekule configurations.
\subref{fig:FvsDeltaKekule-Polar} 
Free energy as a function of $\Delta_{\mathrm{OSP}}$, the amplitude of the $p$-Kekule order parameter $d_z = \Delta_{\mathrm{OSP}} \left( \Phi^K -\Phi^{K'} \right)$.
\subref{fig:FvsDeltaKekule-Cut}
Free energy as a function of the relative phase $\theta$ in the order parameter $d_z = \Delta_{\mathrm{OSP}} \left( \Phi^K + e^{i\theta} \Phi^{K'} \right)$ at optimal $\Delta_{\mathrm{OSP}}$.
Data shown are at $x=1,V=2.65t$. The energy barrier between the three equivalent $p$-Kekule configurations is less than a mK for $t \sim 1$\,eV, and is expected to decrease with increasing $E_g$.
}
\end{figure}

\para{}
In the $p$-Kekule SC, the low-energy degrees of freedom are still effectively decoupled into valleys, resulting in a U(1)$\times$U(1) symmetry corresponding to charge conservation on each valley. 
However, this is only a symmetry of the low energy effective Hamiltonian. 
Unlike the ESP ground state, the U(1) symmetry corresponding to valley charge conservation is broken by the higher-energy fermionic modes which couple the two condensates. 
As a result, on a lattice, the free energy corresponding to the pair potential $d_z=\Delta_\mathrm{OSP} \left( \Phi^K + e^{i\theta} \Phi^{K'} \right) $ does have a weak dependence on the relative phase $\theta$ between the condensates, as shown in Fig.~\ref{fig:FvsDeltaKekule}(b) which breaks the degeneracy between the $p$-Kekule [$\theta=2\pi (n+1/2)/3$] and $s$-Kekule ($\theta=2\pi n/3$) form factors \cite{roy-prb-2010}.
Here, $n\in \bbZ$.

\subsection{Understanding the non-trivial topology}
\label{sec:swaveToP+IP}

\para{}
Having understood the novel SC states in momentum space, we are now in a position to intuitively understand the reason for their non-trivial topology as captured by the Chern number $\tilde{\calC}$ and the $\bbZ_2$ index $\tilde{\nu}$.

\para{}
At $x=\frac{1}{2}$, the low energy dispersion is dominated by spin-polarized Dirac cones at the two valleys
\begin{align}
    \calH^{\mathrm{eff}}_{\rm KM}
    &=
    -t \sum_{\bfk}
    \gamma(K+\bfk) c_{K+\bfk \rmA \up}^\dagger c_{K+\bfk \rmB\up}^{\phantom\dagger} +\mathrm{H.c.}
    \nonumber\\
    &\quad
    -t \sum_{\bfk} \gamma(K'+\bfk) c_{K'+\bfk \rmA \dn}^\dagger c_{K'+\bfk \rmB \dn}^{\phantom\dagger}
    + \mathrm{H.c.}
\end{align}
where $\gamma(K+\bfk)=\sum_{\boldsymbol{\delta}_{l}} e^{i (K+\bfk)\cdot \boldsymbol{\delta}_{l}}=-3 (k_x+ik_y)/2+O(k^2)$ and $\gamma(K'+\bfk)=\sum_{\boldsymbol{\delta}_{l}} e^{i (K'+\bfk) \cdot \boldsymbol{\delta}_{l}}=-3 (k_x-ik_y)/2+O(k^2)$. 
This is diagonalized by a unitary transformation to the band eigenstates $a_{\bfk\eta}=\sum_{\tau}S_{\eta\tau}c_{\bfk\tau\sigma}$ with $\tau=\rmA,\rmB$ and $\eta=\pm$.
\begin{align}
    \calH^{\mathrm{eff}}_{\rm KM}
    &=
    \frac{3 t k}{2}
    \sum_{\bfk} 
    \big(
        a_{K+\bfk,+,\up}^\dagger a_{K+\bfk,+,\up} +
        a_{K'+\bfk,+,\dn}^\dagger a_{K'+\bfk,+,\dn}
    \nonumber\\
    &\qquad
        - a_{K+\bfk,-,\up}^\dagger a_{K+\bfk,-,\up} +
        a_{K'+\bfk,-,\dn}^\dagger a_{K'+\bfk,-,\dn}
    \big)
\end{align}
The effective mean-field Hamiltonian for equal-spin pairing
is
\begin{align}
    \calH^{\mathrm{eff}}_{\rm BdG}
    &=
    \calH^{\mathrm{eff}}_{\rm KM}
    + \sum_{\bfk} \left(
        \Delta_{\up\up}  c_{K-\bfk \rmB \up}^\dagger c_{K+\bfk \rmA \up}^\dagger + \textrm{H.c.}
    \right)
    \nonumber\\
    &\qquad\qquad
    + \left(
        \Delta_{\dn\dn}
        c_{K'-\bfk \rmB \dn}^\dagger c_{K'+\bfk \rmA \dn}^\dagger + \textrm{H.c.}
    \right)
\end{align}
corresponding to a pair potential that is uniform near the valleys.
In terms of the band eigenstates $a_{\bfk\eta}$, the effective Hamiltonian involves ($p\pm i p$)-wave pairing
\begin{align}
    \calH^{\mathrm{eff}}_{\rm BdG}
    &=
    \calH^{\mathrm{eff}}_{\rm KM}
    +\sum_{\bfk\eta} \left(
        \Delta_{\up\up}
        \frac{k_x-i k_y}{k}
        a_{K-\bfk \eta\up}^\dagger a_{K+\bfk\eta\up}^\dagger
        + \textrm{H.c.}
    \right)
    \nonumber\\
    &\qquad\qquad
    + \left( 
        \Delta_{\dn\dn}
        \frac{k_x+i k_y}{k}
        a_{K'-\bfk \eta\dn}^\dagger a_{K'+\bfk \eta\dn}^\dagger
        + \textrm{H.c.} 
    \right).
\end{align}
The order parameter in each spin sector has a Chern number that reflects the chirality of the pair potential.
This results in a helical SC with a non-trivial $\mathbb{Z}_2$ topological index. The net Chern number is 0 as required by time-reversal invariance.

\para{}
It is now easy to see why the purple region in Fig.~\ref{fig:pd-nn} of the main text is a chiral SC.
It has ESP on one valley with a charactersitic chirality and a nonzero Chern number and OSP on the other.
OSP entails twice the Chern number characteristic of the valley it pairs in, since there are two bands with the same winding involved.
This state is $\calT$ breaking and has a net Chern number of $\pm 1$.
Uniform pairing within an odd number of Dirac cones turns out to be the crucial ingredient for a topological superconductor in Dirac systems.

\para{}
The $\calT$-breaking SC has both ESP and OSP pairing in both valleys, and is topologically trivial.

\section{Transformation of order parameters under symmetry operations}
\label{sec:symop}

\begin{table}[]
\centering
\caption{%
Transformations of spin-triplet pairing order parameters with finite center-of-mass momentum $K$.
$\Delta_{\sigma\sigma}^{\bfQ}$ (and $d_z^{\bfQ}$) is a shorthand for pairing order parameter with form factor
$\Delta_{i\sigma;j\sigma} = \Phi^{\bfQ}_{ij}$ [and $(\Delta_{i\up;j\dn} + \Delta_{i\dn;j\up})/2 = \Phi^{\bfQ}_{ij}$], with $\Phi^{\bfQ}$ defined in Table~\ref{tab:phases-nn}.
The symmetry operations are defined as follows.
$E$: identity operation;
$C_{3}^{\rmP}$ and $C_{3}^{\rmA}$: $120^{\circ}$ rotations about the center of a plaquette and about a vertex in sublattice $A$;
$C_{2}'$: rotation about the $y$ axis that passes through a vertex;
$\sigma_{\rmh}$: mirror operation about $x$-$y$ plane;
$t_{\bfa_{1}}$: translation by a lattice constant $\bfa_{1}$.
}
\label{tab:app-op-transform}
{\renewcommand{\arraystretch}{1.8}
\begin{ruledtabular}
\begin{tabular}{rrrrrr}
$E$        & $C_3^{\rmP}$ &  $C_3^{\rmA}$ & $C_2'$ & $\sigma_{\rmh}$ &$t_{\bfa_1}$ \\
\hline
  $\Delta_{\up\up}^{K}$ 
    & $\omega^2 \; \Delta_{\up\up}^{K}$ 
    & $            \Delta_{\up\up}^{K}$ 
    & $            \Delta_{\dn\dn}^{K'}$
    & $           -\Delta_{\up\up}^{K}$ 
    & $\omega^2 \; \Delta_{\up\up}^{K}$
\\
  $d_{z}^{K}$
    & $            d_{z}^{K}$ 
    & $\omega   \; d_{z}^{K}$
    & $           -d_{z}^{K'}$ 
    & $            d_{z}^{K}$
    & $\omega^2 \; d_{z}^{K}$ 
\\
  $\Delta_{\dn\dn}^{K}$
    & $\omega   \; \Delta_{\dn\dn}^{K}$
    & $\omega^2 \; \Delta_{\dn\dn}^{K}$
    & $            \Delta_{\dn\dn}^{K'}$
    & $           -\Delta_{\dn\dn}^{K}$
    & $\omega^2 \; \Delta_{\dn\dn}^{K}$
\end{tabular}
\end{ruledtabular}
}
\end{table}

\para{}
In most of the superconducting phases we have identified in our calculation, the order parameters show non-trivial spatial and spin structures.
Typically, the unconventional nature of a superconducting phase (e.g. $p$ wave, $d$ wave, etc.) can be better understood by studying the transformation of the order parameters under point-group-symmetry operation, and symmetry classifying them according to the irreducible representations~\cite{sigrist-rmp-1991}.
This essentially captures the angular momentum of a Cooper pair.
In addition to angular momentum, in our case, the pairing order parameters are allowed to have nonzero momenta $K$ or $K'$, i.e., the order parameters may transform non-trivially under lattice translations as well.
For example, in the topological helical SC phase, the spin-triplet order parameters $\Delta_{\up\up}$ and $\Delta_{\dn\dn}$ have different momenta
($\Delta_{i\up;j\up} \sim \Phi^{K}_{ij}$ and $\Delta_{i\dn;j\dn} \sim \Phi^{K'}_{ij}$), and therefore transform differently under lattice translation:
\begin{align}
    \Delta_{i\up;j\up} &\rightarrow \Delta_{i\up;j\dn} e^{i K \cdot 2\bfa_1}, \text{ and } &
    \Delta_{i\dn;j\dn} &\rightarrow \Delta_{i\dn;j\dn} e^{-i K \cdot 2\bfa_1}
\end{align}
under $\bfr \rightarrow \bfr + \bfa_1$.
In the $p$-Kekule state, on the other hand, the pairing order parameter breaks translation symmetry with its amplitude modulation;
such symmetry-breaking order parameter can be understood as a superposition of two different irreducible representations.
Table~\ref{tab:app-op-transform} summarizes the transformation of the spin-triplet order parameters with momentum $K$.

\medskip

\section{Direct first-order transitions to topological superconductivity}
\label{sec:firstorder}

\begin{figure*}%
\centering%
\subfigure[\label{fig:1stOrder-a}]{\includegraphics[width=2.36in]{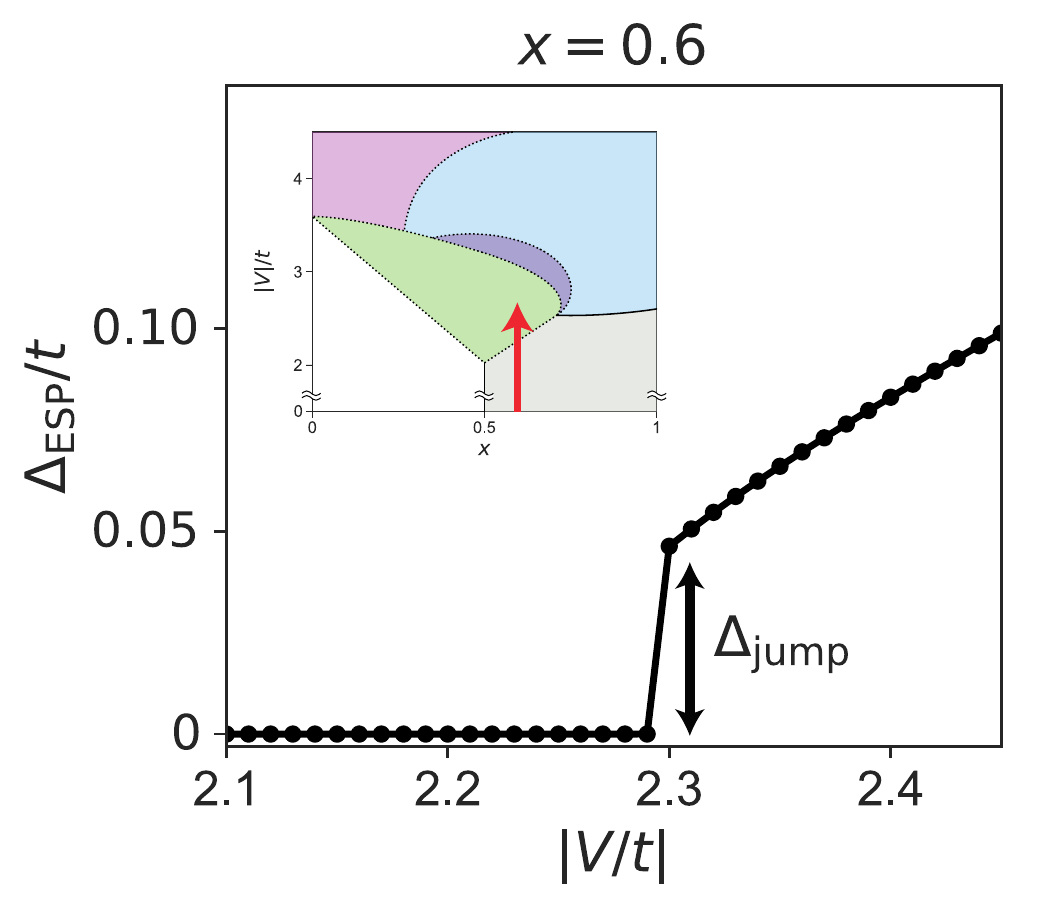}}
\subfigure[\label{fig:1stOrder-b}]{\includegraphics[width=2.34in]{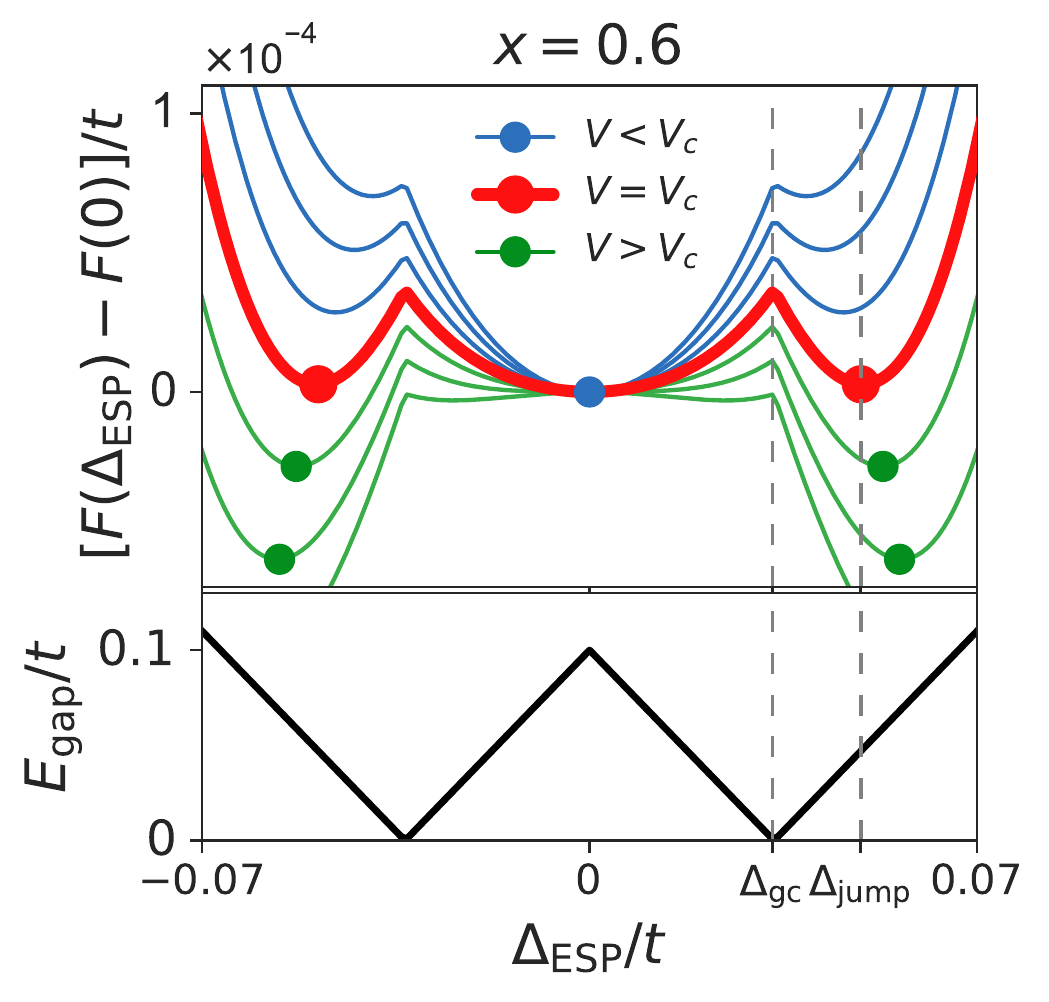}}
\subfigure[\label{fig:1stOrder-c}]{\includegraphics[width=2.28in]{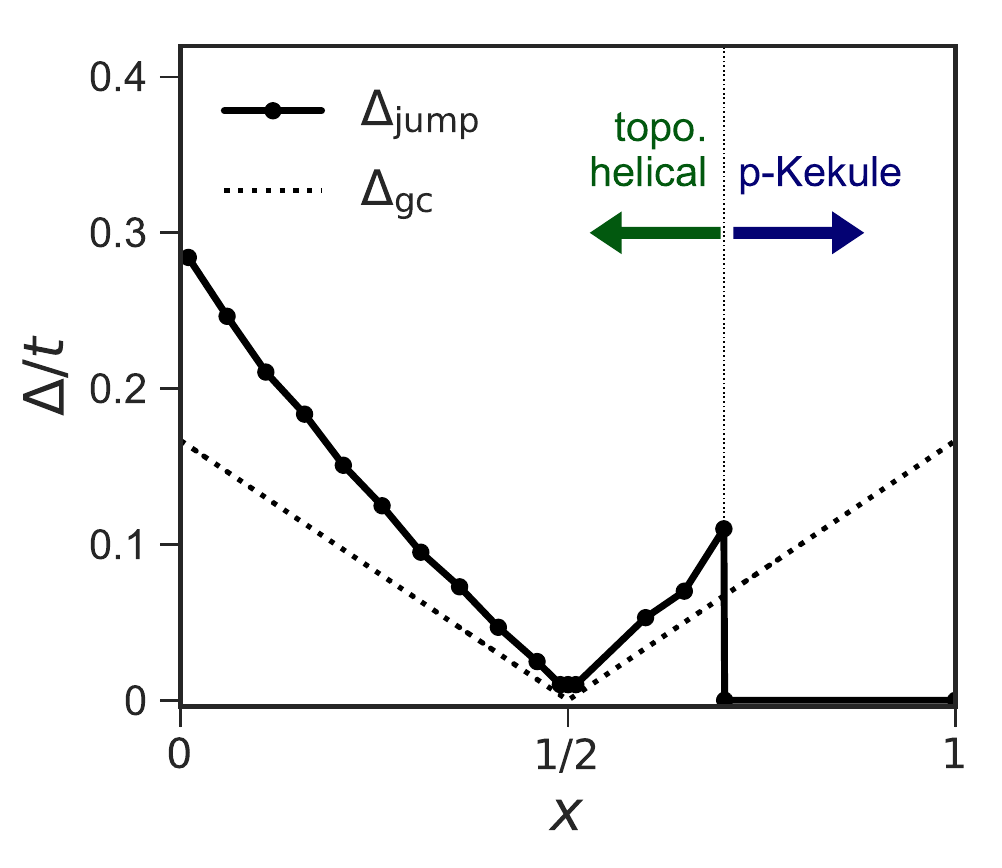}}
\caption{\label{fig:1stOrder}%
Direct (first-order) transition from an insulator to a topological superconductor.
\subref{fig:1stOrder-a}
Evolution of the strength of the pairing potential $\Delta_{\mathrm{ESP}}\equiv \tilde{\Delta}_{K,\up\up}=\tilde{\Delta}_{K',\dn\dn}$ corresponding to the helical SC, as a function of interaction strength $V$, evaluated at $x=0.6$ (i.e. along the vertical cut indicated by the red arrow on the phase diagram shown in the inset),
which clearly shows a first-order jump in the order parameter $\Delta_{\mathrm{jump}} = 0.043t$ at $V=V_{c}=2.3t$.
\subref{fig:1stOrder-b}
Upper panel: free energy as a function of pairing potential $\Delta_{\mathrm{ESP}}$ for a range of $V$ across $V_c$ in steps of $0.01t$.
The markers indicate the locations of global minima.
The first-order transition is highlighted in red, and involves a discontinuous jump in $\Delta_{\mathrm{ESP}}$.
Lower panel: the jump in the order parameter (marked by $\Delta_{\mathrm{jump}}$) exceeds the value of $\Delta_{\mathrm{ESP}}$ required to close and reopen the gap in the Bogoliubov quasiparticles spectrum (marked by $\Delta_{\mathrm{gc}}$), if $\Delta_{\mathrm{ESP}}$ was to increase continuously from $0$.
The topological index $\tilde{\nu}$ changes across the gap closing.
This establishes a direct discontinuous transition from an insulator to a topological superconductor at $x=0.6$.
\subref{fig:1stOrder-c}
The first-order jump in $\Delta_{\mathrm{ESP}}$ exceeds $\Delta_{\mathrm{gc}}$ across the range of $x$ where we find a transition to helical superconductor.
Beyond $x \sim 0.7 $, there is a continuous transition into the topologically trivial $p$-Kekule SC. 
}
\end{figure*}

\para{}
In the phase diagram for nearest-neighbor density-density attraction, for a large range of $x$, we find a direct transition from insulator to topological helical superconductor. 
Following the arguments of the main text (see ``onsite attraction $U$''), we know this is not allowed for a continuous insulator-to-superconductor transition.
Is there really a first-order transition from an insulator to a topological superconductor?
For $x=0.6$, we show the discontinuous jump in $\Delta_{\mathrm{ESP}} \equiv \tilde{\Delta}_{K,\up\up}=\tilde{\Delta}_{K',\dn\dn}=0$ in Fig.~\ref{fig:1stOrder-a}. 
The first-order transition is clearly seen in the free-energy landscape: the insulating state ($\Delta_{\mathrm{ESP}}=0$) remains a local minimum [Fig.~\ref{fig:1stOrder-b}] even as the global minimum shifts to finite $\Delta_{\mathrm{ESP}}$.
Since the topological index $\tilde{\nu}$ cannot be changed by an adiabatic change of parameters, we expect, as we smoothly increase $\Delta_{\mathrm{ESP}}$ from $0$, that the gap in the Bogoliubov quasiparticle spectrum will close at some $\Delta_{\mathrm{gc}}$, as in Fig.~\ref{fig:1stOrder-b}, after which $\tilde{\nu}$ changes. Figure~\ref{fig:1stOrder-c} shows that the jump in $\Delta$ at the first-order SC transition is always greater than $\Delta_{\text{gc}}$, which establishes a direct transition from insulator to the helical topological superconductor across the range of $x$ where there is a insulator to ESP superconductor. We have checked that all other order parameters aside from $\tilde{\Delta}_{K,\up\up},\tilde{\Delta}_{K',\dn\dn}$ are zero near this transition.

\section{Spatial and spin structure of trivial \texorpdfstring{$\calT$}{T}-breaking SC}
\label{sec:tbreakingphase}

\begin{figure}%
\centering%
\includegraphics[height=2.8in]{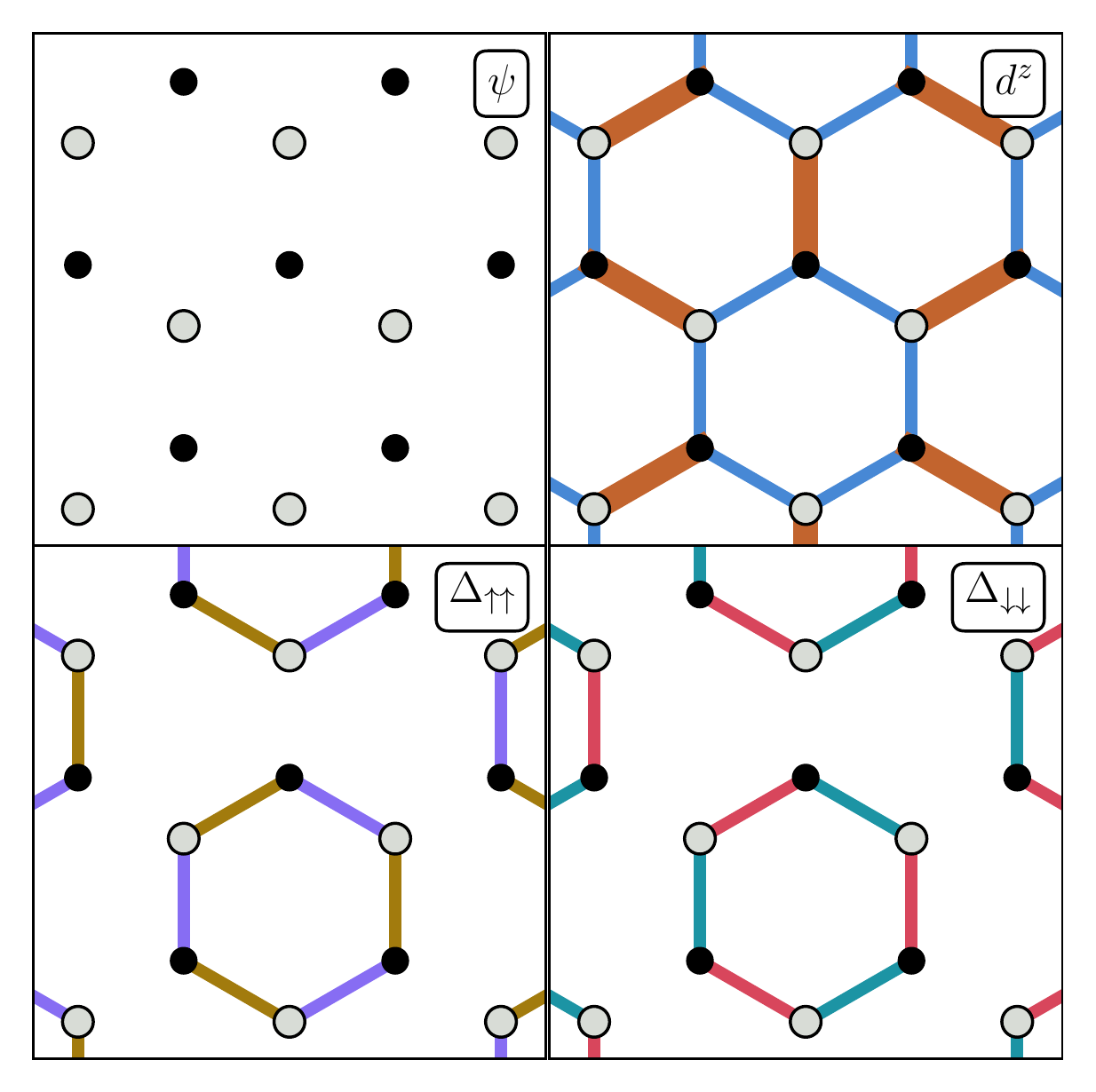}%
\includegraphics[height=1.4in]{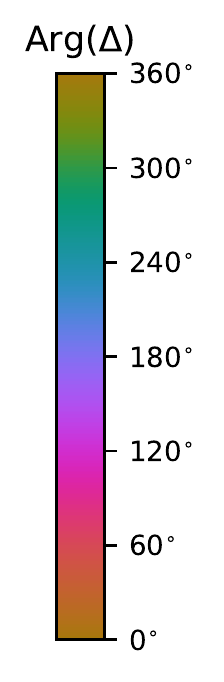}
\caption{\label{fig:nn-pairing-pattern-tbreaking}%
Pairing order parameter in the trivial $\calT$-breaking triplet SC phase [pink region in Fig.~\ref{fig:pd-nn}].
The thickness and the color of a bond indicates the magnitude and the phase angle of the order parameter on the bond.
}
\end{figure}

\para{}
The trivial $\calT$-breaking superconducting phase, which we have not discussed much in the main text, contains both $p$-Kekule and $s$-Kekule patterns, in different spin channels.
Figure~\ref{fig:nn-pairing-pattern-tbreaking} shows the spatial structures of all of the spin components of the pairing order parameter:
singlet component is zero, and only the three spin-triplet components are nonzero.
The $d^z$ component ($S=1, S_z=0$) has $s$-Kekule pattern,
while $\Delta_{\up\up}$  ($S=1, S_z=1$) and $\Delta_{\dn\dn}$ ($S=1, S_z=-1$) show $p$-Kekule pattern.


\begin{thebibliography}{52}%
\makeatletter
\providecommand \@ifxundefined [1]{%
 \@ifx{#1\undefined}
}%
\providecommand \@ifnum [1]{%
 \ifnum #1\expandafter \@firstoftwo
 \else \expandafter \@secondoftwo
 \fi
}%
\providecommand \@ifx [1]{%
 \ifx #1\expandafter \@firstoftwo
 \else \expandafter \@secondoftwo
 \fi
}%
\providecommand \natexlab [1]{#1}%
\providecommand \enquote  [1]{``#1''}%
\providecommand \bibnamefont  [1]{#1}%
\providecommand \bibfnamefont [1]{#1}%
\providecommand \citenamefont [1]{#1}%
\providecommand \href@noop [0]{\@secondoftwo}%
\providecommand \href [0]{\begingroup \@sanitize@url \@href}%
\providecommand \@href[1]{\@@startlink{#1}\@@href}%
\providecommand \@@href[1]{\endgroup#1\@@endlink}%
\providecommand \@sanitize@url [0]{\catcode `\\12\catcode `\$12\catcode
  `\&12\catcode `\#12\catcode `\^12\catcode `\_12\catcode `\%12\relax}%
\providecommand \@@startlink[1]{}%
\providecommand \@@endlink[0]{}%
\providecommand \url  [0]{\begingroup\@sanitize@url \@url }%
\providecommand \@url [1]{\endgroup\@href {#1}{\urlprefix }}%
\providecommand \urlprefix  [0]{URL }%
\providecommand \Eprint [0]{\href }%
\providecommand \doibase [0]{http://dx.doi.org/}%
\providecommand \selectlanguage [0]{\@gobble}%
\providecommand \bibinfo  [0]{\@secondoftwo}%
\providecommand \bibfield  [0]{\@secondoftwo}%
\providecommand \translation [1]{[#1]}%
\providecommand \BibitemOpen [0]{}%
\providecommand \bibitemStop [0]{}%
\providecommand \bibitemNoStop [0]{.\EOS\space}%
\providecommand \EOS [0]{\spacefactor3000\relax}%
\providecommand \BibitemShut  [1]{\csname bibitem#1\endcsname}%
\let\auto@bib@innerbib\@empty
\bibitem [{\citenamefont {Qi}\ and\ \citenamefont {Zhang}(2011)}]{qi-rmp-2011}%
  \BibitemOpen
  \bibfield  {author} {\bibinfo {author} {\bibfnamefont {Xiao-Liang}\
  \bibnamefont {Qi}}\ and\ \bibinfo {author} {\bibfnamefont {Shou-Cheng}\
  \bibnamefont {Zhang}},\ }\bibfield  {title} {\enquote {\bibinfo {title}
  {Topological insulators and superconductors},}\ }\href {\doibase
  10.1103/RevModPhys.83.1057} {\bibfield  {journal} {\bibinfo  {journal} {Rev.
  Mod. Phys.}\ }\textbf {\bibinfo {volume} {83}},\ \bibinfo {pages} {1057}
  (\bibinfo {year} {2011})}\BibitemShut {NoStop}%
\bibitem [{\citenamefont {Sato}\ and\ \citenamefont
  {Ando}(2017)}]{sato-rpp-2017}%
  \BibitemOpen
  \bibfield  {author} {\bibinfo {author} {\bibfnamefont {Masatoshi}\
  \bibnamefont {Sato}}\ and\ \bibinfo {author} {\bibfnamefont {Yoichi}\
  \bibnamefont {Ando}},\ }\bibfield  {title} {\enquote {\bibinfo {title}
  {Topological superconductors: a review},}\ }\href {\doibase
  10.1088/1361-6633/aa6ac7} {\bibfield  {journal} {\bibinfo  {journal} {Rep.
  Prog. Phys.}\ }\textbf {\bibinfo {volume} {80}},\ \bibinfo {pages} {076501}
  (\bibinfo {year} {2017})}\BibitemShut {NoStop}%
\bibitem [{\citenamefont {Mourik}\ \emph {et~al.}(2012)\citenamefont {Mourik},
  \citenamefont {Zuo}, \citenamefont {Frolov}, \citenamefont {Plissard},
  \citenamefont {Bakkers},\ and\ \citenamefont {Kouwenhoven}}]{mourik-s-2012}%
  \BibitemOpen
  \bibfield  {author} {\bibinfo {author} {\bibfnamefont {V.}~\bibnamefont
  {Mourik}}, \bibinfo {author} {\bibfnamefont {K.}~\bibnamefont {Zuo}},
  \bibinfo {author} {\bibfnamefont {S.~M.}\ \bibnamefont {Frolov}}, \bibinfo
  {author} {\bibfnamefont {S.~R.}\ \bibnamefont {Plissard}}, \bibinfo {author}
  {\bibfnamefont {E.~P. A.~M.}\ \bibnamefont {Bakkers}}, \ and\ \bibinfo
  {author} {\bibfnamefont {L.~P.}\ \bibnamefont {Kouwenhoven}},\ }\bibfield
  {title} {\enquote {\bibinfo {title} {Signatures of {Majorana} fermions in
  hybrid superconductor-semiconductor nanowire devices},}\ }\href {\doibase
  10.1126/science.1222360} {\bibfield  {journal} {\bibinfo  {journal}
  {Science}\ }\textbf {\bibinfo {volume} {336}},\ \bibinfo {pages} {1003}
  (\bibinfo {year} {2012})}\BibitemShut {NoStop}%
\bibitem [{\citenamefont {Nadj-Perge}\ \emph {et~al.}(2014)\citenamefont
  {Nadj-Perge}, \citenamefont {Drozdov}, \citenamefont {Li}, \citenamefont
  {Chen}, \citenamefont {Jeon}, \citenamefont {Seo}, \citenamefont {MacDonald},
  \citenamefont {Bernevig},\ and\ \citenamefont {Yazdani}}]{nadj-perge-s-2014}%
  \BibitemOpen
  \bibfield  {author} {\bibinfo {author} {\bibfnamefont {Stevan}\ \bibnamefont
  {Nadj-Perge}}, \bibinfo {author} {\bibfnamefont {Ilya~K.}\ \bibnamefont
  {Drozdov}}, \bibinfo {author} {\bibfnamefont {Jian}\ \bibnamefont {Li}},
  \bibinfo {author} {\bibfnamefont {Hua}\ \bibnamefont {Chen}}, \bibinfo
  {author} {\bibfnamefont {Sangjun}\ \bibnamefont {Jeon}}, \bibinfo {author}
  {\bibfnamefont {Jungpil}\ \bibnamefont {Seo}}, \bibinfo {author}
  {\bibfnamefont {Allan~H.}\ \bibnamefont {MacDonald}}, \bibinfo {author}
  {\bibfnamefont {B.~Andrei}\ \bibnamefont {Bernevig}}, \ and\ \bibinfo
  {author} {\bibfnamefont {Ali}\ \bibnamefont {Yazdani}},\ }\bibfield  {title}
  {\enquote {\bibinfo {title} {Observation of {Majorana} fermions in
  ferromagnetic atomic chains on a superconductor},}\ }\href {\doibase
  10.1126/science.1259327} {\bibfield  {journal} {\bibinfo  {journal}
  {Science}\ }\textbf {\bibinfo {volume} {346}},\ \bibinfo {pages} {602}
  (\bibinfo {year} {2014})}\BibitemShut {NoStop}%
\bibitem [{\citenamefont {Lian}\ \emph {et~al.}(2018)\citenamefont {Lian},
  \citenamefont {Sun}, \citenamefont {Vaezi}, \citenamefont {Qi},\ and\
  \citenamefont {Zhang}}]{lian-pnas-2018}%
  \BibitemOpen
  \bibfield  {author} {\bibinfo {author} {\bibfnamefont {Biao}\ \bibnamefont
  {Lian}}, \bibinfo {author} {\bibfnamefont {Xiao-Qi}\ \bibnamefont {Sun}},
  \bibinfo {author} {\bibfnamefont {Abolhassan}\ \bibnamefont {Vaezi}},
  \bibinfo {author} {\bibfnamefont {Xiao-Liang}\ \bibnamefont {Qi}}, \ and\
  \bibinfo {author} {\bibfnamefont {Shou-Cheng}\ \bibnamefont {Zhang}},\
  }\bibfield  {title} {\enquote {\bibinfo {title} {Topological quantum
  computation based on chiral {Majorana} fermions},}\ }\href {\doibase
  10.1073/pnas.1810003115} {\bibfield  {journal} {\bibinfo  {journal} {Proc.
  Natl. Acad. Sci. U.S.A.}\ }\textbf {\bibinfo {volume} {115}},\ \bibinfo
  {pages} {10938--10942} (\bibinfo {year} {2018})}\BibitemShut {NoStop}%
\bibitem [{\citenamefont {He}\ \emph {et~al.}(2017)\citenamefont {He},
  \citenamefont {Pan}, \citenamefont {Stern}, \citenamefont {Burks},
  \citenamefont {Che}, \citenamefont {Yin}, \citenamefont {Wang}, \citenamefont
  {Lian}, \citenamefont {Zhou}, \citenamefont {Choi}, \citenamefont {Murata},
  \citenamefont {Kou}, \citenamefont {Chen}, \citenamefont {Nie}, \citenamefont
  {Shao}, \citenamefont {Fan}, \citenamefont {Zhang}, \citenamefont {Liu},
  \citenamefont {Xia},\ and\ \citenamefont {Wang}}]{he-science-2017}%
  \BibitemOpen
  \bibfield  {author} {\bibinfo {author} {\bibfnamefont {Qing~Lin}\
  \bibnamefont {He}}, \bibinfo {author} {\bibfnamefont {Lei}\ \bibnamefont
  {Pan}}, \bibinfo {author} {\bibfnamefont {Alexander~L.}\ \bibnamefont
  {Stern}}, \bibinfo {author} {\bibfnamefont {Edward~C.}\ \bibnamefont
  {Burks}}, \bibinfo {author} {\bibfnamefont {Xiaoyu}\ \bibnamefont {Che}},
  \bibinfo {author} {\bibfnamefont {Gen}\ \bibnamefont {Yin}}, \bibinfo
  {author} {\bibfnamefont {Jing}\ \bibnamefont {Wang}}, \bibinfo {author}
  {\bibfnamefont {Biao}\ \bibnamefont {Lian}}, \bibinfo {author} {\bibfnamefont
  {Quan}\ \bibnamefont {Zhou}}, \bibinfo {author} {\bibfnamefont {Eun~Sang}\
  \bibnamefont {Choi}}, \bibinfo {author} {\bibfnamefont {Koichi}\ \bibnamefont
  {Murata}}, \bibinfo {author} {\bibfnamefont {Xufeng}\ \bibnamefont {Kou}},
  \bibinfo {author} {\bibfnamefont {Zhijie}\ \bibnamefont {Chen}}, \bibinfo
  {author} {\bibfnamefont {Tianxiao}\ \bibnamefont {Nie}}, \bibinfo {author}
  {\bibfnamefont {Qiming}\ \bibnamefont {Shao}}, \bibinfo {author}
  {\bibfnamefont {Yabin}\ \bibnamefont {Fan}}, \bibinfo {author} {\bibfnamefont
  {Shou-Cheng}\ \bibnamefont {Zhang}}, \bibinfo {author} {\bibfnamefont {Kai}\
  \bibnamefont {Liu}}, \bibinfo {author} {\bibfnamefont {Jing}\ \bibnamefont
  {Xia}}, \ and\ \bibinfo {author} {\bibfnamefont {Kang~L.}\ \bibnamefont
  {Wang}},\ }\bibfield  {title} {\enquote {\bibinfo {title} {Chiral
  {{Majorana}} fermion modes in a quantum anomalous {{Hall}}
  insulator\textendash{}superconductor structure},}\ }\href {\doibase
  10.1126/science.aag2792} {\bibfield  {journal} {\bibinfo  {journal}
  {Science}\ }\textbf {\bibinfo {volume} {357}},\ \bibinfo {pages} {294--299}
  (\bibinfo {year} {2017})}\BibitemShut {NoStop}%
\bibitem [{\citenamefont {M\'enard}\ \emph {et~al.}(2017)\citenamefont
  {M\'enard}, \citenamefont {Guissart}, \citenamefont {Brun}, \citenamefont
  {Leriche}, \citenamefont {Trif}, \citenamefont {Debontridder}, \citenamefont
  {Demaille}, \citenamefont {Roditchev}, \citenamefont {Simon},\ and\
  \citenamefont {Cren}}]{menard-nc-2017}%
  \BibitemOpen
  \bibfield  {author} {\bibinfo {author} {\bibfnamefont {Gerbold~C.}\
  \bibnamefont {M\'enard}}, \bibinfo {author} {\bibfnamefont {S\'ebastien}\
  \bibnamefont {Guissart}}, \bibinfo {author} {\bibfnamefont {Christophe}\
  \bibnamefont {Brun}}, \bibinfo {author} {\bibfnamefont {Rapha\"el~T.}\
  \bibnamefont {Leriche}}, \bibinfo {author} {\bibfnamefont {Mircea}\
  \bibnamefont {Trif}}, \bibinfo {author} {\bibfnamefont {Fran{\c c}ois}\
  \bibnamefont {Debontridder}}, \bibinfo {author} {\bibfnamefont {Dominique}\
  \bibnamefont {Demaille}}, \bibinfo {author} {\bibfnamefont {Dimitri}\
  \bibnamefont {Roditchev}}, \bibinfo {author} {\bibfnamefont {Pascal}\
  \bibnamefont {Simon}}, \ and\ \bibinfo {author} {\bibfnamefont {Tristan}\
  \bibnamefont {Cren}},\ }\bibfield  {title} {\enquote {\bibinfo {title}
  {Two-dimensional topological superconductivity in {Pb/Co/Si}(111)},}\ }\href
  {\doibase 10.1038/s41467-017-02192-x} {\bibfield  {journal} {\bibinfo
  {journal} {Nat. Comm.}\ }\textbf {\bibinfo {volume} {8}},\ \bibinfo {pages}
  {2040} (\bibinfo {year} {2017})}\BibitemShut {NoStop}%
\bibitem [{\citenamefont {{Palacio-Morales}}\ \emph {et~al.}(2018)\citenamefont
  {{Palacio-Morales}}, \citenamefont {Mascot}, \citenamefont {Cocklin},
  \citenamefont {Kim}, \citenamefont {Rachel}, \citenamefont {Morr},\ and\
  \citenamefont {Wiesendanger}}]{palacio-morales-a-2018}%
  \BibitemOpen
  \bibfield  {author} {\bibinfo {author} {\bibfnamefont {Alexandra}\
  \bibnamefont {{Palacio-Morales}}}, \bibinfo {author} {\bibfnamefont {Eric}\
  \bibnamefont {Mascot}}, \bibinfo {author} {\bibfnamefont {Sagen}\
  \bibnamefont {Cocklin}}, \bibinfo {author} {\bibfnamefont {Howon}\
  \bibnamefont {Kim}}, \bibinfo {author} {\bibfnamefont {Stephan}\ \bibnamefont
  {Rachel}}, \bibinfo {author} {\bibfnamefont {Dirk~K.}\ \bibnamefont {Morr}},
  \ and\ \bibinfo {author} {\bibfnamefont {Roland}\ \bibnamefont
  {Wiesendanger}},\ }\href@noop {} {\bibfield  {journal} {\bibinfo  {journal}
  {arXiv:1809.04503 [cond-mat]}\ } (\bibinfo {year} {2018})},\ \Eprint
  {http://arxiv.org/abs/1809.04503} {arXiv:1809.04503 [cond-mat]} \BibitemShut
  {NoStop}%
\bibitem [{\citenamefont {Yin}\ \emph {et~al.}(2015)\citenamefont {Yin},
  \citenamefont {Wu}, \citenamefont {Wang}, \citenamefont {Ye}, \citenamefont
  {Gong}, \citenamefont {Hou}, \citenamefont {Shan}, \citenamefont {Li},
  \citenamefont {Liang}, \citenamefont {Wu}, \citenamefont {Li}, \citenamefont
  {Ting}, \citenamefont {Wang}, \citenamefont {Hu}, \citenamefont {Hor},
  \citenamefont {Ding},\ and\ \citenamefont {Pan}}]{yin-np-2015}%
  \BibitemOpen
  \bibfield  {author} {\bibinfo {author} {\bibfnamefont {J.-X.}\ \bibnamefont
  {Yin}}, \bibinfo {author} {\bibfnamefont {Zheng}\ \bibnamefont {Wu}},
  \bibinfo {author} {\bibfnamefont {J.-H.}\ \bibnamefont {Wang}}, \bibinfo
  {author} {\bibfnamefont {Z.-Y.}\ \bibnamefont {Ye}}, \bibinfo {author}
  {\bibfnamefont {Jing}\ \bibnamefont {Gong}}, \bibinfo {author} {\bibfnamefont
  {X.-Y.}\ \bibnamefont {Hou}}, \bibinfo {author} {\bibfnamefont {Lei}\
  \bibnamefont {Shan}}, \bibinfo {author} {\bibfnamefont {Ang}\ \bibnamefont
  {Li}}, \bibinfo {author} {\bibfnamefont {X.-J.}\ \bibnamefont {Liang}},
  \bibinfo {author} {\bibfnamefont {X.-X.}\ \bibnamefont {Wu}}, \bibinfo
  {author} {\bibfnamefont {Jian}\ \bibnamefont {Li}}, \bibinfo {author}
  {\bibfnamefont {C.-S.}\ \bibnamefont {Ting}}, \bibinfo {author}
  {\bibfnamefont {Z.-Q.}\ \bibnamefont {Wang}}, \bibinfo {author}
  {\bibfnamefont {J.-P.}\ \bibnamefont {Hu}}, \bibinfo {author} {\bibfnamefont
  {P.-H.}\ \bibnamefont {Hor}}, \bibinfo {author} {\bibfnamefont
  {H.}~\bibnamefont {Ding}}, \ and\ \bibinfo {author} {\bibfnamefont {S.~H.}\
  \bibnamefont {Pan}},\ }\bibfield  {title} {\enquote {\bibinfo {title}
  {Observation of a robust zero-energy bound state in iron-based superconductor
  {Fe}({Te},{Se})},}\ }\href {\doibase 10.1038/nphys3371} {\bibfield  {journal}
  {\bibinfo  {journal} {Nat. Phys.}\ }\textbf {\bibinfo {volume} {11}},\
  \bibinfo {pages} {543} (\bibinfo {year} {2015})}\BibitemShut {NoStop}%
\bibitem [{\citenamefont {Fang}\ \emph {et~al.}(2015)\citenamefont {Fang},
  \citenamefont {Kuate~Defo}, \citenamefont {Shirodkar}, \citenamefont {Lieu},
  \citenamefont {Tritsaris},\ and\ \citenamefont {Kaxiras}}]{fang-prb-2015}%
  \BibitemOpen
  \bibfield  {author} {\bibinfo {author} {\bibfnamefont {Shiang}\ \bibnamefont
  {Fang}}, \bibinfo {author} {\bibfnamefont {Rodrick}\ \bibnamefont
  {Kuate~Defo}}, \bibinfo {author} {\bibfnamefont {Sharmila~N.}\ \bibnamefont
  {Shirodkar}}, \bibinfo {author} {\bibfnamefont {Simon}\ \bibnamefont {Lieu}},
  \bibinfo {author} {\bibfnamefont {Georgios~A.}\ \bibnamefont {Tritsaris}}, \
  and\ \bibinfo {author} {\bibfnamefont {Efthimios}\ \bibnamefont {Kaxiras}},\
  }\bibfield  {title} {\enquote {\bibinfo {title} {\textit{Ab initio}
  tight-binding hamiltonian for transition metal dichalcogenides},}\ }\href
  {\doibase 10.1103/PhysRevB.92.205108} {\bibfield  {journal} {\bibinfo
  {journal} {Phys. Rev. B}\ }\textbf {\bibinfo {volume} {92}},\ \bibinfo
  {pages} {205108} (\bibinfo {year} {2015})}\BibitemShut {NoStop}%
\bibitem [{\citenamefont {Xiao}\ \emph {et~al.}(2012)\citenamefont {Xiao},
  \citenamefont {Liu}, \citenamefont {Feng}, \citenamefont {Xu},\ and\
  \citenamefont {Yao}}]{xiao-prl-2012}%
  \BibitemOpen
  \bibfield  {author} {\bibinfo {author} {\bibfnamefont {Di}~\bibnamefont
  {Xiao}}, \bibinfo {author} {\bibfnamefont {Gui-Bin}\ \bibnamefont {Liu}},
  \bibinfo {author} {\bibfnamefont {Wanxiang}\ \bibnamefont {Feng}}, \bibinfo
  {author} {\bibfnamefont {Xiaodong}\ \bibnamefont {Xu}}, \ and\ \bibinfo
  {author} {\bibfnamefont {Wang}\ \bibnamefont {Yao}},\ }\bibfield  {title}
  {\enquote {\bibinfo {title} {Coupled spin and valley physics in monolayers of
  $\mathrm{MoS_{2}}$ and other group-{VI} dichalcogenides},}\ }\href {\doibase
  10.1103/PhysRevLett.108.196802} {\bibfield  {journal} {\bibinfo  {journal}
  {Phys. Rev. Lett.}\ }\textbf {\bibinfo {volume} {108}},\ \bibinfo {pages}
  {196802} (\bibinfo {year} {2012})}\BibitemShut {NoStop}%
\bibitem [{\citenamefont {Ye}\ \emph {et~al.}(2012)\citenamefont {Ye},
  \citenamefont {Zhang}, \citenamefont {Akashi}, \citenamefont {Bahramy},
  \citenamefont {Arita},\ and\ \citenamefont {Iwasa}}]{ye-s-2012}%
  \BibitemOpen
  \bibfield  {author} {\bibinfo {author} {\bibfnamefont {J.~T.}\ \bibnamefont
  {Ye}}, \bibinfo {author} {\bibfnamefont {Y.~J.}\ \bibnamefont {Zhang}},
  \bibinfo {author} {\bibfnamefont {R.}~\bibnamefont {Akashi}}, \bibinfo
  {author} {\bibfnamefont {M.~S.}\ \bibnamefont {Bahramy}}, \bibinfo {author}
  {\bibfnamefont {R.}~\bibnamefont {Arita}}, \ and\ \bibinfo {author}
  {\bibfnamefont {Y.}~\bibnamefont {Iwasa}},\ }\bibfield  {title} {\enquote
  {\bibinfo {title} {Superconducting dome in a gate-tuned band insulator},}\
  }\href {\doibase 10.1126/science.1228006} {\bibfield  {journal} {\bibinfo
  {journal} {Science}\ }\textbf {\bibinfo {volume} {338}},\ \bibinfo {pages}
  {1193} (\bibinfo {year} {2012})}\BibitemShut {NoStop}%
\bibitem [{\citenamefont {Lu}\ \emph {et~al.}(2015)\citenamefont {Lu},
  \citenamefont {Zheliuk}, \citenamefont {Leermakers}, \citenamefont {Yuan},
  \citenamefont {Zeitler}, \citenamefont {Law},\ and\ \citenamefont
  {Ye}}]{lu-s-2015}%
  \BibitemOpen
  \bibfield  {author} {\bibinfo {author} {\bibfnamefont {J.~M.}\ \bibnamefont
  {Lu}}, \bibinfo {author} {\bibfnamefont {O.}~\bibnamefont {Zheliuk}},
  \bibinfo {author} {\bibfnamefont {I.}~\bibnamefont {Leermakers}}, \bibinfo
  {author} {\bibfnamefont {N.~F.~Q.}\ \bibnamefont {Yuan}}, \bibinfo {author}
  {\bibfnamefont {U.}~\bibnamefont {Zeitler}}, \bibinfo {author} {\bibfnamefont
  {K.~T.}\ \bibnamefont {Law}}, \ and\ \bibinfo {author} {\bibfnamefont
  {J.~T.}\ \bibnamefont {Ye}},\ }\bibfield  {title} {\enquote {\bibinfo {title}
  {Evidence for two-dimensional {Ising} superconductivity in gated
  $\mathrm{MoS_2}$},}\ }\href {\doibase 10.1126/science.aab2277} {\bibfield
  {journal} {\bibinfo  {journal} {Science}\ }\textbf {\bibinfo {volume}
  {350}},\ \bibinfo {pages} {1353} (\bibinfo {year} {2015})}\BibitemShut
  {NoStop}%
\bibitem [{\citenamefont {Lu}\ \emph {et~al.}(2018)\citenamefont {Lu},
  \citenamefont {Zheliuk}, \citenamefont {Chen}, \citenamefont {Leermakers},
  \citenamefont {Hussey}, \citenamefont {Zeitler},\ and\ \citenamefont
  {Ye}}]{lu-pnas-2018}%
  \BibitemOpen
  \bibfield  {author} {\bibinfo {author} {\bibfnamefont {Jianming}\
  \bibnamefont {Lu}}, \bibinfo {author} {\bibfnamefont {Oleksandr}\
  \bibnamefont {Zheliuk}}, \bibinfo {author} {\bibfnamefont {Qihong}\
  \bibnamefont {Chen}}, \bibinfo {author} {\bibfnamefont {Inge}\ \bibnamefont
  {Leermakers}}, \bibinfo {author} {\bibfnamefont {Nigel~E.}\ \bibnamefont
  {Hussey}}, \bibinfo {author} {\bibfnamefont {Uli}\ \bibnamefont {Zeitler}}, \
  and\ \bibinfo {author} {\bibfnamefont {Jianting}\ \bibnamefont {Ye}},\
  }\bibfield  {title} {\enquote {\bibinfo {title} {Full superconducting dome of
  strong {Ising} protection in gated monolayer $\mathrm{WS_2}$},}\ }\href
  {\doibase 10.1073/pnas.1716781115} {\bibfield  {journal} {\bibinfo  {journal}
  {Proc. Nat. Acad. Sci.}\ }\textbf {\bibinfo {volume} {115}},\ \bibinfo
  {pages} {3551} (\bibinfo {year} {2018})}\BibitemShut {NoStop}%
\bibitem [{\citenamefont {Fei}\ \emph {et~al.}(2017)\citenamefont {Fei},
  \citenamefont {Palomaki}, \citenamefont {Wu}, \citenamefont {Zhao},
  \citenamefont {Cai}, \citenamefont {Sun}, \citenamefont {Nguyen},
  \citenamefont {Finney}, \citenamefont {Xu},\ and\ \citenamefont
  {Cobden}}]{fei-np-2017}%
  \BibitemOpen
  \bibfield  {author} {\bibinfo {author} {\bibfnamefont {Zaiyao}\ \bibnamefont
  {Fei}}, \bibinfo {author} {\bibfnamefont {Tauno}\ \bibnamefont {Palomaki}},
  \bibinfo {author} {\bibfnamefont {Sanfeng}\ \bibnamefont {Wu}}, \bibinfo
  {author} {\bibfnamefont {Wenjin}\ \bibnamefont {Zhao}}, \bibinfo {author}
  {\bibfnamefont {Xinghan}\ \bibnamefont {Cai}}, \bibinfo {author}
  {\bibfnamefont {Bosong}\ \bibnamefont {Sun}}, \bibinfo {author}
  {\bibfnamefont {Paul}\ \bibnamefont {Nguyen}}, \bibinfo {author}
  {\bibfnamefont {Joseph}\ \bibnamefont {Finney}}, \bibinfo {author}
  {\bibfnamefont {Xiaodong}\ \bibnamefont {Xu}}, \ and\ \bibinfo {author}
  {\bibfnamefont {David~H.}\ \bibnamefont {Cobden}},\ }\bibfield  {title}
  {\enquote {\bibinfo {title} {Edge conduction in monolayer
  $\mathrm{WTe_2}$},}\ }\href {\doibase 10.1038/nphys4091} {\bibfield
  {journal} {\bibinfo  {journal} {Nat. Phys.}\ }\textbf {\bibinfo {volume}
  {13}},\ \bibinfo {pages} {677} (\bibinfo {year} {2017})}\BibitemShut
  {NoStop}%
\bibitem [{\citenamefont {Kang}\ \emph {et~al.}(2015)\citenamefont {Kang},
  \citenamefont {Zhou}, \citenamefont {Yi}, \citenamefont {Yang}, \citenamefont
  {Guo}, \citenamefont {Shi}, \citenamefont {Zhang}, \citenamefont {Wang},
  \citenamefont {Zhang}, \citenamefont {Jiang}, \citenamefont {Li},
  \citenamefont {Yang}, \citenamefont {Wu}, \citenamefont {Zhang},
  \citenamefont {Sun},\ and\ \citenamefont {Zhao}}]{kang-nc-2015}%
  \BibitemOpen
  \bibfield  {author} {\bibinfo {author} {\bibfnamefont {Defen}\ \bibnamefont
  {Kang}}, \bibinfo {author} {\bibfnamefont {Yazhou}\ \bibnamefont {Zhou}},
  \bibinfo {author} {\bibfnamefont {Wei}\ \bibnamefont {Yi}}, \bibinfo {author}
  {\bibfnamefont {Chongli}\ \bibnamefont {Yang}}, \bibinfo {author}
  {\bibfnamefont {Jing}\ \bibnamefont {Guo}}, \bibinfo {author} {\bibfnamefont
  {Youguo}\ \bibnamefont {Shi}}, \bibinfo {author} {\bibfnamefont {Shan}\
  \bibnamefont {Zhang}}, \bibinfo {author} {\bibfnamefont {Zhe}\ \bibnamefont
  {Wang}}, \bibinfo {author} {\bibfnamefont {Chao}\ \bibnamefont {Zhang}},
  \bibinfo {author} {\bibfnamefont {Sheng}\ \bibnamefont {Jiang}}, \bibinfo
  {author} {\bibfnamefont {Aiguo}\ \bibnamefont {Li}}, \bibinfo {author}
  {\bibfnamefont {Ke}~\bibnamefont {Yang}}, \bibinfo {author} {\bibfnamefont
  {Qi}~\bibnamefont {Wu}}, \bibinfo {author} {\bibfnamefont {Guangming}\
  \bibnamefont {Zhang}}, \bibinfo {author} {\bibfnamefont {Liling}\
  \bibnamefont {Sun}}, \ and\ \bibinfo {author} {\bibfnamefont {Zhongxian}\
  \bibnamefont {Zhao}},\ }\bibfield  {title} {\enquote {\bibinfo {title}
  {Superconductivity emerging from a suppressed large magnetoresistant state in
  tungsten ditelluride},}\ }\href {\doibase 10.1038/ncomms8804} {\bibfield
  {journal} {\bibinfo  {journal} {Nat. Commun.}\ }\textbf {\bibinfo {volume}
  {6}},\ \bibinfo {pages} {7804} (\bibinfo {year} {2015})}\BibitemShut
  {NoStop}%
\bibitem [{\citenamefont {Pan}\ \emph {et~al.}(2015)\citenamefont {Pan},
  \citenamefont {Chen}, \citenamefont {Liu}, \citenamefont {Feng},
  \citenamefont {Wei}, \citenamefont {Zhou}, \citenamefont {Chi}, \citenamefont
  {Pi}, \citenamefont {Yen}, \citenamefont {Song}, \citenamefont {Wan},
  \citenamefont {Yang}, \citenamefont {Wang}, \citenamefont {Wang},\ and\
  \citenamefont {Zhang}}]{pan-nc-2015}%
  \BibitemOpen
  \bibfield  {author} {\bibinfo {author} {\bibfnamefont {Xing-Chen}\
  \bibnamefont {Pan}}, \bibinfo {author} {\bibfnamefont {Xuliang}\ \bibnamefont
  {Chen}}, \bibinfo {author} {\bibfnamefont {Huimei}\ \bibnamefont {Liu}},
  \bibinfo {author} {\bibfnamefont {Yanqing}\ \bibnamefont {Feng}}, \bibinfo
  {author} {\bibfnamefont {Zhongxia}\ \bibnamefont {Wei}}, \bibinfo {author}
  {\bibfnamefont {Yonghui}\ \bibnamefont {Zhou}}, \bibinfo {author}
  {\bibfnamefont {Zhenhua}\ \bibnamefont {Chi}}, \bibinfo {author}
  {\bibfnamefont {Li}~\bibnamefont {Pi}}, \bibinfo {author} {\bibfnamefont
  {Fei}\ \bibnamefont {Yen}}, \bibinfo {author} {\bibfnamefont {Fengqi}\
  \bibnamefont {Song}}, \bibinfo {author} {\bibfnamefont {Xiangang}\
  \bibnamefont {Wan}}, \bibinfo {author} {\bibfnamefont {Zhaorong}\
  \bibnamefont {Yang}}, \bibinfo {author} {\bibfnamefont {Baigeng}\
  \bibnamefont {Wang}}, \bibinfo {author} {\bibfnamefont {Guanghou}\
  \bibnamefont {Wang}}, \ and\ \bibinfo {author} {\bibfnamefont {Yuheng}\
  \bibnamefont {Zhang}},\ }\bibfield  {title} {\enquote {\bibinfo {title}
  {Pressure-driven dome-shaped superconductivity and electronic structural
  evolution in tungsten ditelluride},}\ }\href {\doibase 10.1038/ncomms8805}
  {\bibfield  {journal} {\bibinfo  {journal} {Nat. Commun.}\ }\textbf {\bibinfo
  {volume} {6}},\ \bibinfo {pages} {7805} (\bibinfo {year} {2015})}\BibitemShut
  {NoStop}%
\bibitem [{\citenamefont {Sajadi}\ \emph {et~al.}(2018)\citenamefont {Sajadi},
  \citenamefont {Palomaki}, \citenamefont {Fei}, \citenamefont {Zhao},
  \citenamefont {Bement}, \citenamefont {Olsen}, \citenamefont {Luescher},
  \citenamefont {Xu}, \citenamefont {Folk},\ and\ \citenamefont
  {Cobden}}]{sajadi-s-2018}%
  \BibitemOpen
  \bibfield  {author} {\bibinfo {author} {\bibfnamefont {Ebrahim}\ \bibnamefont
  {Sajadi}}, \bibinfo {author} {\bibfnamefont {Tauno}\ \bibnamefont
  {Palomaki}}, \bibinfo {author} {\bibfnamefont {Zaiyao}\ \bibnamefont {Fei}},
  \bibinfo {author} {\bibfnamefont {Wenjin}\ \bibnamefont {Zhao}}, \bibinfo
  {author} {\bibfnamefont {Philip}\ \bibnamefont {Bement}}, \bibinfo {author}
  {\bibfnamefont {Christian}\ \bibnamefont {Olsen}}, \bibinfo {author}
  {\bibfnamefont {Silvia}\ \bibnamefont {Luescher}}, \bibinfo {author}
  {\bibfnamefont {Xiaodong}\ \bibnamefont {Xu}}, \bibinfo {author}
  {\bibfnamefont {Joshua~A.}\ \bibnamefont {Folk}}, \ and\ \bibinfo {author}
  {\bibfnamefont {David~H.}\ \bibnamefont {Cobden}},\ }\bibfield  {title}
  {\enquote {\bibinfo {title} {Gate-induced superconductivity in a monolayer
  topological insulator},}\ }\href {\doibase 10.1126/science.aar4426}
  {\bibfield  {journal} {\bibinfo  {journal} {Science}\ }\textbf {\bibinfo
  {volume} {362}},\ \bibinfo {pages} {922} (\bibinfo {year}
  {2018})}\BibitemShut {NoStop}%
\bibitem [{\citenamefont {Fatemi}\ \emph {et~al.}(2018)\citenamefont {Fatemi},
  \citenamefont {Wu}, \citenamefont {Cao}, \citenamefont {Bretheau},
  \citenamefont {Gibson}, \citenamefont {Watanabe}, \citenamefont {Taniguchi},
  \citenamefont {Cava},\ and\ \citenamefont {Jarillo-Herrero}}]{fatemi-s-2018}%
  \BibitemOpen
  \bibfield  {author} {\bibinfo {author} {\bibfnamefont {Valla}\ \bibnamefont
  {Fatemi}}, \bibinfo {author} {\bibfnamefont {Sanfeng}\ \bibnamefont {Wu}},
  \bibinfo {author} {\bibfnamefont {Yuan}\ \bibnamefont {Cao}}, \bibinfo
  {author} {\bibfnamefont {Landry}\ \bibnamefont {Bretheau}}, \bibinfo {author}
  {\bibfnamefont {Quinn~D.}\ \bibnamefont {Gibson}}, \bibinfo {author}
  {\bibfnamefont {Kenji}\ \bibnamefont {Watanabe}}, \bibinfo {author}
  {\bibfnamefont {Takashi}\ \bibnamefont {Taniguchi}}, \bibinfo {author}
  {\bibfnamefont {Robert~J.}\ \bibnamefont {Cava}}, \ and\ \bibinfo {author}
  {\bibfnamefont {Pablo}\ \bibnamefont {Jarillo-Herrero}},\ }\bibfield  {title}
  {\enquote {\bibinfo {title} {Electrically tunable low-density
  superconductivity in a monolayer topological insulator},}\ }\href {\doibase
  10.1126/science.aar4642} {\bibfield  {journal} {\bibinfo  {journal}
  {Science}\ }\textbf {\bibinfo {volume} {362}},\ \bibinfo {pages} {926}
  (\bibinfo {year} {2018})}\BibitemShut {NoStop}%
\bibitem [{\citenamefont {Cao}\ \emph {et~al.}(2018)\citenamefont {Cao},
  \citenamefont {Fatemi}, \citenamefont {Fang}, \citenamefont {Watanabe},
  \citenamefont {Taniguchi}, \citenamefont {Kaxiras},\ and\ \citenamefont
  {Jarillo-Herrero}}]{cao-n-2018}%
  \BibitemOpen
  \bibfield  {author} {\bibinfo {author} {\bibfnamefont {Yuan}\ \bibnamefont
  {Cao}}, \bibinfo {author} {\bibfnamefont {Valla}\ \bibnamefont {Fatemi}},
  \bibinfo {author} {\bibfnamefont {Shiang}\ \bibnamefont {Fang}}, \bibinfo
  {author} {\bibfnamefont {Kenji}\ \bibnamefont {Watanabe}}, \bibinfo {author}
  {\bibfnamefont {Takashi}\ \bibnamefont {Taniguchi}}, \bibinfo {author}
  {\bibfnamefont {Efthimios}\ \bibnamefont {Kaxiras}}, \ and\ \bibinfo {author}
  {\bibfnamefont {Pablo}\ \bibnamefont {Jarillo-Herrero}},\ }\bibfield  {title}
  {\enquote {\bibinfo {title} {Unconventional superconductivity in magic-angle
  graphene superlattices},}\ }\href {\doibase 10.1038/nature26160} {\bibfield
  {journal} {\bibinfo  {journal} {Nature}\ }\textbf {\bibinfo {volume} {556}},\
  \bibinfo {pages} {43} (\bibinfo {year} {2018})}\BibitemShut {NoStop}%
\bibitem [{\citenamefont {Po}\ \emph {et~al.}(2018)\citenamefont {Po},
  \citenamefont {Zou}, \citenamefont {Vishwanath},\ and\ \citenamefont
  {Senthil}}]{po-prx-2018}%
  \BibitemOpen
  \bibfield  {author} {\bibinfo {author} {\bibfnamefont {Hoi~Chun}\
  \bibnamefont {Po}}, \bibinfo {author} {\bibfnamefont {Liujun}\ \bibnamefont
  {Zou}}, \bibinfo {author} {\bibfnamefont {Ashvin}\ \bibnamefont
  {Vishwanath}}, \ and\ \bibinfo {author} {\bibfnamefont {T.}~\bibnamefont
  {Senthil}},\ }\bibfield  {title} {\enquote {\bibinfo {title} {Origin of
  {Mott} insulating behavior and superconductivity in twisted bilayer
  graphene},}\ }\href {\doibase 10.1103/PhysRevX.8.031089} {\bibfield
  {journal} {\bibinfo  {journal} {Phys. Rev. X}\ }\textbf {\bibinfo {volume}
  {8}},\ \bibinfo {pages} {031089} (\bibinfo {year} {2018})}\BibitemShut
  {NoStop}%
\bibitem [{\citenamefont {Yuan}\ and\ \citenamefont
  {Fu}(2018)}]{yuan-prb-2018}%
  \BibitemOpen
  \bibfield  {author} {\bibinfo {author} {\bibfnamefont {Noah F.~Q.}\
  \bibnamefont {Yuan}}\ and\ \bibinfo {author} {\bibfnamefont {Liang}\
  \bibnamefont {Fu}},\ }\bibfield  {title} {\enquote {\bibinfo {title} {Model
  for the metal-insulator transition in graphene superlattices and beyond},}\
  }\href {\doibase 10.1103/PhysRevB.98.045103} {\bibfield  {journal} {\bibinfo
  {journal} {Phys. Rev. B}\ }\textbf {\bibinfo {volume} {98}},\ \bibinfo
  {pages} {045103} (\bibinfo {year} {2018})}\BibitemShut {NoStop}%
\bibitem [{\citenamefont {Kang}\ and\ \citenamefont
  {Vafek}(2018)}]{kang-prx-2018}%
  \BibitemOpen
  \bibfield  {author} {\bibinfo {author} {\bibfnamefont {Jian}\ \bibnamefont
  {Kang}}\ and\ \bibinfo {author} {\bibfnamefont {Oskar}\ \bibnamefont
  {Vafek}},\ }\bibfield  {title} {\enquote {\bibinfo {title} {Symmetry,
  maximally localized {Wannier} states, and a low-energy model for twisted
  bilayer graphene narrow bands},}\ }\href {\doibase 10.1103/PhysRevX.8.031088}
  {\bibfield  {journal} {\bibinfo  {journal} {Phys. Rev. X}\ }\textbf {\bibinfo
  {volume} {8}},\ \bibinfo {pages} {031088} (\bibinfo {year}
  {2018})}\BibitemShut {NoStop}%
\bibitem [{\citenamefont {Marrazzo}\ \emph {et~al.}(2018)\citenamefont
  {Marrazzo}, \citenamefont {Gibertini}, \citenamefont {Campi}, \citenamefont
  {Mounet},\ and\ \citenamefont {Marzari}}]{marrazzo-prl-2018}%
  \BibitemOpen
  \bibfield  {author} {\bibinfo {author} {\bibfnamefont {Antimo}\ \bibnamefont
  {Marrazzo}}, \bibinfo {author} {\bibfnamefont {Marco}\ \bibnamefont
  {Gibertini}}, \bibinfo {author} {\bibfnamefont {Davide}\ \bibnamefont
  {Campi}}, \bibinfo {author} {\bibfnamefont {Nicolas}\ \bibnamefont {Mounet}},
  \ and\ \bibinfo {author} {\bibfnamefont {Nicola}\ \bibnamefont {Marzari}},\
  }\bibfield  {title} {\enquote {\bibinfo {title} {Prediction of a large-gap
  and switchable {Kane}-{Mele} quantum spin {Hall} insulator},}\ }\href
  {\doibase 10.1103/PhysRevLett.120.117701} {\bibfield  {journal} {\bibinfo
  {journal} {Phys. Rev. Lett.}\ }\textbf {\bibinfo {volume} {120}},\ \bibinfo
  {pages} {117701} (\bibinfo {year} {2018})}\BibitemShut {NoStop}%
\bibitem [{\citenamefont {Kandrai}\ \emph {et~al.}(2019)\citenamefont
  {Kandrai}, \citenamefont {Kukucska}, \citenamefont {Vancs\'o}, \citenamefont
  {Koltai}, \citenamefont {Baranka}, \citenamefont {Horv\'ath}, \citenamefont
  {Hoffmann}, \citenamefont {Vymazalov\'a}, \citenamefont {Tapaszt\'o},\ and\
  \citenamefont {{Nemes-Incze}}}]{kandra-a-2019}%
  \BibitemOpen
  \bibfield  {author} {\bibinfo {author} {\bibfnamefont {Konr\'ad}\
  \bibnamefont {Kandrai}}, \bibinfo {author} {\bibfnamefont {Gerg{\H o}}\
  \bibnamefont {Kukucska}}, \bibinfo {author} {\bibfnamefont {P\'eter}\
  \bibnamefont {Vancs\'o}}, \bibinfo {author} {\bibfnamefont {J\'anos}\
  \bibnamefont {Koltai}}, \bibinfo {author} {\bibfnamefont {Gy\"orgy}\
  \bibnamefont {Baranka}}, \bibinfo {author} {\bibfnamefont {Zsolt~E.}\
  \bibnamefont {Horv\'ath}}, \bibinfo {author} {\bibfnamefont {\'Akos}\
  \bibnamefont {Hoffmann}}, \bibinfo {author} {\bibfnamefont {Anna}\
  \bibnamefont {Vymazalov\'a}}, \bibinfo {author} {\bibfnamefont {Levente}\
  \bibnamefont {Tapaszt\'o}}, \ and\ \bibinfo {author} {\bibfnamefont
  {P\'eter}\ \bibnamefont {{Nemes-Incze}}},\ }\bibfield  {title} {\enquote
  {\bibinfo {title} {Evidence for room temperature quantum spin {Hall} state in
  the layered mineral {Jacutingaite}},}\ }\href@noop {} {\bibfield  {journal}
  {\bibinfo  {journal} {arXiv:1903.02458 [cond-mat]}\ } (\bibinfo {year}
  {2019})},\ \Eprint {http://arxiv.org/abs/1903.02458} {arXiv:1903.02458
  [cond-mat]} \BibitemShut {NoStop}%
\bibitem [{\citenamefont {Wu}\ \emph {et~al.}(2018)\citenamefont {Wu},
  \citenamefont {Fink}, \citenamefont {Hanke}, \citenamefont {Thomale},\ and\
  \citenamefont {Di~Sante}}]{wu-a-2018}%
  \BibitemOpen
  \bibfield  {author} {\bibinfo {author} {\bibfnamefont {Xianxin}\ \bibnamefont
  {Wu}}, \bibinfo {author} {\bibfnamefont {Mario}\ \bibnamefont {Fink}},
  \bibinfo {author} {\bibfnamefont {Werner}\ \bibnamefont {Hanke}}, \bibinfo
  {author} {\bibfnamefont {Ronny}\ \bibnamefont {Thomale}}, \ and\ \bibinfo
  {author} {\bibfnamefont {Domenico}\ \bibnamefont {Di~Sante}},\ }\bibfield
  {title} {\enquote {\bibinfo {title} {Unconventional superconductivity in a
  doped quantum spin {Hall} insulator},}\ }\href@noop {} {\bibfield  {journal}
  {\bibinfo  {journal} {arXiv:1811.01746 [cond-mat]}\ } (\bibinfo {year}
  {2018})},\ \Eprint {http://arxiv.org/abs/1811.01746} {arXiv:1811.01746
  [cond-mat]} \BibitemShut {NoStop}%
\bibitem [{\citenamefont {Kane}\ and\ \citenamefont
  {Mele}(2005)}]{kane-prl-2005}%
  \BibitemOpen
  \bibfield  {author} {\bibinfo {author} {\bibfnamefont {C.~L.}\ \bibnamefont
  {Kane}}\ and\ \bibinfo {author} {\bibfnamefont {E.~J.}\ \bibnamefont
  {Mele}},\ }\bibfield  {title} {\enquote {\bibinfo {title} {${Z_2}$
  topological order and the quantum spin {{Hall}} effect},}\ }\href {\doibase
  10.1103/PhysRevLett.95.146802} {\bibfield  {journal} {\bibinfo  {journal}
  {Phys. Rev. Lett.}\ }\textbf {\bibinfo {volume} {95}},\ \bibinfo {pages}
  {146802} (\bibinfo {year} {2005})}\BibitemShut {NoStop}%
\bibitem [{\citenamefont {Schnyder}\ \emph {et~al.}(2008)\citenamefont
  {Schnyder}, \citenamefont {Ryu}, \citenamefont {Furusaki},\ and\
  \citenamefont {Ludwig}}]{schnyder-prb-2008}%
  \BibitemOpen
  \bibfield  {author} {\bibinfo {author} {\bibfnamefont {Andreas~P.}\
  \bibnamefont {Schnyder}}, \bibinfo {author} {\bibfnamefont {Shinsei}\
  \bibnamefont {Ryu}}, \bibinfo {author} {\bibfnamefont {Akira}\ \bibnamefont
  {Furusaki}}, \ and\ \bibinfo {author} {\bibfnamefont {Andreas W.~W.}\
  \bibnamefont {Ludwig}},\ }\bibfield  {title} {\enquote {\bibinfo {title}
  {Classification of topological insulators and superconductors in three
  spatial dimensions},}\ }\href {\doibase 10.1103/PhysRevB.78.195125}
  {\bibfield  {journal} {\bibinfo  {journal} {Phys. Rev. B}\ }\textbf {\bibinfo
  {volume} {78}},\ \bibinfo {pages} {195125} (\bibinfo {year}
  {2008})}\BibitemShut {NoStop}%
\bibitem [{\citenamefont {Kitaev}(2009)}]{kitaev-acp-2009}%
  \BibitemOpen
  \bibfield  {author} {\bibinfo {author} {\bibfnamefont {Alexei}\ \bibnamefont
  {Kitaev}},\ }\bibfield  {title} {\enquote {\bibinfo {title} {Periodic table
  for topological insulators and superconductors},}\ }in\ \href {\doibase
  10.1063/1.3149495} {\emph {\bibinfo {booktitle} {Advances in Theoretical
  Physics: Landau Memorial Conference}}},\ \bibinfo {series} {AIP Conf. Proc.},
  Vol.\ \bibinfo {volume} {1134},\ \bibinfo {editor} {edited by\ \bibinfo
  {editor} {\bibfnamefont {V.}~\bibnamefont {Lebedev}}\ and\ \bibinfo {editor}
  {\bibfnamefont {M.}~\bibnamefont {Feigel'man}}}\ (\bibinfo {organization}
  {AIP},\ \bibinfo {address} {New York},\ \bibinfo {year} {2009})\ p.~\bibinfo
  {pages} {22}\BibitemShut {NoStop}%
\bibitem [{\citenamefont {Altland}\ and\ \citenamefont
  {Zirnbauer}(1997)}]{altland-prb-1997}%
  \BibitemOpen
  \bibfield  {author} {\bibinfo {author} {\bibfnamefont {Alexander}\
  \bibnamefont {Altland}}\ and\ \bibinfo {author} {\bibfnamefont {Martin~R.}\
  \bibnamefont {Zirnbauer}},\ }\bibfield  {title} {\enquote {\bibinfo {title}
  {Nonstandard symmetry classes in mesoscopic normal-superconducting hybrid
  structures},}\ }\href {\doibase 10.1103/PhysRevB.55.1142} {\bibfield
  {journal} {\bibinfo  {journal} {Phys. Rev. B}\ }\textbf {\bibinfo {volume}
  {55}},\ \bibinfo {pages} {1142} (\bibinfo {year} {1997})}\BibitemShut
  {NoStop}%
\bibitem [{\citenamefont {Sato}(2003)}]{sato-plb-2003}%
  \BibitemOpen
  \bibfield  {author} {\bibinfo {author} {\bibfnamefont {Masatoshi}\
  \bibnamefont {Sato}},\ }\bibfield  {title} {\enquote {\bibinfo {title}
  {Non-{Abelian} statistics of axion strings},}\ }\href {\doibase
  10.1016/j.physletb.2003.09.047} {\bibfield  {journal} {\bibinfo  {journal}
  {Phys. Lett. B}\ }\textbf {\bibinfo {volume} {575}},\ \bibinfo {pages} {126}
  (\bibinfo {year} {2003})}\BibitemShut {NoStop}%
\bibitem [{\citenamefont {Fu}\ and\ \citenamefont {Kane}(2006)}]{fu-prb-2006}%
  \BibitemOpen
  \bibfield  {author} {\bibinfo {author} {\bibfnamefont {Liang}\ \bibnamefont
  {Fu}}\ and\ \bibinfo {author} {\bibfnamefont {C.~L.}\ \bibnamefont {Kane}},\
  }\bibfield  {title} {\enquote {\bibinfo {title} {Time reversal polarization
  and a {$Z_2$} adiabatic spin pump},}\ }\href {\doibase
  10.1103/PhysRevB.74.195312} {\bibfield  {journal} {\bibinfo  {journal} {Phys.
  Rev. B}\ }\textbf {\bibinfo {volume} {74}},\ \bibinfo {pages} {195312}
  (\bibinfo {year} {2006})}\BibitemShut {NoStop}%
\bibitem [{\citenamefont {Roy}\ and\ \citenamefont
  {Herbut}(2010)}]{roy-prb-2010}%
  \BibitemOpen
  \bibfield  {author} {\bibinfo {author} {\bibfnamefont {Bitan}\ \bibnamefont
  {Roy}}\ and\ \bibinfo {author} {\bibfnamefont {Igor~F.}\ \bibnamefont
  {Herbut}},\ }\bibfield  {title} {\enquote {\bibinfo {title} {Unconventional
  superconductivity on honeycomb lattice: {Theory} of {Kekule} order
  parameter},}\ }\href {\doibase 10.1103/PhysRevB.82.035429} {\bibfield
  {journal} {\bibinfo  {journal} {Phys. Rev. B}\ }\textbf {\bibinfo {volume}
  {82}},\ \bibinfo {pages} {035429} (\bibinfo {year} {2010})}\BibitemShut
  {NoStop}%
\bibitem [{\citenamefont {Tsuchiya}\ \emph {et~al.}(2016)\citenamefont
  {Tsuchiya}, \citenamefont {Goryo}, \citenamefont {Arahata},\ and\
  \citenamefont {Sigrist}}]{tsuchiya-prb-2016}%
  \BibitemOpen
  \bibfield  {author} {\bibinfo {author} {\bibfnamefont {Shunji}\ \bibnamefont
  {Tsuchiya}}, \bibinfo {author} {\bibfnamefont {Jun}\ \bibnamefont {Goryo}},
  \bibinfo {author} {\bibfnamefont {Emiko}\ \bibnamefont {Arahata}}, \ and\
  \bibinfo {author} {\bibfnamefont {Manfred}\ \bibnamefont {Sigrist}},\
  }\bibfield  {title} {\enquote {\bibinfo {title} {Cooperon condensation and
  intravalley pairing states in honeycomb {Dirac} systems},}\ }\href {\doibase
  10.1103/PhysRevB.94.104508} {\bibfield  {journal} {\bibinfo  {journal} {Phys.
  Rev. B}\ }\textbf {\bibinfo {volume} {94}},\ \bibinfo {pages} {104508}
  (\bibinfo {year} {2016})}\BibitemShut {NoStop}%
\bibitem [{\citenamefont {Qi}\ \emph {et~al.}(2009)\citenamefont {Qi},
  \citenamefont {Hughes}, \citenamefont {Raghu},\ and\ \citenamefont
  {Zhang}}]{qi-prl-2009}%
  \BibitemOpen
  \bibfield  {author} {\bibinfo {author} {\bibfnamefont {Xiao-Liang}\
  \bibnamefont {Qi}}, \bibinfo {author} {\bibfnamefont {Taylor~L.}\
  \bibnamefont {Hughes}}, \bibinfo {author} {\bibfnamefont {S.}~\bibnamefont
  {Raghu}}, \ and\ \bibinfo {author} {\bibfnamefont {Shou-Cheng}\ \bibnamefont
  {Zhang}},\ }\bibfield  {title} {\enquote {\bibinfo {title}
  {Time-reversal-invariant topological superconductors and superfluids in two
  and three dimensions},}\ }\href {\doibase 10.1103/PhysRevLett.102.187001}
  {\bibfield  {journal} {\bibinfo  {journal} {Phys. Rev. Lett.}\ }\textbf
  {\bibinfo {volume} {102}},\ \bibinfo {pages} {187001} (\bibinfo {year}
  {2009})}\BibitemShut {NoStop}%
\bibitem [{\citenamefont {Haldane}(1988)}]{haldane-prl-1988}%
  \BibitemOpen
  \bibfield  {author} {\bibinfo {author} {\bibfnamefont {F.~D.~M.}\
  \bibnamefont {Haldane}},\ }\bibfield  {title} {\enquote {\bibinfo {title}
  {Model for a quantum hall effect without landau levels: Condensed-matter
  realization of the "parity anomaly"},}\ }\href {\doibase
  10.1103/PhysRevLett.61.2015} {\bibfield  {journal} {\bibinfo  {journal}
  {Phys. Rev. Lett.}\ }\textbf {\bibinfo {volume} {61}},\ \bibinfo {pages}
  {2015} (\bibinfo {year} {1988})}\BibitemShut {NoStop}%
\bibitem [{\citenamefont {Jotzu}\ \emph {et~al.}(2014)\citenamefont {Jotzu},
  \citenamefont {Messer}, \citenamefont {Desbuquois}, \citenamefont {Lebrat},
  \citenamefont {Uehlinger}, \citenamefont {Greif},\ and\ \citenamefont
  {Esslinger}}]{jotzu-n-2014}%
  \BibitemOpen
  \bibfield  {author} {\bibinfo {author} {\bibfnamefont {Gregor}\ \bibnamefont
  {Jotzu}}, \bibinfo {author} {\bibfnamefont {Michael}\ \bibnamefont {Messer}},
  \bibinfo {author} {\bibfnamefont {R{\'e}mi}\ \bibnamefont {Desbuquois}},
  \bibinfo {author} {\bibfnamefont {Martin}\ \bibnamefont {Lebrat}}, \bibinfo
  {author} {\bibfnamefont {Thomas}\ \bibnamefont {Uehlinger}}, \bibinfo
  {author} {\bibfnamefont {Daniel}\ \bibnamefont {Greif}}, \ and\ \bibinfo
  {author} {\bibfnamefont {Tilman}\ \bibnamefont {Esslinger}},\ }\bibfield
  {title} {\enquote {\bibinfo {title} {Experimental realization of the
  topological haldane model with ultracold fermions},}\ }\href {\doibase
  10.1038/nature13915} {\bibfield  {journal} {\bibinfo  {journal} {Nature}\
  }\textbf {\bibinfo {volume} {515}},\ \bibinfo {pages} {237} (\bibinfo {year}
  {2014})}\BibitemShut {NoStop}%
\bibitem [{\citenamefont {Anisimovas}\ \emph {et~al.}(2016)\citenamefont
  {Anisimovas}, \citenamefont {Ra\ifmmode \check{c}\else
  \v{c}\fi{}i\ifmmode~\bar{u}\else \={u}\fi{}nas}, \citenamefont {Str\"ater},
  \citenamefont {Eckardt}, \citenamefont {Spielman},\ and\ \citenamefont
  {Juzeli\ifmmode~\bar{u}\else \={u}\fi{}nas}}]{anisimovas-pra-2016}%
  \BibitemOpen
  \bibfield  {author} {\bibinfo {author} {\bibfnamefont {E.}~\bibnamefont
  {Anisimovas}}, \bibinfo {author} {\bibfnamefont {M.}~\bibnamefont {Ra\ifmmode
  \check{c}\else \v{c}\fi{}i\ifmmode~\bar{u}\else \={u}\fi{}nas}}, \bibinfo
  {author} {\bibfnamefont {C.}~\bibnamefont {Str\"ater}}, \bibinfo {author}
  {\bibfnamefont {A.}~\bibnamefont {Eckardt}}, \bibinfo {author} {\bibfnamefont
  {I.~B.}\ \bibnamefont {Spielman}}, \ and\ \bibinfo {author} {\bibfnamefont
  {G.}~\bibnamefont {Juzeli\ifmmode~\bar{u}\else \={u}\fi{}nas}},\ }\bibfield
  {title} {\enquote {\bibinfo {title} {Semisynthetic zigzag optical lattice for
  ultracold bosons},}\ }\href {\doibase 10.1103/PhysRevA.94.063632} {\bibfield
  {journal} {\bibinfo  {journal} {Phys. Rev. A}\ }\textbf {\bibinfo {volume}
  {94}},\ \bibinfo {pages} {063632} (\bibinfo {year} {2016})}\BibitemShut
  {NoStop}%
\bibitem [{\citenamefont {Fu}\ and\ \citenamefont {Kane}(2008)}]{fu-prl-2008}%
  \BibitemOpen
  \bibfield  {author} {\bibinfo {author} {\bibfnamefont {Liang}\ \bibnamefont
  {Fu}}\ and\ \bibinfo {author} {\bibfnamefont {C.~L.}\ \bibnamefont {Kane}},\
  }\bibfield  {title} {\enquote {\bibinfo {title} {Superconducting proximity
  effect and {{Majorana}} fermions at the surface of a topological
  insulator},}\ }\href {\doibase 10.1103/PhysRevLett.100.096407} {\bibfield
  {journal} {\bibinfo  {journal} {Phys. Rev. Lett.}\ }\textbf {\bibinfo
  {volume} {100}},\ \bibinfo {pages} {096407} (\bibinfo {year}
  {2008})}\BibitemShut {NoStop}%
\bibitem [{\citenamefont {Kitaev}(2001)}]{kitaev-pu-2001}%
  \BibitemOpen
  \bibfield  {author} {\bibinfo {author} {\bibfnamefont {A.~Yu.}\ \bibnamefont
  {Kitaev}},\ }\bibfield  {title} {\enquote {\bibinfo {title} {Unpaired
  {{Majorana}} fermions in quantum wires},}\ }\href {\doibase
  10.1070/1063-7869/44/10S/S29} {\bibfield  {journal} {\bibinfo  {journal}
  {Phys. Uspekhi}\ }\textbf {\bibinfo {volume} {44}},\ \bibinfo {pages} {131}
  (\bibinfo {year} {2001})}\BibitemShut {NoStop}%
\bibitem [{\citenamefont {Yuan}\ \emph {et~al.}(2014)\citenamefont {Yuan},
  \citenamefont {Mak},\ and\ \citenamefont {Law}}]{yuan-prl-2014}%
  \BibitemOpen
  \bibfield  {author} {\bibinfo {author} {\bibfnamefont {Noah F.~Q.}\
  \bibnamefont {Yuan}}, \bibinfo {author} {\bibfnamefont {Kin~Fai}\
  \bibnamefont {Mak}}, \ and\ \bibinfo {author} {\bibfnamefont {K.~T.}\
  \bibnamefont {Law}},\ }\bibfield  {title} {\enquote {\bibinfo {title}
  {Possible topological superconducting phases of $\mathrm{MoS_2}$},}\ }\href
  {\doibase 10.1103/PhysRevLett.113.097001} {\bibfield  {journal} {\bibinfo
  {journal} {Phys. Rev. Lett.}\ }\textbf {\bibinfo {volume} {113}},\ \bibinfo
  {pages} {097001} (\bibinfo {year} {2014})}\BibitemShut {NoStop}%
\bibitem [{\citenamefont {Hsu}\ \emph {et~al.}(2017)\citenamefont {Hsu},
  \citenamefont {Vaezi}, \citenamefont {Fischer},\ and\ \citenamefont
  {Kim}}]{hsu-nc-2017}%
  \BibitemOpen
  \bibfield  {author} {\bibinfo {author} {\bibfnamefont {Yi-Ting}\ \bibnamefont
  {Hsu}}, \bibinfo {author} {\bibfnamefont {Abolhassan}\ \bibnamefont {Vaezi}},
  \bibinfo {author} {\bibfnamefont {Mark~H.}\ \bibnamefont {Fischer}}, \ and\
  \bibinfo {author} {\bibfnamefont {Eun-Ah}\ \bibnamefont {Kim}},\ }\bibfield
  {title} {\enquote {\bibinfo {title} {Topological superconductivity in
  monolayer transition metal dichalcogenides},}\ }\href {\doibase
  10.1038/ncomms14985} {\bibfield  {journal} {\bibinfo  {journal} {Nat.
  Commun.}\ }\textbf {\bibinfo {volume} {8}},\ \bibinfo {pages} {14985}
  (\bibinfo {year} {2017})}\BibitemShut {NoStop}%
\bibitem [{\citenamefont {Kapitulnik}\ \emph {et~al.}(2009)\citenamefont
  {Kapitulnik}, \citenamefont {Xia}, \citenamefont {Schemm},\ and\
  \citenamefont {Palevski}}]{kapitulnik-njp-2009}%
  \BibitemOpen
  \bibfield  {author} {\bibinfo {author} {\bibfnamefont {Aharon}\ \bibnamefont
  {Kapitulnik}}, \bibinfo {author} {\bibfnamefont {Jing}\ \bibnamefont {Xia}},
  \bibinfo {author} {\bibfnamefont {Elizabeth}\ \bibnamefont {Schemm}}, \ and\
  \bibinfo {author} {\bibfnamefont {Alexander}\ \bibnamefont {Palevski}},\
  }\bibfield  {title} {\enquote {\bibinfo {title} {Polar {Kerr} effect as probe
  for time-reversal symmetry breaking in unconventional superconductors},}\
  }\href {\doibase 10.1088/1367-2630/11/5/055060} {\bibfield  {journal}
  {\bibinfo  {journal} {New J. Phys.}\ }\textbf {\bibinfo {volume} {11}},\
  \bibinfo {pages} {055060} (\bibinfo {year} {2009})}\BibitemShut {NoStop}%
\bibitem [{\citenamefont {Hamidian}\ \emph {et~al.}(2016)\citenamefont
  {Hamidian}, \citenamefont {Edkins}, \citenamefont {Joo}, \citenamefont
  {Kostin}, \citenamefont {Eisaki}, \citenamefont {Uchida}, \citenamefont
  {Lawler}, \citenamefont {Kim}, \citenamefont {Mackenzie}, \citenamefont
  {Fujita}, \citenamefont {Lee},\ and\ \citenamefont
  {Davis}}]{hamidian-n-2016}%
  \BibitemOpen
  \bibfield  {author} {\bibinfo {author} {\bibfnamefont {M.~H.}\ \bibnamefont
  {Hamidian}}, \bibinfo {author} {\bibfnamefont {S.~D.}\ \bibnamefont
  {Edkins}}, \bibinfo {author} {\bibfnamefont {Sang~Hyun}\ \bibnamefont {Joo}},
  \bibinfo {author} {\bibfnamefont {A.}~\bibnamefont {Kostin}}, \bibinfo
  {author} {\bibfnamefont {H.}~\bibnamefont {Eisaki}}, \bibinfo {author}
  {\bibfnamefont {S.}~\bibnamefont {Uchida}}, \bibinfo {author} {\bibfnamefont
  {M.~J.}\ \bibnamefont {Lawler}}, \bibinfo {author} {\bibfnamefont {E.-A.}\
  \bibnamefont {Kim}}, \bibinfo {author} {\bibfnamefont {A.~P.}\ \bibnamefont
  {Mackenzie}}, \bibinfo {author} {\bibfnamefont {K.}~\bibnamefont {Fujita}},
  \bibinfo {author} {\bibfnamefont {Jinho}\ \bibnamefont {Lee}}, \ and\
  \bibinfo {author} {\bibfnamefont {J.~C.~S{é}amus}\ \bibnamefont {Davis}},\
  }\bibfield  {title} {\enquote {\bibinfo {title} {Detection of a {Cooper}-pair
  density wave in {Bi$_2$Sr$_2$CaCu$_2$O$_{8+x}$}},}\ }\href {\doibase
  10.1038/nature17411} {\bibfield  {journal} {\bibinfo  {journal} {Nature}\
  }\textbf {\bibinfo {volume} {532}},\ \bibinfo {pages} {343} (\bibinfo {year}
  {2016})}\BibitemShut {NoStop}%
\bibitem [{\citenamefont {Fukui}\ \emph {et~al.}(2005)\citenamefont {Fukui},
  \citenamefont {Hatsugai},\ and\ \citenamefont {Suzuki}}]{fukui-jpsj-2005}%
  \BibitemOpen
  \bibfield  {author} {\bibinfo {author} {\bibfnamefont {Takahiro}\
  \bibnamefont {Fukui}}, \bibinfo {author} {\bibfnamefont {Yasuhiro}\
  \bibnamefont {Hatsugai}}, \ and\ \bibinfo {author} {\bibfnamefont {Hiroshi}\
  \bibnamefont {Suzuki}},\ }\bibfield  {title} {\enquote {\bibinfo {title}
  {Chern numbers in discretized {Brillouin} zone: Efficient method of computing
  (spin) {{Hall}} conductances},}\ }\href {\doibase 10.1143/JPSJ.74.1674}
  {\bibfield  {journal} {\bibinfo  {journal} {J. Phys. Soc. Jpn.}\ }\textbf
  {\bibinfo {volume} {74}},\ \bibinfo {pages} {1674} (\bibinfo {year}
  {2005})}\BibitemShut {NoStop}%
\bibitem [{\citenamefont {Fukui}\ and\ \citenamefont
  {Hatsugai}(2007)}]{fukui-prb-2007}%
  \BibitemOpen
  \bibfield  {author} {\bibinfo {author} {\bibfnamefont {Takahiro}\
  \bibnamefont {Fukui}}\ and\ \bibinfo {author} {\bibfnamefont {Yasuhiro}\
  \bibnamefont {Hatsugai}},\ }\bibfield  {title} {\enquote {\bibinfo {title}
  {Topological aspects of the quantum spin-{{Hall}} effect in graphene: ${Z_2}$
  topological order and spin {Chern} number},}\ }\href {\doibase
  10.1103/PhysRevB.75.121403} {\bibfield  {journal} {\bibinfo  {journal} {Phys.
  Rev. B}\ }\textbf {\bibinfo {volume} {75}},\ \bibinfo {pages} {121403}
  (\bibinfo {year} {2007})}\BibitemShut {NoStop}%
\bibitem [{\citenamefont {Thouless}\ \emph {et~al.}(1982)\citenamefont
  {Thouless}, \citenamefont {Kohmoto}, \citenamefont {Nightingale},\ and\
  \citenamefont {den Nijs}}]{thouless-prl-1982}%
  \BibitemOpen
  \bibfield  {author} {\bibinfo {author} {\bibfnamefont {D.~J.}\ \bibnamefont
  {Thouless}}, \bibinfo {author} {\bibfnamefont {M.}~\bibnamefont {Kohmoto}},
  \bibinfo {author} {\bibfnamefont {M.~P.}\ \bibnamefont {Nightingale}}, \ and\
  \bibinfo {author} {\bibfnamefont {M.}~\bibnamefont {den Nijs}},\ }\bibfield
  {title} {\enquote {\bibinfo {title} {Quantized {Hall} conductance in a
  two-dimensional periodic potential},}\ }\href {\doibase
  10.1103/PhysRevLett.49.405} {\bibfield  {journal} {\bibinfo  {journal} {Phys.
  Rev. Lett.}\ }\textbf {\bibinfo {volume} {49}},\ \bibinfo {pages} {405}
  (\bibinfo {year} {1982})}\BibitemShut {NoStop}%
\bibitem [{\citenamefont {Fu}\ and\ \citenamefont {Kane}(2007)}]{fu-prb-2007}%
  \BibitemOpen
  \bibfield  {author} {\bibinfo {author} {\bibfnamefont {Liang}\ \bibnamefont
  {Fu}}\ and\ \bibinfo {author} {\bibfnamefont {C.~L.}\ \bibnamefont {Kane}},\
  }\bibfield  {title} {\enquote {\bibinfo {title} {Topological insulators with
  inversion symmetry},}\ }\href {\doibase 10.1103/PhysRevB.76.045302}
  {\bibfield  {journal} {\bibinfo  {journal} {Phys. Rev. B}\ }\textbf {\bibinfo
  {volume} {76}},\ \bibinfo {pages} {045302} (\bibinfo {year}
  {2007})}\BibitemShut {NoStop}%
\bibitem [{\citenamefont {Kohmoto}(1985)}]{kohmoto-ap-1985}%
  \BibitemOpen
  \bibfield  {author} {\bibinfo {author} {\bibfnamefont {Mahito}\ \bibnamefont
  {Kohmoto}},\ }\bibfield  {title} {\enquote {\bibinfo {title} {Topological
  invariant and the quantization of the {Hall} conductance},}\ }\href {\doibase
  10.1016/0003-4916(85)90148-4} {\bibfield  {journal} {\bibinfo  {journal}
  {Ann. Phys.}\ }\textbf {\bibinfo {volume} {160}},\ \bibinfo {pages} {343}
  (\bibinfo {year} {1985})}\BibitemShut {NoStop}%
\bibitem [{\citenamefont {Larkin}(1964)}]{larkin-spj-1965}%
  \BibitemOpen
  \bibfield  {author} {\bibinfo {author} {\bibfnamefont {Yu.N.}\ \bibnamefont
  {Larkin}, \bibfnamefont {A.I.~Ovchinnikov}},\ }\bibfield  {title} {\enquote
  {\bibinfo {title} {Inhomogeneous state of superconductors},}\ }\href@noop {}
  {\bibfield  {journal} {\bibinfo  {journal} {Zh. Eksp. Teor. Fiz.}\ }\textbf
  {\bibinfo {volume} {47}},\ \bibinfo {pages} {1136} (\bibinfo {year}
  {1964})},\ \translation{Sov. Phys. JETP \textbf{20}, 762 (1965)}\BibitemShut
  {NoStop}%
\bibitem [{\citenamefont {Fulde}\ and\ \citenamefont
  {Ferrell}(1964)}]{fulde-pr-1964}%
  \BibitemOpen
  \bibfield  {author} {\bibinfo {author} {\bibfnamefont {Peter}\ \bibnamefont
  {Fulde}}\ and\ \bibinfo {author} {\bibfnamefont {Richard~A.}\ \bibnamefont
  {Ferrell}},\ }\bibfield  {title} {\enquote {\bibinfo {title}
  {Superconductivity in a strong spin-exchange field},}\ }\href {\doibase
  10.1103/PhysRev.135.A550} {\bibfield  {journal} {\bibinfo  {journal} {Phys.
  Rev.}\ }\textbf {\bibinfo {volume} {135}},\ \bibinfo {pages} {A550} (\bibinfo
  {year} {1964})}\BibitemShut {NoStop}%
\bibitem [{\citenamefont {Sigrist}\ and\ \citenamefont
  {Ueda}(1991)}]{sigrist-rmp-1991}%
  \BibitemOpen
  \bibfield  {author} {\bibinfo {author} {\bibfnamefont {Manfred}\ \bibnamefont
  {Sigrist}}\ and\ \bibinfo {author} {\bibfnamefont {Kazuo}\ \bibnamefont
  {Ueda}},\ }\bibfield  {title} {\enquote {\bibinfo {title} {Phenomenological
  theory of unconventional superconductivity},}\ }\href {\doibase
  10.1103/RevModPhys.63.239} {\bibfield  {journal} {\bibinfo  {journal} {Rev.
  Mod. Phys.}\ }\textbf {\bibinfo {volume} {63}},\ \bibinfo {pages} {239}
  (\bibinfo {year} {1991})}\BibitemShut {NoStop}%
\end{thebibliography}
\end{document}